\documentclass[USenglish,aps,pra,twocolumn,preprintnumbers,amsmath,amssymb,superscriptaddress]{revtex4-2}

\usepackage[dvipsnames]{xcolor}
\usepackage{graphicx}
\usepackage{dcolumn}
\usepackage{bm}
\usepackage[breaklinks=true]{hyperref}
\usepackage{mathrsfs}
\usepackage{soul}
\usepackage{extarrows}
\usepackage{physics}

\usepackage{babel}
\usepackage{isodate}

\hypersetup{
  colorlinks   = true, 
  urlcolor     = TealBlue, 
  linkcolor    = MidnightBlue, 
  citecolor   = BrickRed 
}

\begin{document}

\selectlanguage{USenglish}


\title{Bell correlations in a split two-mode-squeezed Bose-Einstein condensate}

\author{Jonas Kitzinger}
\thanks{These authors contributed equally}
\affiliation{State Key Laboratory of Precision Spectroscopy, School of Physical and Material Sciences, East China Normal University, Shanghai 200062, China} 
\affiliation{New York University Shanghai, 1555 Century Ave, Pudong New District, Shanghai 200122, China} 
\affiliation{Humboldt-Universit\"at zu Berlin, Institut f\"ur Physik, Newtonstra{\ss}e 15, 12489 Berlin, Germany} 

\author{Xin Meng}
\thanks{These authors contributed equally}
\affiliation{State Key Laboratory of Precision Spectroscopy, School of Physical and Material Sciences, East China Normal University, Shanghai 200062, China} 

\author{Matteo Fadel}
\affiliation{Department of Physics, University of Basel, Klingelbergstrasse 82, 4056 Basel, Switzerland}

\author{Valentin Ivannikov}
\affiliation{New York University Shanghai, 1555 Century Ave, Pudong New District, Shanghai 200122, China} 

\author{Kae Nemoto}
\affiliation{National Institute of Informatics, 2-1-2 Hitotsubashi, Chiyoda-ku, Tokyo 101-8430, Japan}

\author{William J. Munro}
\affiliation{NTT Research Center for Theoretical Quantum Physics, NTT Corporation, 3-1 Morinosato-Wakamiya, Atsugi, Kanagawa, 243-0198, Japan}
\affiliation{National Institute of Informatics, 2-1-2 Hitotsubashi, Chiyoda-ku, Tokyo 101-8430, Japan}

\author{Tim Byrnes}
\email{tim.byrnes@nyu.edu}
\affiliation{State Key Laboratory of Precision Spectroscopy, School of Physical and Material Sciences, East China Normal University, Shanghai 200062, China}
\affiliation{New York University Shanghai, 1555 Century Ave, Pudong New District, Shanghai 200122, China}  
\affiliation{NYU-ECNU Institute of Physics at NYU Shanghai, 3663 Zhongshan Road North, Shanghai 200062, China}
\affiliation{National Institute of Informatics, 2-1-2 Hitotsubashi, Chiyoda-ku, Tokyo 101-8430, Japan}
\affiliation{Department of Physics, New York University, New York, NY 10003, USA}


\begin{abstract}
We propose and analyze a protocol for observing a violation of the Clauser-Horne-Shimony-Holt (CHSH) Bell inequality using two spatially separated Bose-Einstein condensates (BECs). To prepare the Bell correlated state, spin-changing collisions are used to ﬁrst prepare a two-mode squeezed BEC. This is then split into two BECs by controlling the spatial wave function, e.g., by modifying the trapping potential. Finally, spin-changing collisions are also exploited locally, to compensate local squeezing terms. The correlators appearing in the inequality are evaluated using three different approaches. In the ﬁrst approach, correlators are estimated using normalized expectation values of number operators, in a similar way to evaluating continuous-variable Bell inequalities. An improvement to this approach is developed using the sign binning of total spin measurements, which allows for the construction of two-outcome measurements and violations of the CHSH inequality without auxiliary assumptions. Finally, we show a third approach where maximal violations of the CH inequality can be obtained by assigning zero values to local vacua outcomes under a no-enhancement assumption. The effect of loss and imperfect detection efﬁciency is investigated, and the observed violations are found to be robust to noise.
\end{abstract}

            
\maketitle

\section{Introduction}

Entanglement constitutes an essential resource for quantum technologies such as quantum computing,  quantum communication,  and  quantum metrology \cite{horodecki2009QuantumEntanglement}. From a fundamental point of view,  the challenge that entanglement poses to local-realistic theories was famously highlighted by Einstein, Podolsky, and Rosen (EPR) in their 1935 thought experiment \cite{einstein1935CanQuantumMechanicalDescription},  but it was not until Bell's seminal work \cite{bell1964EinsteinPodolskyRosen} that the EPR paradox was turned into an experimentally testable criterion.  Bell's work inspired extensive theoretical research into nonlocality \cite{brunner2014BellNonlocality}, as well as the experimental search for violations of Bell inequalities. In a quantum information context, Bell correlations are considered the most correlated class of quantum states in a hierarchy including quantum coherence, quantum discord, and EPR-steerability \cite{adesso2016MeasuresApplicationsQuantum, ma2019OperationalAdvantageBasisindependent,baumgratz2014QuantifyingCoherence,ollivier2001QuantumDiscordMeasure,
radhakrishnan2017QuantumCoherenceHeisenberg,radhakrishnan2020MultipartiteGeneralizationQuantum,wiseman2007SteeringEntanglementNonlocality}. 
Following the introduction of the Clauser-Horne-Shimony-Holt (CHSH) Bell inequality \cite{clauser1969ProposedExperimentTest}, first experiments confirmed that entangled photon pairs are able to result in its violation \cite{freedman1972ExperimentalTestLocal, aspect1982ExperimentalTestBell, ou1988ViolationBellInequality, munro1993ViolationBellInequality}, albeit with a number of loopholes that posed a conceptual challenge on the conclusions that can be drawn from this violation \cite{larsson2014LoopholesBellInequality}. Since then, Bell inequality violations have also been reported using other systems as diverse as ions \cite{rowe2001ExperimentalViolationBell}, electron spins \cite{hensen2015LoopholefreeBellInequality}, transmon qubits \cite{white2016PreservingEntanglementWeak}, or ultracold atoms \cite{shin2019BellCorrelationsSpatially}. During the last decades the different loopholes have been successively addressed and closed, finally culminating in the observation of loophole-free Bell inequality violations \cite{hensen2015LoopholefreeBellInequality, shalm2015StrongLoopholeFreeTest, giustina2015SignificantLoopholeFreeTestBell, rosenfeld2017EventReadyBellTest}.

As Bell correlations (i.e. nonlocality) represent the most profound departure of quantum from classical physics, their observation in macroscopic objects, which typically behave classically, is of extreme interest. For this reason, intense efforts have been put into trying to detect such correlations in multipartite and massive systems, such as optomechanical devices \cite{galland2014HeraldedSinglePhononPreparation,qian2012QuantumSignaturesOptomechanical,safavi-naeini2014TwoDimensionalPhononicPhotonicBand} or Bose-Einstein condensates (BECs) \cite{pu2000CreatingMacroscopicAtomic,duan2000SqueezingEntanglementAtomic}. The latter are also versatile systems for a variety of quantum information processing tasks \cite{byrnes2012MacroscopicQuantumComputation,byrnes2015MacroscopicQuantumInformation,pyrkov2014QuantumTeleportationSpin,pyrkov2014FullBlochsphereTeleportationSpinor}, and for quantum metrology \cite{appel2009MesoscopicAtomicEntanglement,krauter2013DeterministicQuantumTeleportation,moxley2016SagnacInterferometryCoherent,pezze2018QuantumMetrologyNonclassical,ilo-okeke2018RemoteQuantumClock}. In these ensembles, the internal states of the atoms form a collective spin. By controlling atom-atom interactions, a variety of nonclassical collective spin states can be prepared, such as spin-squeezed states \cite{kitagawa1993SqueezedSpinStates}, twin-Fock states \cite{lucke2011TwinMatterWaves}, and two-mode squeezed (TMS) states \cite{baumgarten2008FirstMeasurementHydrogen}. These states, observed in BECs in Ref. \cite{esteve2008SqueezingEntanglementBose,riedel2010AtomchipbasedGenerationEntanglement,krauter2011EntanglementGeneratedDissipation,peise2015SatisfyingEinsteinPodolsky,lucke2014DetectingMultiparticleEntanglement}, were shown to exhibit entanglement and EPR steering \cite{lange2018EntanglementTwoSpatially, kunkel2018SpatiallyDistributedMultipartite, fadel2018SpatialEntanglementPatterns}. Moreover, following the discovery of a multipartite Bell inequality involving only low-order correlators \cite{tura2014DetectingNonlocalityManybody}, Bell correlations have been observed in a BEC \cite{schmied2016BellCorrelationsBoseEinstein} and in a thermal atomic ensemble \cite{engelsen2017BellCorrelationsSpinSqueezed} from the violation of an experimentally practical witness involving only collective spin measurements.

So far, the successful violations of Bell correlation witnesses \cite{aloy2019DeviceIndependentWitnessesEntanglement,wagner2017BellCorrelationsManyBody}  were demonstrated within  single atomic ensembles prepared in a nonclassical spin state. On the other hand, the violation of a CHSH Bell  inequality with such systems has proven to be surprisingly difficult, and a scheme that is both experimentally practical and shows a strong violation is still missing even in the simplest bipartite scenario. For example, in Ref. \cite{oudot2019BipartiteNonlocalityManybody}, it was shown that a split spin-squeezed BEC shows Bell violations, but only when atomic parity measurements can be performed. This requires collective measurements with single atom resolution, which are experimentally challenging.  In Ref. \cite{kitzinger2020TwoaxisTwospinSqueezed}, the two-axis two-spin squeezed state was shown to violate a CHSH inequality using the sign of total spin operators, which is experimentally more accessible. However, the level of violation was found to diminish with increasing total atom number $N$. This mirrors the problems of Bell inequalities for continuous-variable (CV) systems (e.g. , \cite{cavalcanti2007BellInequalitiesContinuousVariable}), which can be seen as the large $N$ limiting case of spin systems. In the CV case, many proposals have prohibitively difficult requirements on either the measurements or the considered states \cite{brunner2014BellNonlocality, braunstein2005QuantumInformationContinuous}. Despite these obstacles, the first experimental violation of a CV Bell inequality as proposed by Ralph {\it et al.} \cite{ralph2000ProposalMeasurementBellType} was recently reported \cite{thearle2018ViolationBellInequality}.

\begin{figure}[t]
\includegraphics[width=\columnwidth]{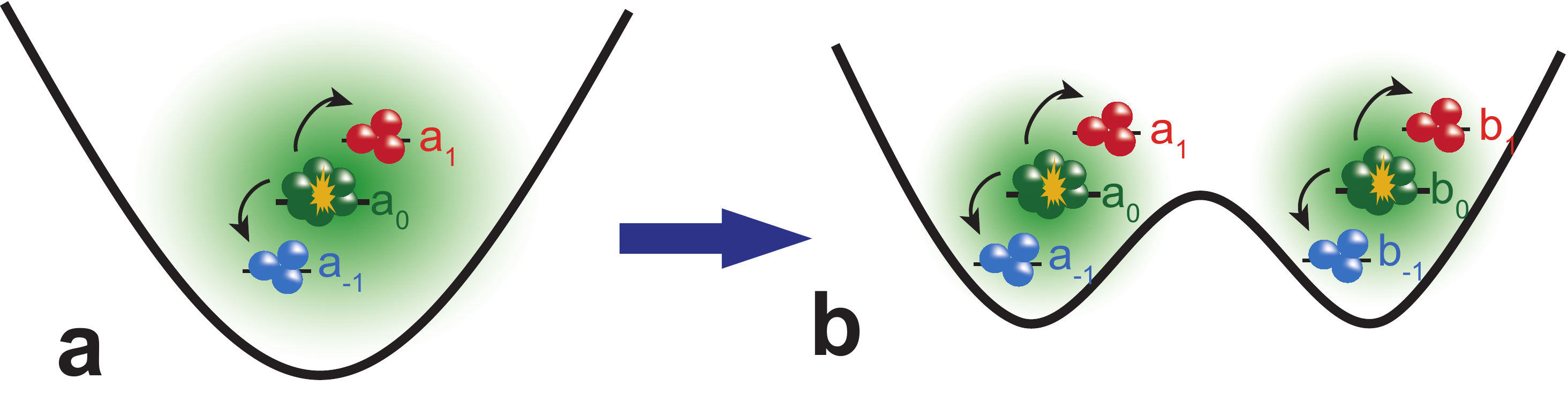}
\caption {Schematic procedure for realizing split two-mode squeezed Bose-Einstein condensates (BECs). (a) A BEC in a single trap prepared in the state $ a_0 $  is two-mode squeezed by spin-changing collisions, scattering into two other spin states $ a_1, a_{-1} $. (b) The atoms are then spatially separated into two ensembles via a beamsplitter interaction. The atoms in the two BECs remain correlated after the split, producing Bell correlations between the two wells.  
\label{fig1}}
\end{figure}

In this paper, we propose a scheme for violating the CHSH Bell inequality with a split two-mode squeezed (TMS) BEC (see Fig. \ref{fig1}), and discuss several strategies for evaluating the necessary correlators. Two spatially separated entangled BECs are generated by spatially splitting a TMS BEC, as experimentally realized in Ref.  \cite{lange2018EntanglementTwoSpatially}. This is analogous to protocols previously considered for split spin-squeezed BECs \cite{fadel2018SpatialEntanglementPatterns,jing2019SplitSpinsqueezedBose,fadel2020RelatingSpinSqueezing}. As for the optical case, where splitting a squeezed state is equivalent to a TMS state with additional local squeezing on the two modes, we show that the same occurs in a split TMS BEC. In the context of detecting Bell correlations, which are enabled by the entanglement present between the two BECs, the local squeezing terms within each BEC give rise to undesirable terms in the wave function. However, these can be eliminated by locally ``desqueezing'' each BEC, as part of the state preparation sequence.  As we show, the resulting state can be used to evaluate the correlators following the approach by Ralph {\it et al.} \cite{ralph2000ProposalMeasurementBellType}, to obtain a maximal violation of the CHSH inequality. However, this method of evaluating correlators does not strictly follow a two-valued probabilistic evaluation of the CHSH correlators, and implicitly discards particular measurement outcomes.  We therefore introduce an alternative way of evaluating Bell correlators which resolves both of these issues. We show that this scheme allows for CHSH violations to be observed in a split TMS BEC even in the case where all measurement outcomes are assigned $ \pm 1 $ values and the effect of losses is taken into account.  We finally show a variation of this same strategy using the CH inequality with the no-enhancement assumption, which allows for larger violations. 

This paper is organized as follows. In Sec. \ref{sec:system} we describe the physical system and the type of operations that are required to prepare the Bell correlations in the system. The split-squeezing method and the full state preparation protocol are presented in  Sec. \ref{sec:stateprep}. In Sec. \ref{sec:ralph} we then use the approach of Ralph {\it et al.}  \cite{ralph2000ProposalMeasurementBellType} (which we call Approach I) to show that a Bell violation can be detected using normalized averages of atom number measurements.  We further improve this approach in Sec. \ref{sec:approachii} by introducing genuine two-valued measurement operators and a way of evaluating the correlators without discarding any measurement outcomes, making it more consistent with the original CHSH inequality.  This Approach II is discussed in Sec. \ref{sec:appii}, and is found to also violate the CHSH inequality, albeit by a smaller amount. A larger violation can be obtained (Approach III, Sec. \ref{sec:appiii}) by using the CH inequality with the no-enhancement assumption, where a ``0'' value is assigned to local vacuum detection.   In Sec. \ref{sec:loss} we discuss the effects of loss and detector efficiency for the three approaches. Finally, in Sec. \ref{sec:conc} we summarize our findings and conclude.

\section{The physical system and required operations}
\label{sec:system}

In this section we describe our model of the physical system underlying this work, as well as the operations that will be needed for the preparation of the Bell correlated state.

\subsection{Spin-changing collisions}

We start by considering a BEC of $N$ atoms in the same internal state. The system is thus in state
\begin{align}
|\psi_0 \rangle=\frac{1}{\sqrt{N!}}(a^{\dagger}_{0})^{N}|0\rangle ,
\label{initialstate}
\end{align}
where $a_0^\dagger $ is the bosonic creation operator for mode $m=0$. Note that all atoms are in the same internal state, as well as in the same spatial wave function. We then consider atom-atom interactions of the form of spin-changing collisions \cite{chang2004ObservationSpinorDynamics}, where the collision between two atoms in spin state $0$ change their spin states to $+1$ and $-1$, respectively. Such interactions can be turned on for a desired duration by applying a magnetic field for a specified time \cite{klempt2010ParametricAmplificationVacuum}.  This results in populating two different modes, which we associate with the bosonic annihilation operators $ a_1 $ and $ a_{-1} $ (see Fig. \ref{fig1}a). The Hamiltonian describing this type of interaction is \cite{law1998QuantumSpinsMixing, gabbrielli2015SpinMixingInterferometryBoseEinstein}
\begin{align}
H=\hbar g(a^{\dagger}_{-1}a^{\dagger}_{1}a_{0}a_{0}+a^{\dagger}_{0}a^{\dagger}_{0}a_{-1}a_{1}) ,
\label{mainham}
\end{align}
where $g$ describes the strength of the scattering process.  Such dynamics were realized in numerous experiments \cite{lucke2011TwinMatterWaves, peise2015SatisfyingEinsteinPodolsky, lange2018EntanglementTwoSpatially} and now represents a standard approach to prepare nonclassical states of atomic ensembles. In general, the modes $\pm 1$ have a nearly identical spatial wave function to mode $0$,  e.g., the ground state of the trapping potential. However, in the case of Ref. \cite{lange2018EntanglementTwoSpatially}, mode $0$ corresponds to the $\vert F=1, m=0 \rangle$ hyperfine state of $ ^{87} \text{Rb} $ atoms in the ground state of the trapping potential, while the states $\pm 1$ correspond to the $\vert F=1, m=\pm 1 \rangle$ states in the spatial first excited state.  

Due to the typically very large number of atoms $ N $ initially in the spin-0 state, and the short evolution time we consider for the Hamiltonian Eq. \eqref{mainham}, we can approximate mode $\vert 0 \rangle$ as a classical field (local oscillator) with amplitude $\sqrt{a_0^\dagger a_0} = \sqrt{N}$, and thus replace the operator $a_0$ with a $c$-number. This results in the effective Hamiltonian
\begin{align}
H\approx \hbar g N( a^{\dagger}_{-1}a^{\dagger}_{1}+a_{-1}a_{1}) ,
\label{ham2m}
\end{align}
which is of the same form as the two-mode squeezing Hamiltonian in quantum optics \cite{carmichael1999StatisticalMethodsQuantum, walls1994QuantumOptics, gerry2004IntroductoryQuantumOptics}. For the approximation to be valid, we require $N \gg 1$ for the atom number and $gt < 1/N$ for the evolution time. An explicit calculation and numerical results showing that the two Hamiltonians lead to equivalent results in the small-squeezing limit can be found in Appendix \ref{app:m=0}.

The above Hamiltonian \eqref{ham2m} has been experimentally demonstrated in BECs to produce entanglement \cite{peise2015SatisfyingEinsteinPodolsky, lange2018EntanglementTwoSpatially}. It induces a unitary transformation given by
\begin{align}
    U_r = e^{-i H t /\hbar }= e^{-i r (a^{\dagger}_{-1}a^{\dagger}_{1}+a_{-1}a_{1}) } ,
\end{align}
where the squeezing parameter is defined as $ r = gNt$.  In the Heisenberg picture this transforms the mode operators as \cite{scully1997QuantumOpticsa, gerry2004IntroductoryQuantumOptics}
\begin{align}
U_r^\dagger a_1 U_r & = a_{1} \cosh r - i a^{\dagger}_{-1} \sinh r \nonumber  \\
U_r^\dagger a_{-1} U_r & = a_{-1} \cosh r - i a^{\dagger}_{1} \sinh r .
\label{tmsstransform}
\end{align}
The state that is generated is thus  \cite{scully1997QuantumOpticsa, gerry2004IntroductoryQuantumOptics}
\begin{align}
 |\psi_r \rangle & = U_r  |\psi_0 \rangle \nonumber \\
 &= \sech r \sum_k (-i)^k \tanh^k r | k,  k \rangle \;,
 \label{twinfockstate}
\end{align}
where we defined the Fock states as 
\begin{align}
    | k, l \rangle = \frac{ (a_1^\dagger)^k  (a_{-1}^\dagger)^l }{\sqrt{k! l!}}  | 0 \rangle .
    \label{fockstatedef}
\end{align}
The state \eqref{twinfockstate} has only equal number Fock states of the  $a_{\pm 1}$ modes, characteristic of a TMS state.

We remind that in the following, under the local oscillator approximation, only the $a_{\pm 1} $ modes are considered as quantum states.  There is always a large population of atoms in the $0$ state, as the transfer of population to the $\pm 1$ states does not deplete it to any significant extent, which allows us to approximate it as a classical field. The large population in the $a_0$ state should be considered implicitly, even when it does not appear in the definitions of states as written in \eqref{fockstatedef}.  Under this approximation, the initial state \eqref{initialstate} therefore corresponds to the vacuum according to \eqref{fockstatedef}.

\subsection{Splitting the BEC}

An important operation that we will perform is to split the BEC into two spatially separated locations, as shown in Fig. \ref{fig1}.  We associate the modes in the left trap with the annihilation operators $a_m$ (Alice's BEC), and the modes in the right trap with $b_m$ (Bob's BEC). The spatial splitting corresponds to applying the splitting operator \cite{jing2019SplitSpinsqueezedBose}
\begin{align}
    U_{W} = e^{-i H_W \pi/4}
    \label{splittingop}
\end{align}

with
\begin{align}
  H_W = \sum_{m=-1,0,1} \left(  -  i a_m^\dagger b_m + i b^\dagger_m a_m  \right) .
\end{align}
Applying the splitting operator, we transform
\begin{align}
U_{W}^\dagger a_m U_{W} = \frac{1}{\sqrt{2}}(a_m+b_m)\label{splittransform} \\
U_{W}^\dagger b_m U_{W} = \frac{1}{\sqrt{2}}(b_m-a_m) . 
\label{bsplit}
\end{align}
This transformation is analogous to a 50/50 beamsplitter in the context of optics.  A possible method for performing the splitting illustrated in Fig. \ref{fig1} is to change a harmonic trapping potential into a double-well potential, and then separate the two wells arbitrarily further apart. A difference between optical and atomic systems, however, is that interference between two initially populated spatial modes $ a_m, b_m $ is more difficult to perform in a controlled way for atoms. We therefore consider only the application of the splitting operator when the $ b_m$ modes are initially unpopulated.  

After a single BEC is split into two wells, the $ N$ atoms are distributed randomly according to a binomial distribution.  To see this, we note that the initial state \eqref{initialstate} containing the bulk of the atoms evolves under the splitting operations as
\begin{align}
    U_W | \psi_0 \rangle &  = \frac{1}{\sqrt{N!}}
    \left( \frac{a_0^\dagger + b_0^\dagger}{\sqrt{2}}   \right)^N | 0 \rangle \nonumber \\
    & = \frac{1}{\sqrt{2^N}} \sum_k \sqrt{N \choose k } (a_0^\dagger)^k  (b_0^\dagger)^{N-k} | 0 \rangle .
    \label{m0_split}
\end{align}
The number of atoms in each well is therefore a conserved number satisfying
\begin{align}
N = N_a + N_b.  
\end{align}

\subsection{Spin-changing collisions after splitting}

After splitting the BEC, spin-changing collisions can also take place locally in each trap. For this, we assume that the mode $m = 0$ is also split coherently and equally according to \eqref{m0_split}. Using the local oscillator approximation as before, the Hamiltonian for the local spin-changing dynamics is described  by 
\begin{align}
        H = \hbar g \frac{N}{2} ( a^{\dagger}_{-1}a^{\dagger}_{1}+a_{-1}a_{1})
        +  \hbar g \frac{N}{2} ( b^{\dagger}_{-1} b^{\dagger}_{1}+b_{-1}b_{1}) ,
        \label{Ham_after_split}
\end{align}
which is shown in Appendix \ref{app:m=0}.
Accordingly we may define the squeezing transformation on each well as
\begin{align}
    U_{r}^{(a)} & = e^{-i \frac{r}{2}  (a^{\dagger}_{-1}a^{\dagger}_{1}+a_{-1}a_{1})  } \nonumber \\
    U_{r}^{(b)} & = e^{-i \frac{r}{2}  (b^{\dagger}_{-1}b^{\dagger}_{1}+b_{-1}b_{1})  },
    \label{uaoperator}
\end{align}
where $ r= g N t $ as before.

\subsection{Local spin rotations}

In the following, we will require applying local spin rotations on Alice and Bob's BECs.  First define local spin operators as \cite{byrnes2021QuantumAtomOptics}
\begin{align}
    S^y_A & = - i  a_{-1}^\dagger a_1 + i a_1^\dagger a_{-1} \nonumber  \\
    S^y_B & = - i  b_{-1}^\dagger b_1 + i b_1^\dagger b_{-1}  .
\end{align}
Then the rotation of the spin around the $ y$-axis of the Bloch sphere is defined as
\begin{align}
    V (\theta_A, \theta_B) & = e^{- i S^y_A \theta_A }  e^{- i S^y_B \theta_B } .
    \label{rotationopera}
\end{align}
Such rotations can be performed by applying a two-photon microwave transition between the $ m = \pm 1$ states \cite{peise2015SatisfyingEinsteinPodolsky}. 

In the state preparation protocol, we will also need to imprint a relative phase between the $ m = -1$ and $  +1$.  We consider to be in the interaction picture, such that the default phase evolution of both the $\pm 1$ states is zero. Then, we introduce a time-dependent controllable energy shift between the states $ \Delta $, which results in the effective Hamiltonian 
\begin{equation}
    H_\phi = \hbar \Delta (a^\dagger_1 a_1 + b^\dagger_1 b_1 ) .
\end{equation}
Experimentally, this can be realized by applying a magnetic field, or producing an AC Stark shift on the $\pm  1$ states.  A relative phase of $ \pi $ can be imprinted on the state by applying this Hamiltonian for a time $ t = \pi/\Delta$.  This corresponds to the $ \pi$-gate
\begin{align}
    U_\pi = e^{-i (a^\dagger_1 a_1 + b^\dagger_1 b_1 ) \pi } .
    \label{upi}
\end{align}

\subsection{Measurement}

Finally, to readout the state of the atoms, the population of the atoms in the  $ a_m, b_m $ states are measured by Alice and Bob.  This can be performed for example by spin-selective absorption imaging \cite{pezze2018QuantumMetrologyNonclassical}.  This corresponds to a projective measurement
\begin{align}
    \Pi_{kl} = | k, l \rangle \langle k, l |
\end{align}
for Alice, and similarly for Bob. The readout values $ k, l $ are then used to evaluate Bell correlations as described in the following sections.

\section{Constructing the Bell correlated state}
\label{sec:stateprep}

\subsection{Split-squeezed state}

Before introducing our protocol for generating the Bell correlated state, it is instructive to write down the wave function of the state that results from applying the two-mode squeezing operation $ U_r $ and then spatially splitting the state with $ U_W$.  Starting from (\ref{twinfockstate}) and splitting the state for $ r \ll 1 $ we may write
\begin{align}
U_W U_r |0 \rangle  \approx|0\rangle - \frac{i r}{2} (a^{\dagger}_{1}a^{\dagger}_{-1}+b^{\dagger}_{1}b^{\dagger}_{-1} +a^{\dagger}_{1}b^{\dagger}_{-1}+b^{\dagger}_{1}a^{\dagger}_{-1})|0\rangle .
\label{loworderexpold}
\end{align}
We see that the last two terms $ a^{\dagger}_{1}b^{\dagger}_{-1}+b^{\dagger}_{1}a^{\dagger}_{-1} $ take the form of a Bell state.  It is tempting to design experimental observables that are sensitive to this part of the wave function, and use the methods of Ref. \cite{ralph2000ProposalMeasurementBellType} to observe a Bell violation. Such operators are in fact not hard to construct,  e.g., any number operator of the form $ n_a n_b$ automatically gives zero for the $ a^{\dagger}_{1}a^{\dagger}_{-1}+b^{\dagger}_{1}b^{\dagger}_{-1} $ terms.  However, this approach is vulnerable to the postselection loophole, as it assigns the value of 0 to the corresponding measurements and thereby effectively discards these outcomes. Such a postselection is known to be problematic as it can distort the measurement outcome statistics and artificially create Bell violations \cite{pearle1970HiddenVariableExampleBased, peres1997BellInequalitiesPostselection,larsson2014LoopholesBellInequality}. Indeed, the problematic nature of such an approach can be seen by observing  that the two-boson terms in \eqref{loworderexpold} can be written as a product state $ (a_1^\dagger + b_1^\dagger)(a_{-1}^\dagger + b_{-1}^\dagger) $, which does not take the form of a Bell state \cite{decaro1994ReliabilityBellinequalityMeasurements}.  For these reasons, we will show in the next section that it is possible to obtain a wave function which exhibits a genuine Bell state for the two-boson terms, by eliminating the $ a^{\dagger}_{1}a^{\dagger}_{-1}+b^{\dagger}_{1}b^{\dagger}_{-1} $ terms.

\subsection{State preparation protocol}
\label{stateprep}

Here we introduce a protocol for generating Bell correlated states in the context of atomic BECs.:
\begin{enumerate}
    \item Starting from state \eqref{initialstate}, apply the squeezing operation $ U_{2r}$.  This can be performed by applying the Hamiltonian (\ref{mainham}) for a time $ 2t $,  resulting in squeezing characterized by the parameter $ 2r = 2gN t$.
    \item Apply a $ \pi $-pulse on the states according to the operator $ U_{\pi} $ [Eq. \eqref{upi}]. 
    \item Split the BEC into two wells, according to the transformation $ U_W $ [Eq. \eqref{splittingop})] 
    \item Squeeze each of the BECs in the wells individually, according to the operator  $ U_{2r}^{(a)} U_{2r}^{(b)} $ [Eq. \eqref{uaoperator}].
\end{enumerate}

To lowest order in $r$, the above procedure results in the state 
\begin{align}
    | \Psi \rangle & =U_{\Psi} | 0 \rangle \label{splitsqueezed} \\
    & \approx | 0 \rangle + i r (a^{\dagger}_{1}b^{\dagger}_{-1}+b^{\dagger}_{1}a^{\dagger}_{-1})|0\rangle ,
    \label{loworderexp}
\end{align}
where we defined the unitary operator for the full state preparation 
\begin{align}
U_{\Psi} = U_{2r}^{(a)} U_{2r}^{(b)} U_W U_{\pi}  U_{2r} . 
\label{upsiseq}
\end{align}
The above state does not contain the unwanted two-boson terms that were present in \eqref{loworderexpold}.  The intuition behind this state preparation protocol is as follows. As can be seen from \eqref{loworderexpold}, the split-squeezing procedure produces terms corresponding to squeezing on each BEC individually (the $ a^{\dagger}_{1}a^{\dagger}_{-1}+b^{\dagger}_{1}b^{\dagger}_{-1} $ terms) and cross-squeezing between the BECs (the $ a^{\dagger}_{1}b^{\dagger}_{-1}+b^{\dagger}_{1}a^{\dagger}_{-1} $ terms).  A similar effect was observed in Ref. \cite{jing2019SplitSpinsqueezedBose}, where splitting a squeezed state produced both local squeezing as well as cross-squeezing between the BECs.  However, first applying a $ \pi $-phase shift to reverse the sign of $ \pi $, and then applying local squeezing operators \eqref{uaoperator}, results in a local ``desqueezing'', meaning that the local squeezing is effectively canceled. The factor of 2 between the squeezing interaction time in Steps 1 and 4 arises because the splitting procedure introduces a factor of 2 as seen in \eqref{loworderexpold}.  

We note that such a procedure is analogous to a well-known procedure in quantum optics, where a two-mode squeezed state is produced by sending two squeezed states into a beamsplitter \cite{braunstein2005QuantumInformationContinuous}.  It is also true in the optics case that if a single squeezed mode is sent into a beamsplitter, the final state is a combination of single and two-mode squeezed states  (see Appendix \ref{app:twomode}).  By interfering two squeezed states, the single-mode squeezing terms can be eliminated, such that the final state is purely a two-mode squeezed state. As explained above, the analog of such a procedure is not practical for BECs as it would involve interfering two split BECs, hence we opt for an alternative strategy where the splitting is performed on a single BEC.

\begin{figure}[t]
\includegraphics[width=\columnwidth]{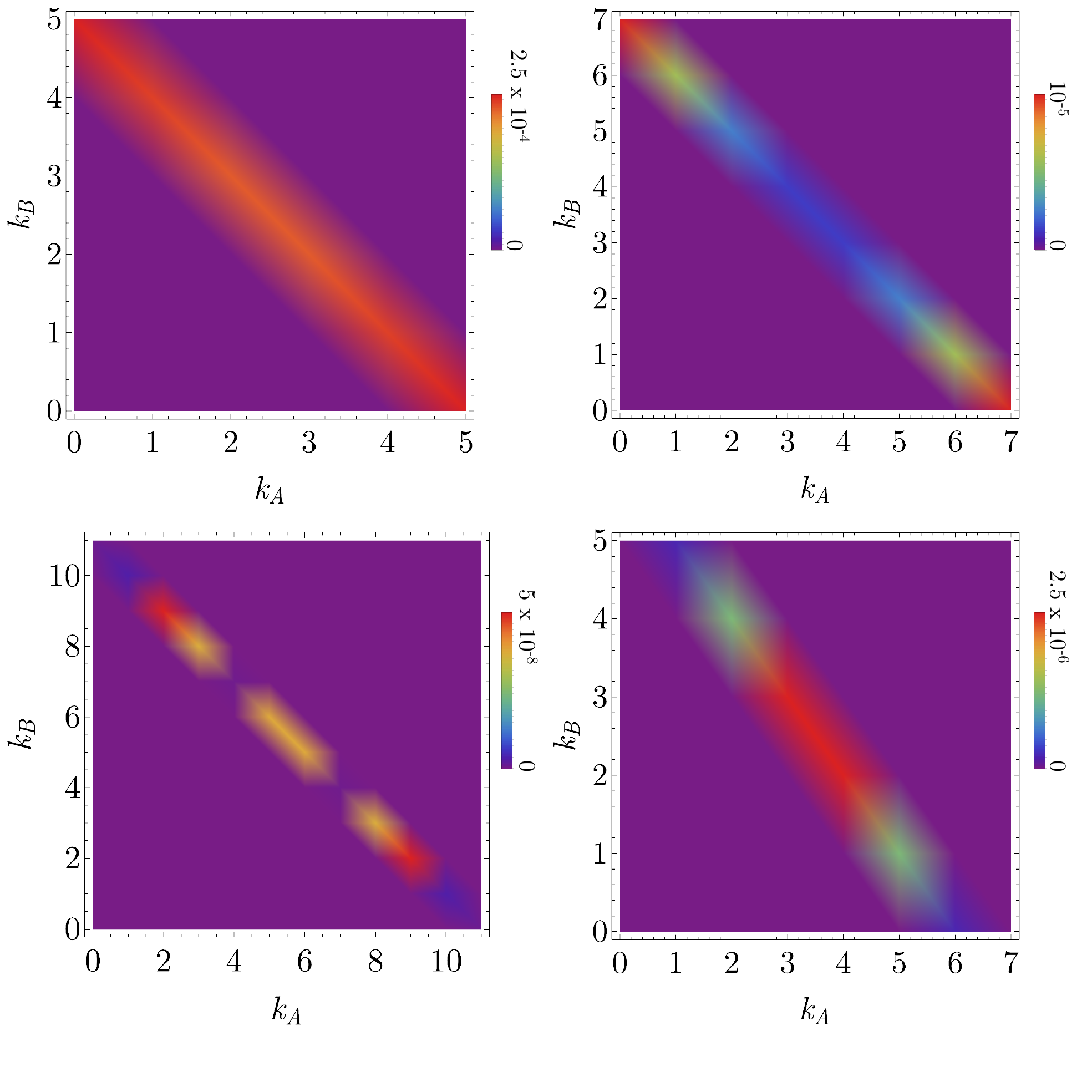}
\caption{The probability distribution \eqref{probplot} for the particle-number sectors (a) $ N_A = N_B = 5$;  (b) $ N_A = N_B = 7$; (c) $ N_A = N_B = 11$; (d) $ N_A = 7, N_B = 5$.  For all plots $ r = 0.5$.   
\label{fig3}}
\end{figure}

In Fig. \ref{fig3} we show the probability distribution of the prepared state in subspaces with definite particle numbers for Alice and Bob, as defined by 
\begin{align}
    N_A & =  a_1^\dagger a_1 + a_{-1}^\dagger a_{-1}  \nonumber \\
    N_B & = b_1^\dagger b_1 + b_{-1}^\dagger b_{-1} .
\label{numberalicebob}
\end{align}
The probability in the number sector $ ( N_A, N_B )$ is defined as
\begin{align}
p_{N_A N_B} (k_A, k_B) = | \langle \Psi | k_A, N_A - k_A \rangle \otimes | k_B, N_B - k_B \rangle |^2.  
\label{probplot}
\end{align}
We see that for particle-number sectors such that  $ |N_A - N_B | $ is even, anticorrelations are observed. Zero probability distributions are obtained when $ |N_A - N_B | $ is an odd integer. This can be understood as a consequence of the fundamental Hamiltonian (\ref{mainham}): Atoms can only scatter into the $+1$ and $-1$ modes in pairs, so that the sum $N_A + N_B$ and therefore also the difference $N_A - N_B$ must be even.
The probabilities diminish for larger $ N_A, N_B$ values for the relatively small squeezing that is shown in Fig. \ref{fig3}.  For the $ r = 0.5$ case that is shown, for small  $ N_A, N_B$, the probabilities are evenly distributed along the diagonal.

\subsection{Alternative methods}
\label{sec:alterna}

As other methods to generate EPR correlated states of the desired form \eqref{splitsqueezed} may be possible, we briefly discuss alternative state preparation strategies.  For example, a similar state to \eqref{loworderexp} could be generated by looking at the momentum or energy degrees of freedom in the atoms. In this case the relevant Hamiltonian would be 
\begin{align}
    H \propto a^\dagger_{-1} b^\dagger_1 + a^\dagger_{1} b^\dagger_{-1}
    + a_{-1} b_1 + a_{1} b_{-1} ,
    \label{momentumham}
\end{align}
where $ a_m$ and $b_m $ label, for example, the positive and negative momentum atoms, respectively. Momentum resolving measurements could be performed by a time-of-flight measurement after releasing the atoms from the trap.    
Then, by performing a spin measurement in a suitable basis, Bell correlations should be present between two groups of atoms with different momenta.  Related approaches have been discussed at the single atom level in Refs.  \cite{bonneau2018CharacterizingTwinparticleEntanglement, shin2019BellCorrelationsSpatially}. Such alternative state preparation strategies are compatible with the measurement techniques of the next section, as long as the lowest order expansions are of the form \eqref{loworderexp}.

\section{Approach I: Evaluating Bell's inequality using normalized averages}
\label{sec:ralph}

In this section we apply the methods of Ref. \cite{ralph2000ProposalMeasurementBellType} to evaluate the Bell inequality using the split-squeezed state \eqref{splitsqueezed}.  Following the notation of  Ref. \cite{ralph2000ProposalMeasurementBellType}, the quantity we wish to evaluate is 
\begin{align}
{\cal B} & = E(\theta_{A},\theta_{B})+E(\theta_{A}',\theta_{B})+E(\theta_{A},\theta_{B}')-E(\theta_{A}',\theta_{B}')  .
\label{bellmain}
\end{align}
For any local realistic description of the system under investigation, the observed correlations must obey \cite{clauser1969ProposedExperimentTest}
\begin{align}
   \vert {\cal  B} \vert \le 2 .
   \label{dabellineq}
\end{align}
In \eqref{bellmain} we defined
\begin{align}
E_{\text{I}} (\theta_{A},\theta_{B})&=P^{++}(\theta_{A},\theta_{B})+P^{--}(\theta_{A},\theta_{B})\nonumber\\&
-P^{+-}(\theta_{A},\theta_{B})-P^{-+}(\theta_{A},\theta_{B}) ,
\label{eoperatordef}
\end{align}
where 
\begin{align}
P^{ij}(\theta_{A},\theta_{B})=\frac{ \langle R^{ij}(\theta_{A},\theta_{B})\rangle}{\sum_{kl=\pm}{\langle}R^{kl}(\theta_{A},\theta_{B})\rangle} .
\label{pvalue}
\end{align}
The correlation operators are 
\begin{align}
 R^{ij}(\theta_{A},\theta_{B})=A^{\dagger}_{i}(\theta_{A})A_{i}(\theta_{A}) B^{\dagger}_j (\theta_{B}) B_j (\theta_{B}) ,
 \label{roperatordef}
\end{align}
where $i,j=\pm$, and the $A_{i}(\theta_{A}), B_j (\theta_{B})$ are annihilation operators on Alice and Bob's subsystems in a basis specified by the angles $ \theta_A, \theta_B$ . In our case, the operators are defined as 
\begin{align}
A_+(\theta_A) & =\cos\theta_{A}a_{1}+\sin\theta_{A}a_{-1} \nonumber \\
A_-(\theta_A) & =\cos\theta_{A}a_{-1}-\sin\theta_{A}a_1 \nonumber \\
B_+(\theta_B) & =\cos\theta_{B}b_{1}+\sin\theta_{B}b_{-1} \nonumber \\
B_-(\theta_B) & =\cos\theta_{B}b_{-1}-\sin\theta_{B}b_1  . \label{bigabtransform}
\end{align}
Using the rotation operators (\ref{rotationopera}) we may then write
\begin{align}
 A_j (\theta_A) & = V^\dagger ( \theta_A, \theta_B) a_j V  ( \theta_A, \theta_B) \nonumber \\
 B_j (\theta_B) & = V^\dagger ( \theta_A, \theta_B) b_j V ( \theta_A, \theta_B) ,
\end{align}
where $ j = \pm $. When considering the state \eqref{splitsqueezed}, the correlators appearing in \eqref{pvalue} can then be written as
\begin{align}
    \langle R^{ij}(\theta_{A},\theta_{B})\rangle & = 
    \langle \Psi | R^{ij} ( \theta_A, \theta_B ) | \Psi \rangle \nonumber \\
    & = \langle 0 | \tilde{A}_i^\dagger \tilde{A}_i \tilde{B}_j^\dagger \tilde{B}_j | 0 \rangle , \label{rcorr}
\end{align}
where
\begin{align}
  \tilde{A}_i =    U_{\Psi}^\dagger  V^\dagger ( \theta_A, \theta_B) a_i
  V ( \theta_A, \theta_B) U_{\Psi} \nonumber \\
    \tilde{B}_i =  U_{\Psi}^\dagger   V^\dagger ( \theta_A, \theta_B) b_i
  V ( \theta_A, \theta_B) U_{\Psi}  .  
\end{align}
Combining the transformations \eqref{tmsstransform}) \eqref{splittransform}, and \eqref{bigabtransform}, we arrive at the explicit expressions
\begin{align}
\tilde{A}_{+} = & \cosh r ( -c_1  \cos \theta_A  + c_{-1} \sin \theta_A ) \nonumber \\
& + i  \sinh r ( d_{-1}^\dagger \cos \theta_A - d_1^\dagger \sin \theta_A ) \nonumber \\
\tilde{B}_{+} = & \cosh r ( d_1  \cos \theta_B  - d_{-1} \sin \theta_B ) \nonumber \\
& - i  \sinh r ( c_{-1}^\dagger \cos \theta_B - c_1^\dagger \sin \theta_B ) \nonumber \\
\tilde{A}_{-} = & \cosh r ( c_{-1} \cos \theta_A  + c_1 \sin \theta_A ) \nonumber \\
& - i  \sinh r ( d_{1}^\dagger \cos \theta_A + d_{-1}^\dagger \sin \theta_A ) \nonumber \\
\tilde{B}_{-} = & -\cosh r ( d_{-1}  \cos \theta_B + d_1 \sin \theta_B ) \nonumber \\
& + i  \sinh r ( c_{1}^\dagger \cos \theta_B + c_{-1}^\dagger \sin \theta_B ) , \nonumber 
\end{align}
where we have defined
\begin{align}
    c_m = \frac{1}{\sqrt{2}} ( a_m + b_m ) \nonumber \\
    d_m = \frac{1}{\sqrt{2}} ( a_m - b_m )
\end{align}
for $ m = \pm 1 $.
Substituting these expressions into \eqref{rcorr}, and evaluating \eqref{bellmain} with the optimal angles
\begin{align}
    \theta_A=\frac{3\pi}{8}, \theta'_{A}=\frac{\pi}{8}, \theta_{B}=\frac{\pi}{4}, 
    \theta'_{B}=0,
    \label{optimalangles}
\end{align}
we obtain
\begin{align}
{\cal B} =\frac{4\sqrt{2} \cosh^{2}r}{3 \cosh2r -1} .
\label{bellsimplecase}
\end{align}
The largest violation is thus obtained in the limit of  $ r \rightarrow 0 $, where $ {\cal B}=2\sqrt{2}$.

\begin{figure}[t]
\includegraphics[width=\columnwidth]{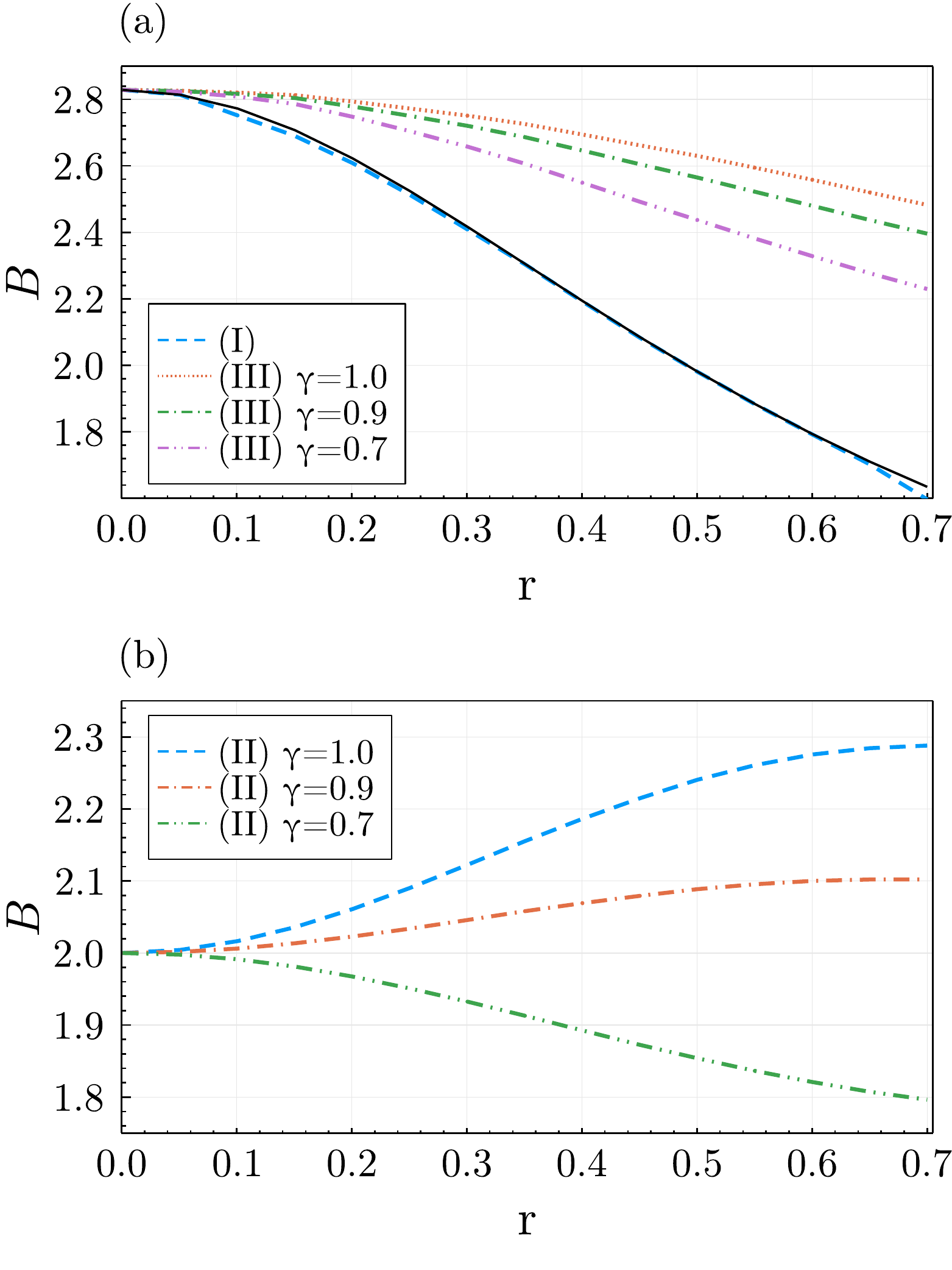}
\caption {The Bell-CHSH quantity $ {\cal B} $ as a function of the dimensionless squeezing parameter $r$ for (a) Approaches I and III and  (b) Approach II. Numerically evaluated lines (dashed/dotted lines) for (a) Approach I and III, corresponding to the methods given in Sec. \ref{sec:ralph} and Sec. \ref{sec:appiii} respectively; and (b) Approach II, given in Sec. \ref{sec:appii} are shown.  The black (solid) line in (a) is the exact result \eqref{bellsimplecase}.  In all cases  $ {\cal B} $ is calculated using \eqref{bellmain}, but using various definitions of $ E(\theta_A, \theta_B)$, where for Approach I Eq. \eqref{origeexpprob}; II Eq. \eqref{newedefprob}; III Eq. \eqref{newedefprob2} is used for the cases without loss $ \gamma = 1 $.  For the cases with loss $ \gamma < 1 $, for Approach II Eq. \eqref{eprimenopos}; III Eq. \eqref{eprimewithpos} is used.  For all numerical calculations  $ k_{\text{cut}} =40$ is used. 
\label{fig2}}
\end{figure}

A plot of \eqref{bellsimplecase} is shown in Fig. \ref{fig2}(a) (solid line).  We see that violations of the Bell inequality are seen for $ r \lesssim 0.49 $. Similarly to what is shown in Ref. \cite{ralph2000ProposalMeasurementBellType}, the level of violation decreases with squeezing.  This occurs since the operators to detect the Bell correlations are constructed with the lowest order terms \eqref{loworderexp} in mind, and larger values of squeezing create additional undesired contributions. We expect that the angle choices \eqref{optimalangles} are optimal only for the $N_A = N_B =1$ particle sector, analogously to what has been found in the optical case of Bell inequalities for states with large photon numbers \cite{munro1993ViolationBellInequality, kitzinger2020TwoaxisTwospinSqueezed}. However, for small $ r $, the dominant terms responsible for the Bell violation are the first-order terms with $N_A = N_B =1$, so that \eqref{optimalangles} should be close to optimal in this regime. This is confirmed by Fig. \ref{fig2}(a).  In practice, due to the presence of experimental imperfections one may require working at finite levels of squeezing such that a sufficient signal-to-noise is obtained.  Hence there is a practical trade-off between maximizing the CHSH violation and working at squeezing levels within an experimentally accessible range.

In the following sections we will compare two other alternative ways of defining the correlators $ E(\theta_A, \theta_B) $.  In order to distinguish which alternative is being used, we refer to the method used in this section as ``Approach I'' henceforth. Quantities that can be evaluated in different ways, such as the correlators, will be labeled by I, II, or III,  to specify which approach is being used.

\section{Evaluating the Bell inequality using two-outcome measurements}
\label{sec:approachii}

\subsection{Evaluation of correlators using Born's rule}

In the derivation of the original CHSH inequality \cite{clauser1969ProposedExperimentTest}, each of the terms $ E( \theta_A, \theta_B) $ in \eqref{bellmain} is calculated by assigning a value of  $\pm 1$ to every measurement outcome, and averaging over the full set of probabilistic outcomes. 
Closer inspection of \eqref{eoperatordef} reveals that this is not precisely the quantity that is being evaluated, following the procedure in Ref. \cite{ralph2000ProposalMeasurementBellType}.  The key point is that \eqref{pvalue} is evaluated using a normalized average of number operators, and is not a probabilistic average over $ \pm 1 $ outcomes.  

It is also possible to understand the evaluation of the correlators $ E ( \theta_A, \theta_B) $ in a different way, which more clearly shows this point.  Although the expression \eqref{eoperatordef} has the appearance of assigning $ \pm 1 $ outcomes based on the probabilities $ P^{ij} $, in fact the same expressions can be written in a way that reveals how it deviates from a simple average over two-valued outcomes. 
To see this, let us first write \eqref{eoperatordef} more explicitly by substituting the expressions \eqref{pvalue} and \eqref{roperatordef}, which gives
\begin{align}
    E_{\text{I}} (\theta_A, \theta_B) = \frac{ \sum_{ij=\pm} ij \langle N_A^i (\theta_A) N_B^j (\theta_B) \rangle}{  \sum_{kl=\pm} \langle N_A^k (\theta_A) N_B^l (\theta_B) \rangle } ,
    \label{enexpr}
\end{align}
where we defined the number operators
\begin{align}
    N_A^j (\theta_A) & = A_j^\dagger (\theta_A) A_j (\theta_A) \nonumber \\
     N_B^j (\theta_B) & = B_j^\dagger (\theta_B) B_j (\theta_B) . 
     \label{numberoperatorsdef}
\end{align}
This can be equally written in a spin language as
\begin{align}
     E_{\text{I}} (\theta_A, \theta_B) = \frac{\langle S_A (\theta_A) S_B (\theta_B)  \rangle }{\langle  N_A N_B  \rangle} ,
     \label{espinexpr}
\end{align}
where we defined the spin operators
\begin{align}
    S_A (\theta_A) & = N_A^+ (\theta_A) - N_A^- (\theta_A) \nonumber \\
   S_B (\theta_A) & = N_B^+ (\theta_A) - N_B^- (\theta_A)
   \label{spinoperatordef}
\end{align}
and total number operators
\begin{align}
       N_A & = N_A^+ (\theta_A) + N_A^- (\theta_A) \nonumber \\
   N_B & = N_B^+ (\theta_A) + N_B^- (\theta_A),
\end{align}
where there is no angular dependence for the total number operators. 

In the one-particle sector $ N_A = N_B = 1 $, the spin operator in the numerator of (\ref{espinexpr}) has two eigenvalues $ \pm 1$, and $ N_A N_B = 1$.  However, in higher particle-number sectors the spin operators $ S_A (\theta_A) $ take eigenvalues between $ -N_A $ and $ N_A $, and similarly for Bob.  Furthermore, it is also clear that \eqref{espinexpr} has no contribution when either Alice or Bob measures a vacuum.  This is because for a vacuum measurement on Alice we have $ N_A =  S_A =0$ and similarly $ N_B =  S_B =0$ when Bob measures his vacuum.  This immediately gives a 0 coefficient for such outcomes and effectively removes these events from the correlations. 
This explains why in Fig. \ref{fig2}(a) the violations are maximal as $ r \rightarrow 0 $.  In this limit, the state is well-approximated by \eqref{loworderexp}, which is a Bell state except for the vacuum term.  By removing the vacuum contribution from the measurements implicitly, only the Bell state is included, yielding the maximal violation.
This shows that the correlators are not calculated as an average over two-outcomed events, which is assumed in the original derivation of the CHSH inequality.

\subsection{Approach II: Alternative definition of the correlators}

We now propose an alternative way to define the correlators such that each measurement outcome is assigned to $ \pm 1$, and the correlators are averaged over the full set of possible outcomes. This not only gives a more direct connection to the original CHSH inequality  \cite{clauser1969ProposedExperimentTest}, but in fact it also improves the level of violation beyond that given by \eqref{bellsimplecase}, under certain assumptions. 
Our proposed modification is to redefine \eqref{eoperatordef} as
\begin{align}
     E_{\text{II}} (\theta_A, \theta_B) = \langle  \text{sgn} [  S_A (\theta_A)] \text{sgn} [ S_B (\theta_B) ] \rangle  ,
     \label{newedef}
\end{align}
where we use a two-valued version of the sign function with definition
\begin{align}
     \text{sgn} (x) = \left\{
     \begin{array}{cc}
     -1 & \text{for } x < 0 \\
     1 &  \text{for } x \ge 0
     \end{array}
     \right.  .
     \label{signfuncdef}
\end{align}
A similar sign-binning approach is regularly considered for homodyne measurements on CV systems \cite{gilchrist1998ContradictionQuantumMechanics, munro1999OptimalStatesBellinequality, wenger2003MaximalViolationBell, nha2004ProposedTestQuantum}. Note that our definition of the sign function bins $x=0$ into the $+1$ outcome, which ensures that two-valued measurement outcomes are constructed without discarding any data. Due to this feature, our sign function is not multiplicative, i.e.  $  \text{sgn} (xy) \ne \text{sgn} (x) \text{sgn} (y) $, which is why we define the expectation values using a product of two sign functions in \eqref{newedef}.  This enables Alice and Bob to evaluate their outcomes independently, without the knowledge of each other's result. We call this Approach II, and label \eqref{newedef} accordingly. 

To be clear on the meaning of \eqref{newedef}, let us rewrite the expression entirely in terms of probabilities.  First, we define the eigenstates of the number operators \eqref{numberoperatorsdef}
\begin{align}
&|k_A, l_A \rangle^{(\theta_A)}   = 
    \frac{1}{\sqrt{k_A! l_A! }} (A_+^\dagger ( \theta_A) )^{k_A} 
    (A_-^\dagger ( \theta_A) )^{l_A} | 0 \rangle  \\
& | k_B, l_B \rangle^{(\theta_B)}  = 
    \frac{1}{\sqrt{  k_B! l_B! }}  (B_+^\dagger ( \theta_B) )^{k_B} 
    (B_-^\dagger ( \theta_B) )^{l_B}   | 0 \rangle  \\
& |k_A, l_A, k_B, l_B \rangle^{(\theta_A, \theta_B)}  \equiv  
|k_A, l_A \rangle^{(\theta_A)} \otimes | k_B, l_B \rangle^{(\theta_B)}
    \label{genfockstate}
\end{align}
which satisfy
\begin{align}
    N_A^+ (\theta_A)  |k_A, l_A  \rangle^{(\theta_A )} & = k_A |k_A, l_A \rangle^{(\theta_A )} \nonumber \\
     N_A^- (\theta_A)  |k_A, l_A  \rangle^{(\theta_A )} & = l_A |k_A, l_A  \rangle^{(\theta_A )} \nonumber \\  
     N_B^+ (\theta_B)  | k_B, l_B \rangle^{(  \theta_B)} & = k_B | k_B, l_B \rangle^{(  \theta_B)} \nonumber \\
     N_B^- (\theta_B)  |  k_B, l_B \rangle^{(  \theta_B)} & = l_B | k_B, l_B \rangle^{( \theta_B)} .
\end{align}
Then the sign operators take the explicit form
\begin{align}
\text{sgn} [  S_A (\theta_A) ] & = \sum_{k_A \ge l_A}  | k_A, l_A \rangle^{(\theta_A)} \langle k_A, l_A |^{(\theta_A)} \nonumber \\
   &   - \sum_{k_A < l_A}  | k_A, l_A \rangle^{(\theta_A)} \langle k_A, l_A |^{(\theta_A)}  \nonumber \\
\text{sgn} [  S_B  (\theta_B)] & = \sum_{k_B \ge l_B}  | k_B, l_B \rangle^{(\theta_B)} \langle k_B, l_B |^{(\theta_B)}  \nonumber \\
   &   - \sum_{k_B < l_B}  | k_B, l_B \rangle^{(\theta_B)} \langle k_B, l_B |^{(\theta_B)} . 
     \label{signoperatorfock}
\end{align}

When a measurement of the state \eqref{splitsqueezed} is made in the basis \eqref{genfockstate}, the probabilities of the corresponding measurement outcomes are given by Born's rule as
\begin{align}
p_{k_A l_A k_B l_B}^{(\theta_A, \theta_B)} = 
| \langle \Psi |k_A, l_A, k_B, l_B \rangle^{(\theta_A, \theta_B)} |^2 .  
\label{probabilities}
\end{align}
Then we can evaluate the quantity \eqref{newedef} explicitly to be
\begin{align}
   & E_{\text{II}} (\theta_A, \theta_B)  =  \nonumber \\
   &   \sum_{k_A l_A k_B l_B  } p_{k_A l_A k_B l_B}^{(\theta_A, \theta_B)} 
     \text{sgn} (k_A - l_A)  \text{sgn}(k_B - l_B) . 
     \label{newedefprob}
\end{align}
It is clear from this expression that for each outcome labeled by $ (k_A, l_A, k_B, l_B )$, an outcome of $ \pm 1$ is assigned, and the average is performed over the full probability distribution for all possible outcomes..

\subsection{Numerical evaluation}
\label{sec:appii}

In this section we discuss the numerical results corresponding to the alternative definition of the correlator \eqref{newedefprob}.   For details regarding the numerical calculation, see Appendix \ref{app:numerics}. 

In Fig. \ref{fig2}(b) we show the Bell quantity $ \cal B $ using Approach II.  We see that in the ideal case ($\gamma =1$), its value starts at 2 and increases with the amount of squeezing. Bell violations are seen for all values of squeezing $ r>0 $. This rather different behavior compared to Approach I calls for an explanation. As we have seen in the low order expansion of the state $ | \Psi \rangle $ \eqref{loworderexp}, for small values of squeezing the state is dominated by the vacuum state, with a small contribution of a Bell state. For the vacuum state, \eqref{newedef} evaluates to $ E (\theta_A, \theta_B) = 1$ for all choices of angles.  
Substituting this into \eqref{bellmain}, we recover the result $ {\cal B} = 2$, which explains the result for $ r = 0 $.  As $ r $ is increased, the Bell state gives a contribution of $ 2 \sqrt{2}$ and increases the overall level of violation. In fact, higher order terms to \eqref{loworderexp} also show analogous anticorrelations in terms of $ S_A(0)$ and $ S_B (0)$, hence the level of violation continues to increase with $ r $ as seen in Fig. 2.

The dependence of the Bell quantity $ \cal B $ on the amount of squeezing in Fig. \ref{fig2}(b) appears to reach an inflection point around $ r \approx 0.4 $, from where it starts to plateau to a finite value. This flattening might however be partly caused by the truncation at $k_\text{cut}$ in our simulation. We expect that the violation will monotonically increase with $ r $, but due to the limits of our numerical methods we were unable to find the limiting value \cite{capraravivoli2015ChallengingPreconceptionsBella}. In Fig. \ref{fig2}(a) we show a comparison of the numerically determined $ \cal B $ and the exact value, both for Approach I.  We see that they are perfectly congruent until at least $ r = 0.6$, from where they start to deviate, with the numerical value underestimating the true value.  For levels of squeezing $r < 0.6$ we expect the numerical results to be reliable, as the numerical calculation of $ \cal B $ with Approach I agrees with the exact result \eqref{bellsimplecase} to within 0.5\%.

At this point it is natural to ask why, contrary to Approach II, Approach I shows an increasing violation for $ r \rightarrow 0 $. The reason for this is that by using the correlators according to  \eqref{espinexpr}, we implicitly discard all vacuum states for which $S_A=N_A=0$ or $S_B=N_B=0$. Since these states do not contribute to the normalization factor in the correlators, this can be considered  an effective postselection of the vacuum. To see this explicitly, we write the correlators \eqref{espinexpr} in terms of probabilities as
\begin{align}
  E_{\text{I}} (\theta_A, \theta_B) = 
    \frac{ \sum_{ k_A l_A k_B l_B } p_{k_A l_A k_B l_B}^{(\theta_A, \theta_B)} 
      (k_A - l_A)(k_B - l_B)}{\sum_{ k_A l_A k_B l_B } p_{k_A l_A k_B l_B}^{(\theta_A, \theta_B)} 
      (k_A + l_A)(k_B + l_B)} . 
      \label{origeexpprob}
\end{align}
Written in this form, it is clear that all vacuum terms with $k_A=l_A=0$ or $k_B=l_B=0$ are excluded from the sums. Also, the probabilities for $ k_A = l_A$ or $ k_B = l_B$ are assigned a factor of zero and do not contribute to the numerator.  We note that a similar decrease in the level of CHSH violation when including the vacuum and higher-order contributions was also observed in Refs. \cite{capraravivoli2015ChallengingPreconceptionsBella, zhao2019HigherAmountsLoopholefree}. 

While for small $r$ the level of violation is lower using Approach II than I, we consider the former to be a more rigorous evaluation of the CHSH inequality, as the full probability distribution over all possible measurement outcomes is used in the calculation, and only $ \pm 1 $ outcomes are assigned. We also note that using Approach II, no auxiliary assumptions such as the fair-sampling or the no-enhancement assumptions \cite{clauser1978BellTheoremExperimentala} are needed to achieve a violation of the CHSH inequality or, equivalently, of the CH inequality \cite{clauser1974ExperimentalConsequencesObjective}. Intuitively, the dependence of the Bell quantity violation on $r$ is reasonable, as one expects that the amount of correlations should increase with the level of squeezing $r$, which also reflects the amount of entanglement that is present in the system. At $ r=0 $ there are no quantum correlations in the system, hence one expects no violation of the CHSH inequality. Therefore, in our opinion, evaluation of the correlators following \eqref{newedefprob} provides a more rigorous approach if one is interested in the evaluation of the original CHSH inequality with as few additional assumptions as possible.

\subsection{Approach III: Evaluation with the CH inequality}
\label{sec:appiii}

As discussed in the previous section, one of the disadvantages of Approach II is that the level of violation is much smaller than the one of Approach I.  The main reason for this difference is that in Approach I measurements of the vacuum by Alice or Bob are assigned to a value of 0, as can be seen from \eqref{espinexpr} or \eqref{origeexpprob}.  Since the vacuum does not contain any correlations, but yet is the dominant term in the expansion \eqref{loworderexp}, removing this contribution allows one to obtain a large violation for small $ r $.  However, the postselection of undesirable measurement outcomes is well-known to carry the risk of opening loopholes in a Bell test, thereby invalidating the conclusion of a Bell inequality violation \cite{pearle1970HiddenVariableExampleBased, peres1997BellInequalitiesPostselection, larsson2014LoopholesBellInequality}. Hence, any such procedure should be carefully analyzed and justified with regard to the physical system on which the Bell test is performed.

To this end, we consider the Clauser-Horne (CH) inequality including the no-enhancement assumption \cite{clauser1974ExperimentalConsequencesObjective, larsson2014LoopholesBellInequality}.  In our notation this reads
\begin{align}
- P^{\forall \forall }  \le &  P^{\sigma \sigma'} ( \theta_A, \theta_B) + P^{\sigma \sigma'} ( \theta_A, \theta_B') + P^{\sigma \sigma'} ( \theta_A', \theta_B) \nonumber \\
& - P^{\sigma \sigma'} ( \theta_A', \theta_B') -  P^{\sigma \forall } ( \theta_A) -  P^{ \forall \sigma'} ( \theta_B) \le 0 
\label{chinequality}
\end{align}
where $ P^{\sigma \sigma'}$ is the probability of detecting an event of type $ \sigma$ at Alice and $ \sigma'$ at Bob, where $ \sigma, \sigma' \in \pm$. The index $ \forall$ denotes that any detection event is counted.  In the case of, e.g., photons with polarization encoding, this corresponds to removing the polarization analyzer and detecting all photons. In our case, we may define this more concretely by
\begin{align}
P^{\forall \forall } & = \sum_{\sigma,\sigma'= \pm} P^{\sigma \sigma'} (\theta_A, \theta_B)  \label{pallall} \\
P^{\sigma \forall } (\theta_A)   & = \sum_{\sigma'= \pm} P^{\sigma \sigma'} (\theta_A, \theta_B)   \label{psigmaall} \\
P^{\forall \sigma }  (\theta_B)   & =   \sum_{\sigma'= \pm} P^{\sigma' \sigma} (\theta_A, \theta_B)   .  \label{pallsigma}
\end{align}
An important aspect of all the probabilities that appear in \eqref{chinequality} is that only outcomes for Alice and Bob both registering a detection are counted.  This means that if Alice or Bob detects the vacuum, they are not counted in any of the probabilities.  The no-enhancement assumption amounts to demanding that the probability of an event of type $ \sigma $ is less than that where any detection is allowed:
\begin{align}
    P^\sigma ( \theta) \le P^\forall
    \label{noenhancement}
\end{align}
for Alice or Bob. We note that the definitions \eqref{pallall}-\eqref{pallsigma} imply that we should define $ P^\forall = \sum_{\sigma= \pm} P^{\sigma } (\theta) $, which would mean that \eqref{noenhancement} will always be satisfied. 

An alternative form of the CH inequality can be derived from \eqref{chinequality} taking the form \cite{larsson2014LoopholesBellInequality}
\begin{align}
    {\cal B} \le 2
\end{align}
by using the definition \eqref{bellmain} and defining the correlators 
\begin{align}
E_{\text{III}} (\theta_A, \theta_B) = 
\frac{1}{P^{\forall \forall}}\Big[  
P^{++} (\theta_A, \theta_B) + P^{--}  (\theta_A, \theta_B) \nonumber \\
- P^{+-} (\theta_A, \theta_B) - P^{-+}  (\theta_A, \theta_B)  \Big] .   \label{chineqappiii}
\end{align}
We show the derivation of this alternative form of the CH inequality in Appendix \ref{app:ch}.  The advantage of the CH inequality with the no-enhancement assumption is apparent in \eqref{chineqappiii}.  All probabilities involve only probability outcomes where Alice and Bob both record nonvacuum events, and 0 is assigned otherwise.  This means that events corresponding to vacuum measurements can be eliminated, and the factor of $ P^{\forall \forall}$ acts as a normalizing factor to account for the small probabilities of the nonvacuum terms.  In the case of atom number measurements, the detection of the vacuum corresponds to all the atoms being detected in the state $ m = 0$.  This is different to the photonic situation where vacuum detection corresponds to ``no click'' and is indistinguishable from no detection.  In the framework of the CH inequality, we may nevertheless assign such vacuum detection events to zero outcomes, which can be considered as a particular binning strategy.

We now explicitly show the expressions for the probabilities that appear in the CH inequality.  The two-outcome probabilities are binned as
\begin{align}
P^{\sigma \sigma'} (\theta_A, \theta_B) = \langle \Pi_A^\sigma (\theta_A) 
\Pi_B^{\sigma'} (\theta_B) \rangle \;,
\label{probexpressionappiii}
\end{align}
with projection operators defined as
\begin{align}
 \Pi_A^+ (\theta_A)  & = \sum_{
 \begin{subarray}{c} k_A \ge  l_A \\
     (\ne k_A=l_A=0)
     \end{subarray} }  | k_A, l_A \rangle^{(\theta_A)}  \langle k_A, l_A |^{(\theta_A)}  \nonumber \\
 \Pi_A^- (\theta_A)  & = \sum_{k_A < l_A}  | k_A, l_A \rangle^{(\theta_A)}  \langle k_A, l_A |^{(\theta_A)}  \nonumber \\
 \Pi_B^+ (\theta_B)& = \sum_{
 \begin{subarray}{c} k_B \ge  l_B \\
     (\ne k_B=l_B=0)
     \end{subarray} } 
     | k_B, l_B \rangle^{(\theta_B)}  \langle k_B, l_B |^{(\theta_B)}  \nonumber \\
 \Pi_B^- (\theta_B)& =  \sum_{k_B < l_B}  | k_B, l_B \rangle^{(\theta_B)}  \langle k_B, l_B |^{(\theta_B)} \label{projectionappiii} .
\end{align}
The four projection operators divide all possible outcomes into four regions, depending on whether $ k_A\ge l_A$ or $ k_A < l_A$, and similarly for Bob.  This is the same strategy as we performed in Approach II, and it involves the same quantities as seen in  \eqref{signoperatorfock}.  The only difference here is that the vacuum outcomes $ k_A = l_A = 0$ and $ k_B = l_B = 0 $ are excluded from the sums, which amounts to assigning them a value of zero.  This is done in accordance to the CH inequality, where only coincidence outcomes (i.e. a particle is detected by Alice and Bob in the $ m = \pm 1 $ states) are included.  

Expressions for the other probabilities are given from the definitions \eqref{pallall}-\eqref{pallsigma}. We have
\begin{align}
P^{\forall \forall}  =  \langle \Pi \rangle 
\end{align}
where
\begin{align}
\Pi = (I_A - \Pi_A^0) (I_B - \Pi_B^0) 
\label{projectordef}
\end{align}
is the projector onto the space where the vacuum state for Alice and Bob is removed. Here, we defined the identity operators $ I_{A}, I_B $,  and the vacuum projectors
\begin{align}
 \Pi_A^0 &  = | 0_A, 0_A \rangle \langle  0_A, 0_A | \nonumber \\
   \Pi_B^0 &  = | 0_B, 0_B \rangle \langle 0_B, 0_B |  
\end{align}
for Alice and Bob's respective subspaces. Although we do not need the expressions for \eqref{psigmaall} and \eqref{pallsigma}, we provide them for completeness:
\begin{align}
P^{\sigma \forall} (\theta_A)  & =  \langle   \Pi^\sigma_A ( \theta_A)  \Pi \rangle    \nonumber \\
P^{ \forall \sigma } (\theta_B) &  =  \langle  \Pi^\sigma_B ( \theta_B)  \Pi \rangle     . 
\end{align}
We point out that the assignment of the 1 or 0 values can be done locally by Alice and Bob since the projections in \eqref{probexpressionappiii} are of a product form. 

In terms of the correlators, using \eqref{chineqappiii}  we may write
\begin{align}
E_{\text{III}} (\theta_A, \theta_B) = \frac{\langle \Pi \text{sgn} [S_A (\theta_A)]  \text{sgn} [S_B (\theta_B)]  \rangle }{ \langle \Pi \rangle } . 
\label{appiiiecorre}
\end{align}
Here we used the fact that 
\begin{align}
\text{sgn} [S_A (\theta)] &  = \Pi^+_A( \theta) - \Pi^-_A( \theta) + \Pi^0_A \nonumber \\
\text{sgn} [S_B (\theta)] &  = \Pi^+_B( \theta) - \Pi^-_B( \theta) + \Pi^0_B  .
\end{align}
In terms of probabilities, the correlators in this case are
\begin{align}
     & E_{\text{III}} (\theta_A, \theta_B) = 
  \frac{1}{\langle  \Pi  \rangle } \nonumber \\
    &  \times 
     \sum_{\begin{subarray}{c} k_A l_A k_B l_B \\
     (\ne k_A=l_A=0, \\
     \ne k_B=l_B=0 )
     \end{subarray} } p_{k_A l_A k_B l_B}^{(\theta_A, \theta_B)} 
     \text{sgn} (k_A - l_A)  \text{sgn}(k_B - l_B) \; ,
     \label{newedefprob2}
\end{align}
where
\begin{align}
\langle  \Pi \rangle = 1 - \sum_{ k_A l_A  } p_{k_A l_A 0 0 }^{(\theta_A, \theta_B)}  - \sum_{ k_B l_B } p_{0 0 k_B l_B }^{(\theta_A, \theta_B)}  + p_{ 0 0 0 0 }^{(\theta_A, \theta_B)} . \label{probnumerator}
\end{align}
In the summation in \eqref{newedefprob2}, all terms where Alice detects the vacuum (i.e. $ k_A = l_A = 0 $), or Bob detects the vacuum (i.e. $ k_B = l_B = 0 $) are excluded.  Written in this form, it is clear that \eqref{appiiiecorre} can be evaluated by assigning a two-valued $ \pm 1 $ outcome for every measurement outcome labeled by $ k_A, l_A, k_B, l_B$, except for local vacuum outcomes.

In Fig. \ref{fig2}(a) we show the Bell quantity using CH correlators \eqref{newedefprob2}, indicated by the Approach III labels.  We observe that the level of violation is significantly greater than that of Approach I for finite values of squeezing $ r$. It also improves upon Approach II, which starts at $ {\cal B} = 2$ due to the included vacuum contribution. 
Even for the largest values of squeezing we calculated, there is a violation of $ {\cal B} \approx 2.48$. Overall, this approach is an improvement to Approach I, since it shows a significantly larger level of violation without discarding any further measurement outcomes.

\section{Effects of loss and imperfect atom number resolution}
\label{sec:loss}

We finally show the robustness to noise of the approaches we considered, by showing the effect of loss and imperfect atom number resolution on the Bell violations previously evaluated.

\subsection{Loss}

The primary mechanism of decoherence in atomic systems is particle losses, where atoms escape from the trap.  We take a relatively simple approach to incorporating losses, where we assume that initially the state \eqref{splitsqueezed} is prepared and the rotation operators \eqref{rotationopera} are applied without decoherence, producing the state 
\begin{align}
    | \Psi(\theta_A, \theta_B) \rangle  = V(\theta_A, \theta_B) U_{\Psi} | 0 \rangle . 
    \label{allstate}
\end{align}
Losses are then considered to take place after state preparation, resulting in the state 
\begin{align}
\rho = & \sum_{n_A m_A n_B m_B} F_{n_A} F_{m_A} F_{n_B} F_{m_B} \nonumber \\
& \times | \Psi(\theta_A, \theta_B \rangle\langle \Psi(\theta_A, \theta_B | F_{m_B}^\dagger F_{n_B}^\dagger F_{m_A}^\dagger F_{n_A}^\dagger
\label{rhodef}
\end{align}
according to the operator-sum representation.  Here the Kraus operators for the loss of $ n $ atoms on a single mode are defined as \cite{nielsen2010QuantumComputationQuantuma}
\begin{align}
F_n=\sum^{\infty}_{k=n}\sqrt{ {k \choose n}}  \sqrt{\gamma ^{k-n}(1-\gamma)^n}|k-n\rangle\langle k| ,
\label{krausdef}
\end{align}
where $ 1- \gamma $ is the probability of losing an atom.  The four operators in \eqref{rhodef} are for the four atom species as given in \eqref{genfockstate}.  

Since we have already taken account of the Bell angle rotations in the state \eqref{allstate}, here the relevant correlations to be evaluated using Approach I are
\begin{align}
    E_{\text{I}}^\gamma (\theta_A, \theta_B) = \frac{\text{Tr} (  S^z_A   S^z_B  \rho  ) }{\text{Tr} ( N_A  N_B  \rho)  } \;,
    \label{newedefrhoapp1}
\end{align}
while for Approach II we have
\begin{align}
    E_{\text{II}}^\gamma (\theta_A, \theta_B) = \text{Tr} ( \text{sgn}[ S^z_A ]  \text{sgn} [S^z_B  ]\rho  ) ,
    \label{newedefrho}
\end{align}
where 
\begin{align}
    S^z_A &  = a_1^\dagger a_1 - a_{-1}^\dagger a_{-1} \nonumber \\
    S^z_B &  = b_1^\dagger b_1 - b_{-1}^\dagger b_{-1} ,
\end{align}
$N_A, N_B $ are defined in \eqref{numberalicebob}, and the sign function is taken as before. 

When evaluating \eqref{newedefrho}, instead of applying the Kraus operators on the state, we find it more convenient to apply them on the observable operator, defining an observable incorporating losses. We defer the derivations to Appendix \ref{app:loss} and present only the final results here.  For Approach I, from the fact that 
\begin{align}
  \sum_n F^\dagger_n a^\dagger a F_n = \gamma a^\dagger a ,
  \label{lossynumber}
\end{align}
all $ \gamma $ factors cancel out and the same expression as \eqref{origeexpprob} is obtained. Hence, for Approach I, the Bell measurements are unchanged under loss.

For Approach II, we obtain
\begin{align}
E_{\text{II}}^\gamma ( \theta_A, \theta_B) =  \sum_{k_A l_A k_B l_B} p_{k_A l_A k_B l_B}^{(\theta_A, \theta_B)} G^\gamma_{k_A l_A} G^\gamma_{k_B l_B} ,
\label{eprimenopos}
\end{align}
where the sign operators including loss are defined as
\begin{align}
G^\gamma_{k l} = 1 - 2 \sum_{n<m} {k \choose k-n}  {l \choose l-m} \gamma^{n+m} (1- \gamma)^{k-n + l - m} .
\label{gcool}
\end{align}

For Approach III, using the expression \eqref{appiiiecorre} we may write
\begin{align}
E_{\text{III}}^\gamma ( \theta_A, \theta_B) = 
\frac{ 1}{\text{Tr} ( \Pi \rho) } \Big[
\text{Tr} ( \text{sgn}[ S^z_A]  \text{sgn} [S^z_B ]\rho  ) \nonumber \\
- \text{Tr} ( \Pi_A^0  \text{sgn} [S^z_B ]\rho  ) 
- \text{Tr} ( \text{sgn} [S^z_A ]  \Pi_B^0 \rho  ) 
+  \langle 0 | \rho | 0 \rangle
\Big]
\label{eprimewithpos}
\end{align}
Here the first term in the numerator is the same as the Approach II correlator \eqref{eprimenopos}.  The remaining terms are given by
\begin{align}
\text{Tr} ( \Pi_A^0  \text{sgn} [S^z_B ]\rho  ) & =  \sum_{k_A l_A k_B l_B} p_{k_A l_A k_B l_B}^{(\theta_A, \theta_B)} H_{k_A l_A}^\gamma G^\gamma_{k_B l_B}  \nonumber \\
\text{Tr} ( \text{sgn} [S^z_A ]\rho  \Pi_B^0  ) & =  \sum_{k_A l_A k_B l_B} p_{k_A l_A k_B l_B}^{(\theta_A, \theta_B)} G^\gamma_{k_A l_A} H_{k_B l_B}^\gamma  \nonumber \\
\langle 0 | \rho | 0 \rangle & = \sum_{k_A l_A k_B l_B} p_{k_A l_A k_B l_B}^{(\theta_A, \theta_B)} H_{k_A l_A}^\gamma H_{k_B l_B}^\gamma  ,
\end{align}
where we defined
\begin{align}
    H_{k l}^\gamma = | \langle 0,0 | F_k F_l | k l \rangle |^2 =  (1- \gamma)^{k+l} .  
\end{align}
Finally, we have
\begin{align}
\text{Tr} ( \Pi \rho) = 1 - \text{Tr} ( \Pi_A^0 \rho) - \text{Tr} ( \Pi_B^0 \rho) + \langle 0 | \rho | 0 \rangle,
\end{align}
where
\begin{align}
\text{Tr} ( \Pi_A^0 \rho) & = \sum_{k_A l_A k_B l_B} p_{k_A l_A k_B l_B}^{(\theta_A, \theta_B)} H_{k_A l_A}^\gamma   \nonumber \\
\text{Tr} ( \Pi_B^0  \rho  ) & =  \sum_{k_A l_A k_B l_B} p_{k_A l_A k_B l_B}^{(\theta_A, \theta_B)}  H_{k_B l_B}^\gamma .
\end{align}

\subsection{Imperfect number resolution}

Atom number detectors in practice have often an imperfect number resolution. 
 We model this as a detector inefficiency, where there is a probability $ 1- \eta$ of failing to detect an atom.  Following Ref. \cite{barnett1998ImperfectPhotodetectionProjection}, we define the probability of obtaining a readout of $ k -n $ for a physical number state $ k $ as
\begin{align}
P_{\eta} (k,n) = {k \choose n } \eta^{k-n} (1- \eta)^n ,
\label{detecprob}
\end{align}
which is a normalized as $ \sum_n P_{\eta} (k,n) = 1 $. With respect to the readout $ k-n$, this distribution has a mean value
\begin{align}
\mu_{\text{det}}  = k\eta
\end{align}
and  standard deviation
\begin{align}
\sigma_{\text{det}} = \sqrt{k (1- \eta)\eta}.  
\end{align}
The distribution is approximately distributed as a Gaussian for $ 1/k \lesssim \eta \lesssim 1- 1/k $. A typical efficiency that may be achieved experimentally  using absorption imaging is a resolution of $ \pm 4 $ atoms for $ 1000 $ atoms \cite{pezze2018QuantumMetrologyNonclassical}. 

Now consider a physical collapse on a number state $ |k  \rangle $.  Then, with probability $ P_{\eta} (k,n)  $ this will be counted as an outcome with $ k-n$ atoms. The  number operator including detector inefficiency is then
\begin{align}
N^{\eta} & = \sum_k \sum_n (k-n) P_{\eta} (k,n) | k \rangle \langle k |  \nonumber \\
& = \eta a^\dagger a  .  
\label{detectionnumber}
\end{align}
In this case, correlators for Approach I are defined as
\begin{align}
E_{\text{I}}^\eta (\theta_A, \theta_B) = \frac{ \langle  ( N_A^{\eta+} - N_A^{\eta-})  ( N_B^{\eta+} - N_B^{\eta-})  \rangle }{ \langle ( N_A^{\eta+} + N_A^{\eta-})  ( N_B^{\eta+} + N_B^{\eta-})  \rangle}
\end{align}
where the expectation values are evaluated with respect to the state \eqref{allstate}.  Due to the linear nature of \eqref{detectionnumber}, as with loss, the factors $ \eta$ cancel out, and the same expression as \eqref{origeexpprob} is obtained.  Hence, for Approach I the Bell measurements are unchanged under imperfect detector efficiency.  

For Approach II, we define a version of the spin sign operator with detection inefficiency as
\begin{align}
\text{sgn}_{\eta} (S^z ) =&  \sum_{k l } \sum_{nm} \text{sgn} ( k - n - l +m ) \nonumber \\
& \times P_{\eta} (k,n) P_{\eta} (l,m)  | k, l \rangle \langle k, l | \nonumber \\
= &  \sum_{k l } G_{kl}^{\eta}  | k, l \rangle \langle k, l | ,
\end{align}
where the function $  G_{kl}^{\eta} $ is the same as that defined in \eqref{gcool} with $ \gamma \rightarrow \eta$.  The correlators for Approach II are in this case defined as
\begin{align}
    E_{\text{II}}^\eta  (\theta_A, \theta_B) = \langle \text{sgn}_\eta [ S^z_A ]  \text{sgn}_\eta [S^z_B  ] \rangle ,
\end{align}
where the expectation values are evaluated with respect to the state \eqref{allstate}. This takes precisely the same form as \eqref{eprimenopos}, since the functional form of the lossy spin sign operators is the same as that for that including detection inefficiency. 

Similarly, for Approach III the correlators give exactly the same result as \eqref{eprimewithpos}.  The calculations are shown in Appendix \ref{app:detineff}.  The reason for this is that in our model, loss occurs after state preparation, and occurs with a probability distribution that is identical to \eqref{detecprob}.  In this case the loss of physical photons is exactly equivalent to the lack of detection of atoms and results in the same modified probability distribution up to the replacement $ \gamma \rightarrow \eta$.  Since the results for loss and imperfect detector efficiency have exactly the same dependence, we henceforth consider only the effect of loss.

\subsection{Numerical results}

In Fig. \ref{fig2}(a) we show numerical results including Approach III with losses.  We see that as expected, the effect of the loss is to diminish the level of violation.  We see that for $ 1- \gamma =  0.1 $ loss probability, the results are almost unchanged from the ideal results, and still show a large violation even for loss probabilities up to $ 1- \gamma =0.3 $. We also do not see a very large change in the $ r $ dependence when including losses. Our results suggest that our observed Bell violations are robust to experimental imperfections.  

For Approach II, we found a more significant dependence on losses, see Fig. \ref{fig2}(b).  At  $ r \rightarrow 0  $, all curves including those with losses start at $ {\cal B} = 2$, which is expected as the presence of losses should increase only the vacuum contribution. When losses are taken into account, the violation decreases for $ r > 0 $, and no violation is seen for any $ r $ when $ 1- \gamma \sim  0.3$. For loss probabilities in the region of  $ 1- \gamma = 0.1 $, however, we conclude that it is still possible to detect Bell violations. Since the level of violation is smaller in this postselection-free approach, we imagine that it could be more difficult to observe experimentally.

Another possible experimental imperfection lies in the interaction time for which the squeezing and ``desqueezing'' are performed. To investigate the effect of non-ideal squeezing times, we numerically calculated the Bell violations for all three approaches using ``desqueezing'' times that were 50\% lower than the ideal times. As expected, the level of Bell violation decreases as interaction times differ from their optimal values. The maximum Bell violations decreased by less than 10\% for $ r<0.2 $. We therefore expect that this source of imperfection can be sufficiently controlled in realistic experimental scenarios.

\section{Summary and conclusions}
\label{sec:conc}

We have proposed a scheme for observing Bell correlations between spatially separated Bose-Einstein condensates using the CHSH inequality.  The state preparation protocol consists of three main steps.  The first step involves exploiting spin-changing collisions in a single BEC to produce the analog of a two-mode squeezed state. In the second step, this system is split into two BECs by controlling,  e.g. the trapping potential. As a third step, spin-changing collisions are exploited locally on each BEC, in order to remove undesired local squeezing terms from the state. In the two-particle sector, this leaves the two BECs in a Bell state.  To evaluate the Bell correlations, three approaches were considered.  In Approach I, the methods of Ref. \cite{ralph2000ProposalMeasurementBellType} was used, where the correlators are evaluated as ratios of number operator expectation values.  Despite the fact that maximal Bell violations are observed using this approach, we find that this method of evaluating the correlators deviates from the original CHSH derivation, since it does not use genuine two-valued measurement outcomes $ \pm 1 $. Moreover, it also implicitly does not sum over the full probability distribution of all possible outcomes, since particular outcomes (notably vacuum measurements by Alice and Bob) are assigned to a zero value and thus do not contribute to the correlators.  We therefore introduced Approach II based on sign functions of spin operators, such that genuine $ \pm 1$ measurement outcomes are constructed and averaged over the full set of possible outcomes. Even though this approach involves no postselection and no auxiliary assumptions, violations are still observed, albeit at a lower level due to the inclusion of the vacuum. In Approach III, we use the CH inequality with the no-enhancement assumption to effectively remove the local vacuum contributions.  Here the level of violation again reaches the maximum value of $ 2 \sqrt{2} $ for small squeezing, and shows larger levels of violation than Approach I.  The results are found to be rather robust in the presence of loss, and violations are observed for reasonable parameters.  

Each of the approaches that we have considered possess advantages and disadvantages.  One major advantage of Approach I is that it is highly robust against loss and detector resolution issues. For the models considered in this work, the results remain completely unchanged.  However, it has the conceptual disadvantage that the correlators are evaluated not strictly following a CHSH evaluation.  Both Approaches II and III are improvements of Approach I, and in this sense they should be the preferred methods if a rigorous evaluation of correlators is a priority. Approach II represents the most rigorous violation of Bell's inequality, in the sense that it does not require any auxiliary assumptions besides those of Bell's theorem. However, it has the practical disadvantage of a smaller level of violation and higher susceptibility to losses. Approach III provides significantly larger levels of violation, but it requires an auxiliary no-enhancement assumption. When losses are taken into account, it additionally requires a fair-sampling assumption, i.e., that losses are independent of the measurement settings. It is thus vulnerable to two more loopholes than Approach II.

The main experimental challenges of our scheme lie in the splitting process during the initial state preparation and the imperfect atomic number resolution in order to evaluate the probability densities \eqref{probabilities}.  With current technology, a resolution at the level of $ \pm 4$ atoms can be performed within $ \sim 1000 $ atoms \cite{pezze2018QuantumMetrologyNonclassical}. One of the advantages of our approach is that the sign-operator approach is relatively insensitive to atom number fluctuations, unlike for example the parity operator \cite{oudot2019BipartiteNonlocalityManybody}.  We do note that for small squeezing the dominant contribution to the Bell violation is in the $ N_A = N_B = 1$ particle sector, for which single atom resolution is required. Hence in the case of limited atom number resolution, it may be advantageous to probe the large squeezing regime where the effective spins are larger. We note that there have been improvements in atom number resolution recently which are beneficial towards the observation of Bell correlations \cite{hume2013AccurateAtomCounting, huper2020NumberresolvedPreparationMesoscopic, qu2020ProbingSpinCorrelations}.  While we primarily considered the split-squeezing approach to generate correlations between spatially separated BECs, alternative methods such as that discussed in Sec. \ref{sec:alterna} could equally be applied to the sign-operator method of Approaches II and III.

\section*{Acknowledgments}

This work is supported by the National Natural Science Foundation of China (62071301), State Council of the People’s Republic of China (D1210036A), NSFC Research Fund for International Young Scientists (11850410426), NYU-ECNU Institute of Physics at NYU Shanghai, Science and Technology Commission of Shanghai Municipality (19XD1423000), China Science and Technology Exchange Center (NGA-16-001), and the NYU Shanghai Boost Fund.

\appendix

\section{Approximated Hamiltonians (\ref{ham2m}) and (\ref{Ham_after_split})}
\label{app:m=0}

At two points [Eq. \eqref{ham2m} and Eq. \eqref{Ham_after_split}] we make use of the local oscillator approximation, i.e., the approximation of the $a_0$ and $b_0$ modes as classical fields. In the following we provide calculations confirming the validity of the approximations in the limits of $r \ll 1$ and $N \gg 1$,  and present additional numerical results which were obtained with the unapproximated Hamiltonian \eqref{mainham}. 

\begin{widetext}
\subsection{First-order calculation confirming the approximated Hamiltonian (\ref{ham2m})}

Given the exact Hamiltonian (2), $H=\hbar g(a^{\dagger}_{-1}a^{\dagger}_{1}a_{0}a_{0}+a^{\dagger}_{0}a^{\dagger}_{0}a_{-1}a_{1})$, its induced unitary transformation $U_r$ now reads
\begin{align}
U_r = e^{-iHt/\hbar} = e^{-\frac{ir}{N}(a^{\dagger}_{-1}a^{\dagger}_{1}a_{0}a_{0}+a^{\dagger}_{0}a^{\dagger}_{0}a_{-1}a_{1})}.
\end{align}
In the following we evaluate each of the steps in the state preparation protocol \eqref{upsiseq} to first order in $r$. 
Up to the linear term in $r$, the state after the first step can be directly calculated to be
\begin{align}|\psi_1\rangle = U_{2r} |\psi_0\rangle  &= \left[1 - i\frac{2r}{N}(a^{\dagger}_{-1}a^{\dagger}_{1}a_{0}a_{0}+a^{\dagger}_{0}a^{\dagger}_{0}a_{-1}a_{1}) \right] \frac{1}{\sqrt{N!}}\left(a_0^{\dagger} \right)^N |0\rangle \nonumber \\ 
&= \left[\frac{1}{\sqrt{N!}}\left(a_0^{\dagger} \right)^N - i2r \frac{\sqrt{N(N-1)}}{N}\frac{1}{\sqrt{(N-2)!}} \left(a_0^{\dagger} \right)^{N-2}a_1^{\dagger}a_{-1}^{\dagger} \right] |0\rangle \;.
\end{align}
The result of the second step can also immediately be evaluated to
\begin{align}
|\psi_2\rangle = U_{\pi}|\psi_1\rangle 
= \left[\frac{1}{\sqrt{N!}}\left(a_0^{\dagger} \right)^N + i2r  \frac{\sqrt{N(N-1)}}{N}\frac{1}{\sqrt{(N-2)!}} \left(a_0^{\dagger} \right)^{N-2}a_1^{\dagger}a_{-1}^{\dagger} \right] |0\rangle \; .
\end{align}
The third step evenly splits all three modes: 
\begin{align}
|\psi_3 \rangle &= U_W |\psi_2 \rangle  \nonumber \\
 &= \left[ \underbrace{\frac{1}{\sqrt{N!}}\left(\frac{a_0^{\dagger} + b_0^{\dagger}}{\sqrt{2}}\right)^N}_\text{(i)} + \underbrace{i2r \frac{\sqrt{N(N-1)}}{N}\frac{1}{\sqrt{(N-2)!}} \left(\frac{a_0^{\dagger} + b_0^{\dagger}}{\sqrt{2}} \right)^{N-2} \left(\frac{a_1^{\dagger} + b_1^{\dagger}}{\sqrt{2}} \right)\left(\frac{a_{-1}^{\dagger} + b_{-1}^{\dagger}}{\sqrt{2}} \right)}_\text{(ii)} \right] |0\rangle \; .
\end{align}
To evaluate the fourth step $|\psi_4\rangle = U_{2r}^{(a)}U_{2r}^{(b)} |\psi_3 \rangle$, we divide the above formula into the two terms labeled by (i) and (ii). Starting with the unitary $U_{2r}^{(a)}$ and the first term (i) in $|\psi_3\rangle$, we can rewrite and evaluate 
\begin{align} U_{2r}^{(a)} \frac{1}{\sqrt{N!}}\left(\frac{a_0^{\dagger} + b_0^{\dagger}}{\sqrt{2}}\right)^N |0\rangle 
&= U_W^a \left(U_W^a \right)^{\dagger} e^{-i\frac{2r}{N}(a_1^{\dagger}a_{-1}^{\dagger}a_0^2 + \text{H.c.})} U_W^a \left(U_W^a \right)^{\dagger} \frac{1}{\sqrt{N!}}\left(\frac{a_0^{\dagger} + b_0^{\dagger}}{\sqrt{2}}\right)^N  U_W^a \left(U_W^a \right)^{\dagger} |0\rangle  \\ 
&= U_W^a e^{-i\frac{2r}{N}\left(a_1^\dagger a_{-1}^\dagger \frac{(a_0 + b_0)^2}{2} + \text{H.c.} \right) } \frac{1}{\sqrt{N!}} \left(a_0^\dagger \right)^N |0\rangle  \\ 
&\approx U_W^a \left( 1 - i\frac{r}{N}\left(a_1^\dagger a_{-1}^\dagger (a_0 + b_0)^2 + (a_0^\dagger + b_0^\dagger)^2 a_1a_{-1} \right) \right) \frac{1}{\sqrt{N!}} \left(a_0^\dagger \right)^N |0\rangle  \\ 
&= U_W^a \left( \frac{1}{\sqrt{N!}} \left(a_0^\dagger \right)^N - ir\frac{\sqrt{N(N-1)}}{N} a_1^\dagger a_{-1}^\dagger \frac{\left(a_0^\dagger \right)^{N-2}}{\sqrt{(N-2)!}} \right) |0\rangle  \\ 
&=  \left[ \frac{1}{\sqrt{N!}} \left(\frac{a_0^{\dagger} + b_0^{\dagger}}{\sqrt{2}} \right)^N - ir\frac{\sqrt{N(N-1)}}{N} a_1^\dagger a_{-1}^\dagger \frac{1}{\sqrt{(N-2)!}}\left(\frac{a_0^{\dagger} + b_0^{\dagger}}{\sqrt{2}}\right)^{N-2} \right] |0\rangle
\end{align}
to first order in $r$, with 
\begin{align}
U_W^a = e^{-\frac{\pi}{4}(a_0^\dagger b_0 + b_0^\dagger a_0)} \; .
\label{a-rotation}
\end{align}
With a completely analogous calculation we also obtain 
\begin{align}
U_{2r}^{(b)} \frac{1}{\sqrt{N!}}\left(\frac{a_0^{\dagger} + b_0^{\dagger}}{\sqrt{2}}\right)^N |0\rangle = \left[ \frac{1}{\sqrt{N!}} \left(\frac{a_0^{\dagger} + b_0^{\dagger}}{\sqrt{2}} \right)^N -  ir\frac{\sqrt{N(N-1)}}{N} b_1^\dagger b_{-1}^\dagger \frac{1}{\sqrt{(N-2)!}}\left(\frac{a_0^{\dagger} + b_0^{\dagger}}{\sqrt{2}}\right)^{N-2} \right] |0\rangle
\end{align}
to first order in $r$.
The action on the second term (ii) can be evaluated in a similar fashion. Performing the same manipulations as before (inserting three copies of $U_W^a(U_W^a)^\dagger$ and expanding the exponential to first order), we arrive at %
\begin{align} &U_{2r}^{(a)} \left[ ir \frac{\sqrt{N(N-1)}}{N}\frac{1}{\sqrt{(N-2)!}} \left(\frac{a_0^{\dagger} + b_0^{\dagger}}{\sqrt{2}} \right)^{N-2}(a_1^{\dagger} + b_1^{\dagger})(a_{-1}^{\dagger} + b_{-1}^{\dagger})\right] |0\rangle \\ &\approx U_W^a \left[1 - ir \left(a_1^\dagger a_{-1}^\dagger(a_0 + b_0)^2 + (a_0^\dagger + b_0^\dagger)^2a_1a_{-1} \right) \right] ir \frac{\sqrt{N(N-1)}}{N}\frac{1}{\sqrt{(N-2)!}} \left(a_0^{\dagger}\right)^{N-2}(a_1^{\dagger} + b_1^{\dagger})(a_{-1}^{\dagger} + b_{-1}^{\dagger}) |0\rangle \\  &=  U_W^a \left[ir \frac{\sqrt{N(N-1)}}{N}\frac{1}{\sqrt{(N-2)!}} \left(a_0^{\dagger}\right)^{N-2}(a_1^{\dagger} + b_1^{\dagger})(a_{-1}^{\dagger} + b_{-1}^{\dagger}) + r^2 (\dots) \right] |0\rangle \; , 
\end{align}
where we omitted the detailed evaluation of terms entering in the second order of $r$. Keeping only terms up to linear order in $r$ and applying the rotation, this is simply 
\begin{equation}
ir \frac{\sqrt{N(N-1)}}{N}(a_1^{\dagger} + b_1^{\dagger})(a_{-1}^{\dagger} + b_{-1}^{\dagger}) \frac{1}{\sqrt{(N-2)!}}\left(\frac{a_0^{\dagger} + b_0^{\dagger}}{\sqrt{2}} \right)^{N-2} |0\rangle  \; .
\end{equation}
When we also apply $U_{2r}^{(b)}$ to this state, only the $U_{2r}^{(b)} \approx 1$ term will be relevant to first order in $r$, so that we finally arrive at
\begin{align} &U_{2r}^{(b)} U_{2r}^{(a)} \left[ ir \frac{\sqrt{N(N-1)}}{N}\frac{1}{\sqrt{(N-2)!}} \left(\frac{a_0^{\dagger} + b_0^{\dagger}}{\sqrt{2}} \right)^{N-2}(a_1^{\dagger} + b_1^{\dagger})(a_{-1}^{\dagger} + b_{-1}^{\dagger})\right] |0\rangle \\ &\approx ir \frac{\sqrt{N(N-1)}}{N}(a_1^{\dagger} + b_1^{\dagger})(a_{-1}^{\dagger} + b_{-1}^{\dagger}) \frac{1}{\sqrt{(N-2)!}} \left(\frac{a_0^{\dagger} + b_0^{\dagger}}{\sqrt{2}} \right)^{N-2} |0\rangle \; . 
\end{align}
Combining the actions on the two terms of $|\psi_3\rangle$ gives the final state
\begin{align} |\psi_4\rangle = U_{2r}^{(b)} U_{2r}^{(a)} |\psi_3\rangle &=  \left[ \frac{1}{\sqrt{N!}} \left(\frac{a_0^{\dagger} + b_0^{\dagger}}{\sqrt{2}}\right)^N -ir \frac{\sqrt{N(N-1)}}{N}(a_1^\dagger a_{-1}^\dagger + b_1^\dagger b_{-1}^\dagger) \frac{1}{\sqrt{(N-2)!}} \left(\frac{a_0^{\dagger} + b_0^{\dagger}}{\sqrt{2}}\right)^{N-2}\right. \nonumber \\ &+ \left. ir \frac{\sqrt{N(N-1)}}{N}(a_1^\dagger + b_1^\dagger)(a_{-1}^\dagger + b_{-1}^\dagger) \frac{1}{\sqrt{(N-2)!}} \left(\frac{a_0^{\dagger} + b_0^{\dagger}}{\sqrt{2}}\right)^{N-2} \right] |0\rangle \\ &= \left[\frac{1}{\sqrt{N!}}\left(\frac{a_0^{\dagger} + b_0^{\dagger}}{\sqrt{2}}\right)^N + ir\frac{\sqrt{N(N-1)}}{N}(a_1^\dagger b_{-1}^\dagger + b_1^\dagger a_{-1}^\dagger) \frac{1}{\sqrt{(N-2)!}}\left(\frac{a_0^{\dagger} + b_0^{\dagger}}{\sqrt{2}}\right)^{N-2} \right] |0\rangle 
\end{align}
to first order in $r$.

We can also write the state in a Fock basis as 
\begin{align}
|\psi_4\rangle = |0, 0, 0, 0, N, 0\rangle 
+ ir \frac{\sqrt{N(N-1)}}{N} (|1,0,1,1, N-2, 0\rangle + |0, 1, 1, 1, N-2, 0\rangle) + \dots \, , 
\end{align}
where we defined the Fock states as 
\begin{align}
|k_A, k_B, N_A, N_B, N_+, N_- \rangle = 
\frac{\left(a_1^\dagger\right)^{k_A}}{\sqrt{k_A!}} \frac{\left(a_{-1}^\dagger \right)^{N_A - k_A}}{\sqrt{(N_A - k_A)!}} \frac{\left(b_1^\dagger\right)^{k_B}}{\sqrt{k_B!}} \frac{\left(b_{-1}^\dagger \right)^{N_B - k_B}}{\sqrt{(N_B - k_B)!}} \frac{1}{\sqrt{N_+!}}
\left(\frac{a_0^{\dagger} + b_0^{\dagger}}{\sqrt{2}}\right)^{N_+} \frac{1}{\sqrt{N_-!}}\left(\frac{a_0^{\dagger} - b_0^{\dagger}}{\sqrt{2}}\right)^{N_-} |0\rangle \; .
\label{fockbasis}
\end{align}
For large $N$, we can approximate $\frac{\sqrt{N(N-1)}}{N} \approx 1$. We can trace out the $N_A, N_B, N_+,$ and $N_-$ modes, leading to the state 
\begin{align}
\rho = |0, 0\rangle \langle 0, 0 | + r^2\left(|1,0\rangle + |0, 1\rangle \right) \left(\langle 1, 0| + \langle 0, 1| \right) + \dots
\end{align}
for large $N$. This is almost the same state as \eqref{loworderexp}: In contrast to \eqref{loworderexp}, it lacks coherences between different particle-number sectors. This does not affect any of the results, however, since all observables in Approaches I, II, and III conserve particle number. 
\end{widetext}

\subsection{Derivation of the Hamiltonian \eqref{Ham_after_split} for spin-changing collisions after splitting}

We now show that the application of the local Hamiltonian for spin-changing collisions \eqref{mainham} to the split initial state \eqref{m0_split} results in the Hamiltonian \eqref{Ham_after_split}. First we apply the local squeezing Hamiltonian \eqref{mainham} for the BEC labeled by $a$ on the split state,
\begin{align}
&e^{-igt(a_1^\dagger a_{-1}^\dagger a_0^2+(a_0^\dagger)^2 a_1a_{-1})}\frac{1}{\sqrt{N!}}\left(\frac{a_0^\dagger+b_0^\dagger}{\sqrt2}\right)^N|0\rangle .
\end{align}
With $U_W^a$ as in \eqref{a-rotation} we can rewrite this state as
\begin{align}
&U_W^a{U_W^a}^\dagger e^{-igt(a_1^\dagger a_{-1}^\dagger a_0^2+(a_0^\dagger)^2 a_1a_{-1})} U_W^a{U_W^a}^\dagger \nonumber\\
&\times\frac{1}{\sqrt{N!}}\left(\frac{a_0^\dagger+b_0^\dagger}{\sqrt2}\right)^N U_W^a{U_W^a}^\dagger |0\rangle\nonumber\\
&=U_W^ae^{-igt(a_1^\dagger a_{-1}^\dagger \frac{(a_0+b_0)^2}{2}+\frac{(a_0^\dagger+b_0^\dagger)^2}{2} a_1a_{-1})}\frac{1}{\sqrt{N!}}(a_0^\dagger)^N|0\rangle .
\label{f2}
\end{align}
Since the state on which the squeezing Hamiltonian acts in \eqref{f2} now contains only atoms in the $a_0$ mode, all terms in the Hamiltonian that directly act on it with a $b_0$, $a_1$ or $a_{-1}$ annihilation operator will evaluate to zero. After expanding the exponential up to second order and simplifying in this way, we obtain 
\begin{align}
& U_W^a \left[\frac{(a_0^{\dagger})^N}{\sqrt{N!}} - \frac{igt\sqrt{N(N-1)}}{2} \frac{(a_0^{\dagger})^{N-2}}{\sqrt{(N-2)!}} a_1^{\dagger}a_{-1}^{\dagger}\right. \\ \nonumber
&- \frac{(gt)^2}{2} \frac{\sqrt{N(N-1)(N-2)(N-3)}}{4} \frac{(a_0^{\dagger})^{N-4}}{\sqrt{(N-4)!}}(a_1^{\dagger}a_{-1}^{\dagger})^2 \\ \nonumber &- \left.\frac{(gt)^2}{2}\frac{N(N-1)}{2}\left(\frac{a_0^{\dagger}+ b_0^{\dagger}}{\sqrt{2}} \right)^2  \frac{(a_0^{\dagger})^{N-2}}{\sqrt{N!}} + \dots  \right] |0\rangle  .
\end{align}
For $N \gg 1$, this can be approximated as
\begin{align}
\approx U_W^a &\left[\frac{(a_0^{\dagger})^N}{\sqrt{N!}} - \frac{igtN}{2} \frac{(a_0^{\dagger})^{N-2}}{\sqrt{(N-2)!}} a_1^{\dagger}a_{-1}^{\dagger}\right. \\ \nonumber
&- \frac{(gtN)^2}{8} \frac{(a_0^{\dagger})^{N-4}}{\sqrt{(N-4)!}}(a_1^{\dagger}a_{-1}^{\dagger})^2 \\ \nonumber &- \left.\frac{(gtN)^2}{4} \left(\frac{a_0^{\dagger}+ b_0^{\dagger}}{\sqrt{2}} \right)^2  \frac{(a_0^{\dagger})^{N-2}}{\sqrt{N!}}  + \dots \right] |0\rangle    .
\end{align}
Now consider the last term in this expansion:
\begin{align}
 &\left(\frac{a_0^{\dagger}+ b_0^{\dagger}}{\sqrt{2}} \right)^2  \frac{(a_0^{\dagger})^{N-2}}{\sqrt{N!}} |0 \rangle  = \frac{1}{2} \Big[ \left( \frac{(a_0^{\dagger})^{N}}{\sqrt{N!}} \right) \nonumber \\
& + \frac{2}{\sqrt{N}} \left( \frac{(a_0^{\dagger})^{N-1} b_0^{\dagger}}{\sqrt{(N-1)!}} \right)\nonumber \\
& + \sqrt{\frac{2 }{N(N-1)}}
\left( \frac{(a_0^{\dagger})^{N-2} (b_0^{\dagger})^2}{ \sqrt{N!} \sqrt{2}}  \right) \Big] | 0 \rangle  .
\end{align}
The terms in the brackets, when applied on the vacuum, give normalized states. For $ N \gg 1$, only the first term is the dominant contribution, hence we drop the remaining terms to give
\begin{align}
\approx U_W^a &\left[\frac{(a_0^{\dagger})^N}{\sqrt{N!}} - \frac{igtN}{2} \frac{(a_0^{\dagger})^{N-2}}{\sqrt{(N-2)!}} a_1^{\dagger}a_{-1}^{\dagger}\right. \\ \nonumber
&- \frac{(gtN)^2}{8} \frac{(a_0^{\dagger})^{N-4}}{\sqrt{(N-4)!}}(a_1^{\dagger}a_{-1}^{\dagger})^2 \\ \nonumber &- \left.\frac{(gtN)^2}{8}  \frac{(a_0^{\dagger})^{N}}{\sqrt{N!}}  + \dots \right] |0\rangle    .
\end{align}
We can also express this state in terms of the Fock basis \eqref{fockbasis} as 
\begin{align}
    &\left( 1 - \frac{(gtN)^2}{8} \right)|0, 0, 0, 0, N, 0\rangle \nonumber \\
    &- \frac{igtN}{2} |1, 0, 2, 0, N-2, 0\rangle \nonumber \\
    &- \frac{(gtN)^2}{4} |2, 0, 4, 0, N-4, 0\rangle + \dots
\end{align}
Like in the previous section, we can trace out the $N_A, N_B, N_+,$ and $N_-$ modes. Noting again that our approach only uses particle-number-conserving observables, we can also treat these modes as implicit and rewrite the above state as
\begin{align}
\left[ 1- \frac{(gtN )^2}{8 } - \frac{igtN}{2}a_1^{\dagger}a_{-1}^{\dagger} - \frac{(gtN )^2}{8} (a_1^{\dagger}a_{-1}^{\dagger})^2 + \dots  \right] |0\rangle ,
\end{align}
which has the same expansion coefficients as
\begin{align}
e^{-igt\frac{N}{2}(a_1^{\dagger}a_{-1}^{\dagger} + a_1a_{-1})}|0\rangle .
\end{align}
The calculation for the local squeezing Hamiltonian acting on the BEC labeled by $b$ gives the same results up to the interchange of $ a_m \rightleftharpoons b_m $. This shows that local squeezing on an equally split BEC with total atom number $N$ is, in the limit of large $N$, equivalent to local squeezing with $N/2$ atoms as given in (\ref{Ham_after_split}).

\subsection{Additional numerical results using the unapproximated Hamiltonian \eqref{mainham}}

\begin{figure}[t]
\includegraphics[width=\columnwidth]{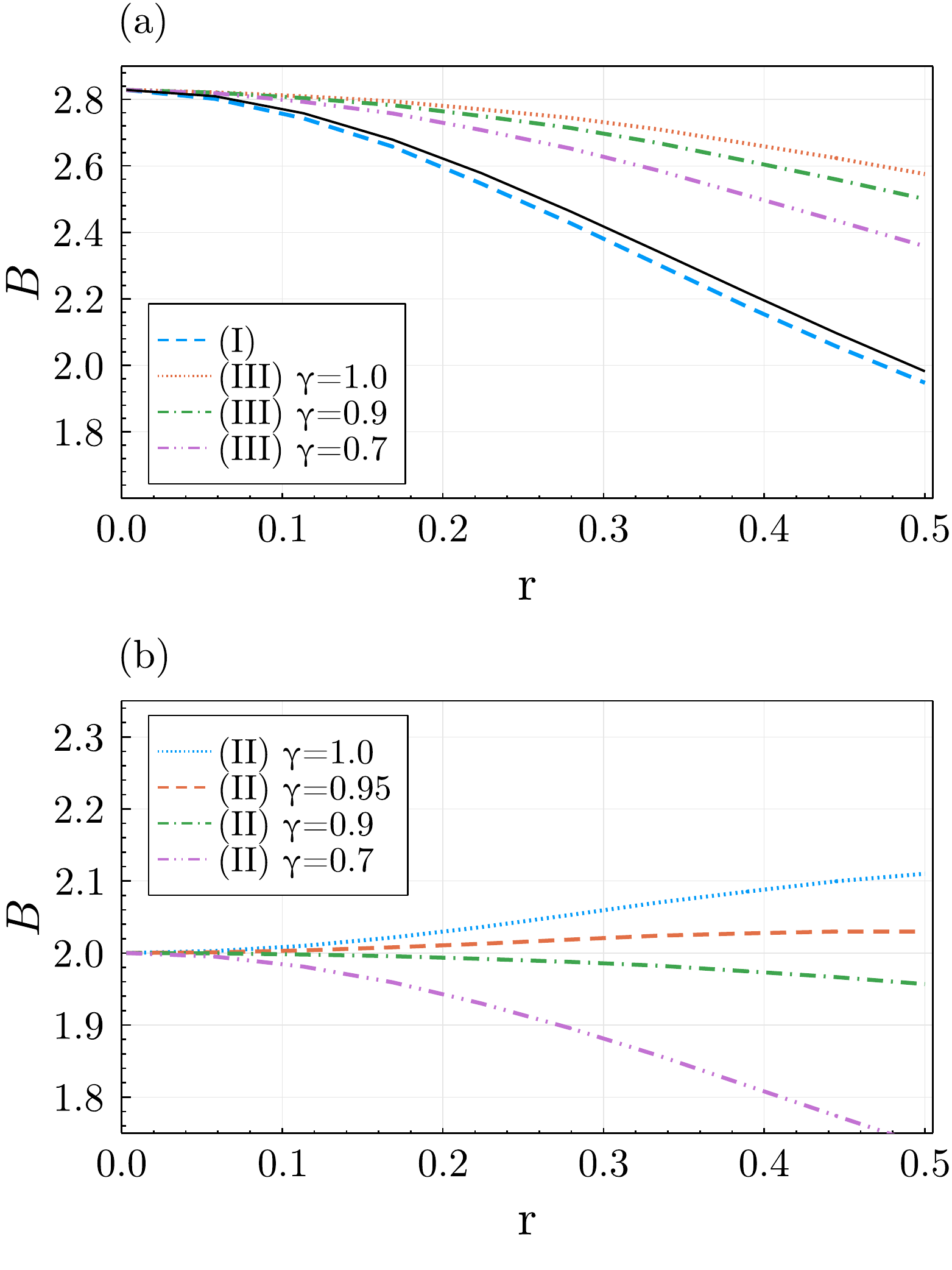}
\caption {Analogous results to Fig. \ref{fig2} obtained using the unapproximated Hamiltonian \eqref{mainham}: the Bell-CHSH quantity $ {\cal B} $ as a function of the dimensionless squeezing parameter $r$ for (a) Approaches I and III and  (b) Approach II.  Layout and calculation of the three different Bell-CHSH quantities are identical to Fig. \ref{fig2}. For the simulation of the unapproximated Hamiltonian \eqref{mainham} a total atom number $N=10$ and a numerical truncation value of $k_{\text{cut}} = 10$ were used.
\label{extrafig}}
\end{figure}

In Fig. \ref{extrafig} we present additional numerical results which are analogous to Fig. \ref{fig2}. The main difference is that we used the unapproximated Hamiltonian  \eqref{mainham} instead of the approximated expression \eqref{ham2m}.  Since the numerical simulation for this case is significantly more resource-intensive, we performed it only with $N=10$ and up to a squeezing parameter of $r=0.5$. This is not yet within the regime $N \gg 1$ in which our approximation is valid. Deviations from Fig. \ref{fig2} are therefore explicable by the very low total atom number $N=10$. Especially for Approach III, however, the results are still in excellent agreement with Fig. \ref{fig2} in the regime of low $r$.
For Approach II, displayed in Fig. \ref{extrafig}(b), we note an overall reduced level of violation and a larger susceptibility to noise. We expect these deviations to disappear when the atom number $N$ becomes sufficiently large.

\section{Generating two-mode squeezed states by interference} 
\label{app:twomode}

In this section we show how a two-mode optical squeezed state can be produced by interfering two single-mode squeezed states.  

First define the squeezing operators for modes $ a, b$
\begin{align}
S^{(a)}_r  &  = e^{\frac{r}{2} ( (a^\dagger)^2 - a^2 ) } \nonumber \\
S^{(b)}_r  &  =e^{- \frac{r}{2} ( (b^\dagger)^2 - b^2) } .
\end{align}
For $ r> 0 $, this generates $ p = (-i a + i a^\dagger)/2$ squeezed states for mode $ a $ and $ x = (b + b^\dagger)/2$ squeezed states for mode $ b$.  

Splitting a single-mode squeezed state produces the state
\begin{align}
    U_W S^{(a)}_r | 0 \rangle \approx | 0 \rangle + \frac{r}{2} ( (a^\dagger)^2 + 2 a^\dagger b^\dagger +  (b^\dagger)^2 ) | 0 \rangle
\end{align}
for $ r  \ll 1 $.
This is a combination of single- and two-mode squeezed states on modes $ a $ and $ b $.  The single-mode squeezing terms can be eliminated by squeezing the $ b $ mode by the same amount
\begin{align}
    U_W S^{(b)}_r S^{(a)}_r | 0 \rangle \approx 
    | 0 \rangle +2r  a^\dagger b^\dagger  | 0 \rangle ,
\end{align}
which is the two-mode squeezed state to leading order.

\section{Numerical calculation}
\label{app:numerics}

Numerical results are obtained by evolving the wave function in Fock space \eqref{genfockstate} with the Schr\"odinger equation, according to the sequence \eqref{upsiseq}, starting from the vacuum state. After the final state is obtained, basis rotations according to \eqref{rotationopera} are made and probabilities are evaluated according to \eqref{probabilities}.  Finally, \eqref{newedefprob} is evaluated for the optimal angles \eqref{optimalangles} and substituted into \eqref{bellmain}. The code for reproducing Figs. \ref{fig2} and \ref{extrafig} is available online \footnote{\url{https://github.com/kitzingj/BEC-Bell-correlations}}.  For our simulation we made use of the \textsc{Julia} package \textsc{DifferentialEquations.jl} \cite{rackauckas2017DifferentialEquationsJlPerformant} and Tsitouras’ 5/4 Runge-Kutta method \cite{tsitouras2011RungeKuttaPairs}.

Due to the numerical evaluation, the Hilbert space must be truncated  within the range $ 0 \le k_A, l_A, k_B, l_B \le k_{\text{cut}} $. The maximal truncation value we use is $k_\text{cut} = 40$. To ensure that the truncation does not affect the physical results, we first check for convergence with $ k_{\text{cut}} $ and evaluate \eqref{espinexpr} to  check that it recovers the exact result (\ref{bellsimplecase}).  For values in the range $ 0 < r < 0.6 $ in Fig. \ref{fig2}(a), our numerical results agree with the exact result \eqref{bellsimplecase} to within 0.5\%.  The numerical estimates underestimate the true values, which is likely caused by neglecting higher order correlations due to the numerical truncation.

\section{CH Inequality}
\label{app:ch}

Here we show how the CH inequality \eqref{chinequality} can be rewritten in a CHSH-like inequality \eqref{bellmain} and \eqref{dabellineq}. 

From \eqref{chinequality}, we can write
\begin{align}
& P^{\sigma \sigma'} ( \theta_A, \theta_B) + P^{\sigma \sigma'} ( \theta_A, \theta_B') + P^{\sigma \sigma'} ( \theta_A' , \theta_B) \nonumber \\
& -P^{\sigma \sigma'} ( \theta_A' , \theta_B')  \le  P^{\sigma \forall} (\theta_A) + P^{\forall\sigma'  } (\theta_B)
\label{plusineqch}
\end{align}
and
\begin{align}
& - \Big[ P^{\sigma \sigma'} ( \theta_A, \theta_B) + P^{\sigma \sigma'} ( \theta_A, \theta_B') + P^{\sigma \sigma'} ( \theta_A' , \theta_B) \nonumber \\
& -P^{\sigma \sigma'} ( \theta_A' , \theta_B')  \Big] \le P^{\forall \forall} - P^{\sigma \forall} (\theta_A) - P^{\forall\sigma'  } (\theta_B) .
\label{minusineqch}
\end{align}
Setting $ \sigma = \sigma' = \pm $ in  \eqref{plusineqch} and $ \sigma = -\sigma' = \pm $ in  \eqref{minusineqch} and adding the four inequalities gives
\begin{align}
& {\cal B } = E_{\text{III}} (\theta_A, \theta_B) +  E_{\text{III}} (\theta_A', \theta_B) + E_{\text{III}} (\theta_A, \theta_B') \nonumber \\
& - E_{\text{III}} (\theta_A', \theta_B') \le 2 \; ,
\end{align}
where we used the definition \eqref{chineqappiii}.

\section{Derivation of the sign operators including loss}
\label{app:loss}

We first show the derivation of \eqref{eprimenopos}. 
Substituting \eqref{rhodef} into 
\begin{align}
 \Tr& ( \text{sgn}[ S^z_A]  \text{sgn} [S^z_B ]\rho  ) \nonumber \\
= &  \sum_{\bm{k} \bm{n} }  \langle \bm{k} | \text{sgn} ( S^z_A)  \text{sgn} ( S^z_B)  F_{n_A} F_{m_A} F_{n_B} F_{m_B} | \Psi(\theta_A, \theta_B) \rangle \nonumber \\
& \times \langle \Psi(\theta_A, \theta_B) | F_{m_B}^\dagger F_{n_B}^\dagger F_{m_A}^\dagger F_{n_A}^\dagger | \bm{k} \rangle  \\
 = &\sum_{\bm{k} \bm{k}'  \bm{n} }  \langle \Psi(\theta_A, \theta_B) |  \bm{k} \rangle \langle  \bm{k} | F_{m_B}^\dagger F_{n_B}^\dagger F_{m_A}^\dagger F_{n_A}^\dagger \nonumber \\
& \times \text{sgn}(S^z_A ) \text{sgn}(S^z_B)  F_{n_A} F_{m_A} F_{n_B} F_{m_B} | \bm{k}' \rangle \langle \bm{k}' | \Psi(\theta_A, \theta_B) \rangle ,
\end{align}
where for brevity we defined $ \bm{n} = (n_A, m_A, n_B, m_B)$ and $ \bm{k} =(k_A, l_A, k_B, l_B) $.  Since the Kraus operators shift the Fock states by a number given by their indices, and the $ S^z_A, S^z_B$ are diagonal, 
we have $ k=k'$ and we can eliminate one of the summations to give
\begin{align}
 \Tr& ( \text{sgn} ( S^z_A)  \text{sgn} (S^z_B )\rho  ) 
= &  \sum_{\bm{k} } p_{k_A l_A k_B l_B}^{(\theta_A, \theta_B)} 
G^\gamma_{k_A l_A} G^\gamma_{k_B l_B} ,
\end{align}
where we defined
\begin{align}
G^\gamma_{kl} =  \sum_{nm} \langle k l  | F_{m}^\dagger F_{n}^\dagger \text{sgn}( S^z ) F_{n} F_{m} | k l  \rangle  .
\end{align}
From the fact that 
\begin{align}
\text{sgn}( S^z ) = 1 - 2 \sum_{k<l}| k l \rangle \langle k l |
\end{align}
and substituting the definitions \eqref{krausdef}, we obtain explicitly the form \eqref{gcool}. 

For the Approach I correlators, similar steps to the above can be used to show that 
\begin{align}
&  \Tr (S^z_A S^z_B\rho  ) 
 = \sum_{\bm{k}  \bm{n} }  | \langle \Psi(\theta_A, \theta_B) |  \bm{k} \rangle |^2 \nonumber \\
& \times  \langle  \bm{k} | F_{m_B}^\dagger F_{n_B}^\dagger F_{m_A}^\dagger F_{n_A}^\dagger S^z_A  S^z_B  F_{n_A} F_{m_A} F_{n_B} F_{m_B} | \bm{k}' \rangle  .
\end{align}
We require the lossy version of the number operators which can be evaluated as 
\begin{align}
\sum_n F^\dagger_n a^\dagger a F_n = \sum_n \sum_{k=0}^n (k-n)  {k \choose n}   \gamma^{k-n} (1- \gamma)^n | k \rangle \langle k |
\end{align}
giving the result \eqref{lossynumber}.

\section{Approach III correlators for imperfect detection efficiency}
\label{app:detineff}

First let us find the probability that the physical state collapses to the  Fock state $ | \bm{k} \rangle $ where $ \bm{k} =(k_A, l_A, k_B, l_B) $, but each of the detectors fails to detect $ \bm{n} = (n_A, m_A, n_B, m_B)$ atoms.  The joint probability of this occurrence is
\begin{align}
p_{ \bm{k} \bm{n}}^{  (\theta_A, \theta_B) } = & 
P_\eta (k_A, n_A) P_\eta (l_A, m_A) \nonumber \\
& \times P_\eta (k_B, n_B) P_\eta (l_B, m_B ) p_{k_A l_A k_B l_B }^{ (\theta_A, \theta_B) } .  
\end{align}
Since local vacua events by Alice and Bob are assigned 0 outcomes, let us define an unnormalized probability-like distribution where these events are removed:
\begin{align}
\tilde{p}_{ \bm{k} \bm{n}}^{  (\theta_A, \theta_B) } =
 p_{ \bm{k} \bm{n}}^{ (\theta_A, \theta_B) } 
- p_{ \bm{k} \bm{n}}^{ (\theta_A, \theta_B) } \delta_{k_A n_A} \delta_{l_A m_A}  \nonumber \\
- p_{ \bm{k} \bm{n}}^{  (\theta_A, \theta_B) } \delta_{k_B n_B} \delta_{l_B m_B} 
+ p_{ \bm{k} \bm{n}}^{  (\theta_A, \theta_B) } \delta_{\bm{k} \bm{n}} .
\end{align}
This distribution is then used to evaluate the average of the sign functions.  Meanwhile the denominator is
\begin{align}
P^{\forall \forall}  & =  1 - \sum_{ \bm{k}}  p_{k_A l_A k_B l_B }^{ (\theta_A, \theta_B) } ( 1- \eta)^{k_A + l_A} 
 \nonumber \\
 & - \sum_{ \bm{k}}  p_{k_A l_A k_B l_B }^{ (\theta_A, \theta_B) } ( 1- \eta)^{k_B + l_B} \nonumber \\
  & + \sum_{ \bm{k}}  p_{k_A l_A k_B l_B }^{ (\theta_A, \theta_B) } ( 1- \eta)^{k_A + l_A+ k_B + l_B}  ,
\end{align}
where we used the fact that $ \sum_n P_\eta (k, n) = 1 $. 
The Approach III correlators can be defined in this context as
\begin{align}
E_{\text{III}}^\gamma ( \theta_A, \theta_B) & = \frac{1}{P^{\forall \forall}}
 \sum_{\bm{k} \bm{n}} \tilde{p}_{ \bm{k} \bm{n}}^{  (\theta_A, \theta_B) } \text{sgn} ( k_A - n_A -l_A + m_A) \nonumber \\
&  \times  \text{sgn} ( k_B - n_B -l_B + m_B) .
\end{align}
Substitution yields exactly the same expression as \eqref{eprimewithpos} with the replacement $ \gamma \rightarrow \eta$.  


\begin{thebibliography}{91}%
\makeatletter
\providecommand \@ifxundefined [1]{%
 \@ifx{#1\undefined}
}%
\providecommand \@ifnum [1]{%
 \ifnum #1\expandafter \@firstoftwo
 \else \expandafter \@secondoftwo
 \fi
}%
\providecommand \@ifx [1]{%
 \ifx #1\expandafter \@firstoftwo
 \else \expandafter \@secondoftwo
 \fi
}%
\providecommand \natexlab [1]{#1}%
\providecommand \enquote  [1]{``#1''}%
\providecommand \bibnamefont  [1]{#1}%
\providecommand \bibfnamefont [1]{#1}%
\providecommand \citenamefont [1]{#1}%
\providecommand \href@noop [0]{\@secondoftwo}%
\providecommand \href [0]{\begingroup \@sanitize@url \@href}%
\providecommand \@href[1]{\@@startlink{#1}\@@href}%
\providecommand \@@href[1]{\endgroup#1\@@endlink}%
\providecommand \@sanitize@url [0]{\catcode `\\12\catcode `\$12\catcode
  `\&12\catcode `\#12\catcode `\^12\catcode `\_12\catcode `\%12\relax}%
\providecommand \@@startlink[1]{}%
\providecommand \@@endlink[0]{}%
\providecommand \url  [0]{\begingroup\@sanitize@url \@url }%
\providecommand \@url [1]{\endgroup\@href {#1}{\urlprefix }}%
\providecommand \urlprefix  [0]{URL }%
\providecommand \Eprint [0]{\href }%
\providecommand \doibase [0]{https://doi.org/}%
\providecommand \selectlanguage [0]{\@gobble}%
\providecommand \bibinfo  [0]{\@secondoftwo}%
\providecommand \bibfield  [0]{\@secondoftwo}%
\providecommand \translation [1]{[#1]}%
\providecommand \BibitemOpen [0]{}%
\providecommand \bibitemStop [0]{}%
\providecommand \bibitemNoStop [0]{.\EOS\space}%
\providecommand \EOS [0]{\spacefactor3000\relax}%
\providecommand \BibitemShut  [1]{\csname bibitem#1\endcsname}%
\let\auto@bib@innerbib\@empty
\bibitem [{\citenamefont {Horodecki}\ \emph {et~al.}(2009)\citenamefont
  {Horodecki}, \citenamefont {Horodecki}, \citenamefont {Horodecki},\ and\
  \citenamefont {Horodecki}}]{horodecki2009QuantumEntanglement}%
  \BibitemOpen
  \bibfield  {author} {\bibinfo {author} {\bibfnamefont {R.}~\bibnamefont
  {Horodecki}}, \bibinfo {author} {\bibfnamefont {P.}~\bibnamefont
  {Horodecki}}, \bibinfo {author} {\bibfnamefont {M.}~\bibnamefont
  {Horodecki}},\ and\ \bibinfo {author} {\bibfnamefont {K.}~\bibnamefont
  {Horodecki}},\ }\href {https://doi.org/10.1103/RevModPhys.81.865} {\bibfield
  {journal} {\bibinfo  {journal} {Rev. Mod. Phys.}\ }\textbf {\bibinfo {volume}
  {81}},\ \bibinfo {pages} {865} (\bibinfo {year} {2009})}\BibitemShut
  {NoStop}%
\bibitem [{\citenamefont {Einstein}\ \emph {et~al.}(1935)\citenamefont
  {Einstein}, \citenamefont {Podolsky},\ and\ \citenamefont
  {Rosen}}]{einstein1935CanQuantumMechanicalDescription}%
  \BibitemOpen
  \bibfield  {author} {\bibinfo {author} {\bibfnamefont {A.}~\bibnamefont
  {Einstein}}, \bibinfo {author} {\bibfnamefont {B.}~\bibnamefont {Podolsky}},\
  and\ \bibinfo {author} {\bibfnamefont {N.}~\bibnamefont {Rosen}},\ }\href
  {https://doi.org/10.1103/PhysRev.47.777} {\bibfield  {journal} {\bibinfo
  {journal} {Phys. Rev.}\ }\textbf {\bibinfo {volume} {47}},\ \bibinfo {pages}
  {777} (\bibinfo {year} {1935})}\BibitemShut {NoStop}%
\bibitem [{\citenamefont {Bell}(1964)}]{bell1964EinsteinPodolskyRosen}%
  \BibitemOpen
  \bibfield  {author} {\bibinfo {author} {\bibfnamefont {J.~S.}\ \bibnamefont
  {Bell}},\ }\href {https://doi.org/10.1103/PhysicsPhysiqueFizika.1.195}
  {\bibfield  {journal} {\bibinfo  {journal} {Physics Physique Fizika}\
  }\textbf {\bibinfo {volume} {1}},\ \bibinfo {pages} {195} (\bibinfo {year}
  {1964})}\BibitemShut {NoStop}%
\bibitem [{\citenamefont {Brunner}\ \emph {et~al.}(2014)\citenamefont
  {Brunner}, \citenamefont {Cavalcanti}, \citenamefont {Pironio}, \citenamefont
  {Scarani},\ and\ \citenamefont {Wehner}}]{brunner2014BellNonlocality}%
  \BibitemOpen
  \bibfield  {author} {\bibinfo {author} {\bibfnamefont {N.}~\bibnamefont
  {Brunner}}, \bibinfo {author} {\bibfnamefont {D.}~\bibnamefont {Cavalcanti}},
  \bibinfo {author} {\bibfnamefont {S.}~\bibnamefont {Pironio}}, \bibinfo
  {author} {\bibfnamefont {V.}~\bibnamefont {Scarani}},\ and\ \bibinfo {author}
  {\bibfnamefont {S.}~\bibnamefont {Wehner}},\ }\href
  {https://doi.org/10.1103/RevModPhys.86.419} {\bibfield  {journal} {\bibinfo
  {journal} {Rev. Mod. Phys.}\ }\textbf {\bibinfo {volume} {86}},\ \bibinfo
  {pages} {419} (\bibinfo {year} {2014})}\BibitemShut {NoStop}%
\bibitem [{\citenamefont {Adesso}\ \emph {et~al.}(2016)\citenamefont {Adesso},
  \citenamefont {Bromley},\ and\ \citenamefont
  {Cianciaruso}}]{adesso2016MeasuresApplicationsQuantum}%
  \BibitemOpen
  \bibfield  {author} {\bibinfo {author} {\bibfnamefont {G.}~\bibnamefont
  {Adesso}}, \bibinfo {author} {\bibfnamefont {T.~R.}\ \bibnamefont
  {Bromley}},\ and\ \bibinfo {author} {\bibfnamefont {M.}~\bibnamefont
  {Cianciaruso}},\ }\href {https://doi.org/10.1088/1751-8113/49/47/473001}
  {\bibfield  {journal} {\bibinfo  {journal} {J. Phys. A: Math. Theor.}\
  }\textbf {\bibinfo {volume} {49}},\ \bibinfo {pages} {473001} (\bibinfo
  {year} {2016})}\BibitemShut {NoStop}%
\bibitem [{\citenamefont {Ma}\ \emph {et~al.}(2019)\citenamefont {Ma},
  \citenamefont {Cui}, \citenamefont {Cao}, \citenamefont {Fei}, \citenamefont
  {Vedral}, \citenamefont {Byrnes},\ and\ \citenamefont
  {Radhakrishnan}}]{ma2019OperationalAdvantageBasisindependent}%
  \BibitemOpen
  \bibfield  {author} {\bibinfo {author} {\bibfnamefont {Z.-H.}\ \bibnamefont
  {Ma}}, \bibinfo {author} {\bibfnamefont {J.}~\bibnamefont {Cui}}, \bibinfo
  {author} {\bibfnamefont {Z.}~\bibnamefont {Cao}}, \bibinfo {author}
  {\bibfnamefont {S.-M.}\ \bibnamefont {Fei}}, \bibinfo {author} {\bibfnamefont
  {V.}~\bibnamefont {Vedral}}, \bibinfo {author} {\bibfnamefont
  {T.}~\bibnamefont {Byrnes}},\ and\ \bibinfo {author} {\bibfnamefont
  {C.}~\bibnamefont {Radhakrishnan}},\ }\href
  {https://doi.org/10.1209/0295-5075/125/50005} {\bibfield  {journal} {\bibinfo
   {journal} {EPL}\ }\textbf {\bibinfo {volume} {125}},\ \bibinfo {pages}
  {50005} (\bibinfo {year} {2019})}\BibitemShut {NoStop}%
\bibitem [{\citenamefont {Baumgratz}\ \emph {et~al.}(2014)\citenamefont
  {Baumgratz}, \citenamefont {Cramer},\ and\ \citenamefont
  {Plenio}}]{baumgratz2014QuantifyingCoherence}%
  \BibitemOpen
  \bibfield  {author} {\bibinfo {author} {\bibfnamefont {T.}~\bibnamefont
  {Baumgratz}}, \bibinfo {author} {\bibfnamefont {M.}~\bibnamefont {Cramer}},\
  and\ \bibinfo {author} {\bibfnamefont {M.~B.}\ \bibnamefont {Plenio}},\
  }\href {https://doi.org/10.1103/PhysRevLett.113.140401} {\bibfield  {journal}
  {\bibinfo  {journal} {Phys. Rev. Lett.}\ }\textbf {\bibinfo {volume} {113}},\
  \bibinfo {pages} {140401} (\bibinfo {year} {2014})}\BibitemShut {NoStop}%
\bibitem [{\citenamefont {Ollivier}\ and\ \citenamefont
  {Zurek}(2001)}]{ollivier2001QuantumDiscordMeasure}%
  \BibitemOpen
  \bibfield  {author} {\bibinfo {author} {\bibfnamefont {H.}~\bibnamefont
  {Ollivier}}\ and\ \bibinfo {author} {\bibfnamefont {W.~H.}\ \bibnamefont
  {Zurek}},\ }\href {https://doi.org/10.1103/PhysRevLett.88.017901} {\bibfield
  {journal} {\bibinfo  {journal} {Phys. Rev. Lett.}\ }\textbf {\bibinfo
  {volume} {88}},\ \bibinfo {pages} {017901} (\bibinfo {year}
  {2001})}\BibitemShut {NoStop}%
\bibitem [{\citenamefont {Radhakrishnan}\ \emph {et~al.}(2017)\citenamefont
  {Radhakrishnan}, \citenamefont {Parthasarathy}, \citenamefont {Jambulingam},\
  and\ \citenamefont {Byrnes}}]{radhakrishnan2017QuantumCoherenceHeisenberg}%
  \BibitemOpen
  \bibfield  {author} {\bibinfo {author} {\bibfnamefont {C.}~\bibnamefont
  {Radhakrishnan}}, \bibinfo {author} {\bibfnamefont {M.}~\bibnamefont
  {Parthasarathy}}, \bibinfo {author} {\bibfnamefont {S.}~\bibnamefont
  {Jambulingam}},\ and\ \bibinfo {author} {\bibfnamefont {T.}~\bibnamefont
  {Byrnes}},\ }\href {https://doi.org/10.1038/s41598-017-13871-6} {\bibfield
  {journal} {\bibinfo  {journal} {Sci Rep}\ }\textbf {\bibinfo {volume} {7}},\
  \bibinfo {pages} {13865} (\bibinfo {year} {2017})}\BibitemShut {NoStop}%
\bibitem [{\citenamefont {Radhakrishnan}\ \emph {et~al.}(2020)\citenamefont
  {Radhakrishnan}, \citenamefont {Lauri{\`e}re},\ and\ \citenamefont
  {Byrnes}}]{radhakrishnan2020MultipartiteGeneralizationQuantum}%
  \BibitemOpen
  \bibfield  {author} {\bibinfo {author} {\bibfnamefont {C.}~\bibnamefont
  {Radhakrishnan}}, \bibinfo {author} {\bibfnamefont {M.}~\bibnamefont
  {Lauri{\`e}re}},\ and\ \bibinfo {author} {\bibfnamefont {T.}~\bibnamefont
  {Byrnes}},\ }\href {https://doi.org/10.1103/PhysRevLett.124.110401}
  {\bibfield  {journal} {\bibinfo  {journal} {Phys. Rev. Lett.}\ }\textbf
  {\bibinfo {volume} {124}},\ \bibinfo {pages} {110401} (\bibinfo {year}
  {2020})}\BibitemShut {NoStop}%
\bibitem [{\citenamefont {Wiseman}\ \emph {et~al.}(2007)\citenamefont
  {Wiseman}, \citenamefont {Jones},\ and\ \citenamefont
  {Doherty}}]{wiseman2007SteeringEntanglementNonlocality}%
  \BibitemOpen
  \bibfield  {author} {\bibinfo {author} {\bibfnamefont {H.~M.}\ \bibnamefont
  {Wiseman}}, \bibinfo {author} {\bibfnamefont {S.~J.}\ \bibnamefont {Jones}},\
  and\ \bibinfo {author} {\bibfnamefont {A.~C.}\ \bibnamefont {Doherty}},\
  }\href {https://doi.org/10.1103/PhysRevLett.98.140402} {\bibfield  {journal}
  {\bibinfo  {journal} {Phys. Rev. Lett.}\ }\textbf {\bibinfo {volume} {98}},\
  \bibinfo {pages} {140402} (\bibinfo {year} {2007})}\BibitemShut {NoStop}%
\bibitem [{\citenamefont {Clauser}\ \emph {et~al.}(1969)\citenamefont
  {Clauser}, \citenamefont {Horne}, \citenamefont {Shimony},\ and\
  \citenamefont {Holt}}]{clauser1969ProposedExperimentTest}%
  \BibitemOpen
  \bibfield  {author} {\bibinfo {author} {\bibfnamefont {J.~F.}\ \bibnamefont
  {Clauser}}, \bibinfo {author} {\bibfnamefont {M.~A.}\ \bibnamefont {Horne}},
  \bibinfo {author} {\bibfnamefont {A.}~\bibnamefont {Shimony}},\ and\ \bibinfo
  {author} {\bibfnamefont {R.~A.}\ \bibnamefont {Holt}},\ }\href
  {https://doi.org/10.1103/PhysRevLett.23.880} {\bibfield  {journal} {\bibinfo
  {journal} {Phys. Rev. Lett.}\ }\textbf {\bibinfo {volume} {23}},\ \bibinfo
  {pages} {880} (\bibinfo {year} {1969})}\BibitemShut {NoStop}%
\bibitem [{\citenamefont {Freedman}\ and\ \citenamefont
  {Clauser}(1972)}]{freedman1972ExperimentalTestLocal}%
  \BibitemOpen
  \bibfield  {author} {\bibinfo {author} {\bibfnamefont {S.~J.}\ \bibnamefont
  {Freedman}}\ and\ \bibinfo {author} {\bibfnamefont {J.~F.}\ \bibnamefont
  {Clauser}},\ }\href {https://doi.org/10.1103/PhysRevLett.28.938} {\bibfield
  {journal} {\bibinfo  {journal} {Phys. Rev. Lett.}\ }\textbf {\bibinfo
  {volume} {28}},\ \bibinfo {pages} {938} (\bibinfo {year} {1972})}\BibitemShut
  {NoStop}%
\bibitem [{\citenamefont {Aspect}\ \emph {et~al.}(1982)\citenamefont {Aspect},
  \citenamefont {Dalibard},\ and\ \citenamefont
  {Roger}}]{aspect1982ExperimentalTestBell}%
  \BibitemOpen
  \bibfield  {author} {\bibinfo {author} {\bibfnamefont {A.}~\bibnamefont
  {Aspect}}, \bibinfo {author} {\bibfnamefont {J.}~\bibnamefont {Dalibard}},\
  and\ \bibinfo {author} {\bibfnamefont {G.}~\bibnamefont {Roger}},\ }\href
  {https://doi.org/10.1103/PhysRevLett.49.1804} {\bibfield  {journal} {\bibinfo
   {journal} {Phys. Rev. Lett.}\ }\textbf {\bibinfo {volume} {49}},\ \bibinfo
  {pages} {1804} (\bibinfo {year} {1982})}\BibitemShut {NoStop}%
\bibitem [{\citenamefont {Ou}\ and\ \citenamefont
  {Mandel}(1988)}]{ou1988ViolationBellInequality}%
  \BibitemOpen
  \bibfield  {author} {\bibinfo {author} {\bibfnamefont {Z.~Y.}\ \bibnamefont
  {Ou}}\ and\ \bibinfo {author} {\bibfnamefont {L.}~\bibnamefont {Mandel}},\
  }\href {https://doi.org/10.1103/PhysRevLett.61.50} {\bibfield  {journal}
  {\bibinfo  {journal} {Phys. Rev. Lett.}\ }\textbf {\bibinfo {volume} {61}},\
  \bibinfo {pages} {50} (\bibinfo {year} {1988})}\BibitemShut {NoStop}%
\bibitem [{\citenamefont {Munro}\ and\ \citenamefont
  {Reid}(1993)}]{munro1993ViolationBellInequality}%
  \BibitemOpen
  \bibfield  {author} {\bibinfo {author} {\bibfnamefont {W.~J.}\ \bibnamefont
  {Munro}}\ and\ \bibinfo {author} {\bibfnamefont {M.~D.}\ \bibnamefont
  {Reid}},\ }\href {https://doi.org/10.1103/PhysRevA.47.4412} {\bibfield
  {journal} {\bibinfo  {journal} {Phys. Rev. A}\ }\textbf {\bibinfo {volume}
  {47}},\ \bibinfo {pages} {4412} (\bibinfo {year} {1993})}\BibitemShut
  {NoStop}%
\bibitem [{\citenamefont {Larsson}(2014)}]{larsson2014LoopholesBellInequality}%
  \BibitemOpen
  \bibfield  {author} {\bibinfo {author} {\bibfnamefont {J.-{\AA}.}\
  \bibnamefont {Larsson}},\ }\href
  {https://doi.org/10.1088/1751-8113/47/42/424003} {\bibfield  {journal}
  {\bibinfo  {journal} {J. Phys. A: Math. Theor.}\ }\textbf {\bibinfo {volume}
  {47}},\ \bibinfo {pages} {424003} (\bibinfo {year} {2014})}\BibitemShut
  {NoStop}%
\bibitem [{\citenamefont {Rowe}\ \emph {et~al.}(2001)\citenamefont {Rowe},
  \citenamefont {Kielpinski}, \citenamefont {Meyer}, \citenamefont {Sackett},
  \citenamefont {Itano}, \citenamefont {Monroe},\ and\ \citenamefont
  {Wineland}}]{rowe2001ExperimentalViolationBell}%
  \BibitemOpen
  \bibfield  {author} {\bibinfo {author} {\bibfnamefont {M.~A.}\ \bibnamefont
  {Rowe}}, \bibinfo {author} {\bibfnamefont {D.}~\bibnamefont {Kielpinski}},
  \bibinfo {author} {\bibfnamefont {V.}~\bibnamefont {Meyer}}, \bibinfo
  {author} {\bibfnamefont {C.~A.}\ \bibnamefont {Sackett}}, \bibinfo {author}
  {\bibfnamefont {W.~M.}\ \bibnamefont {Itano}}, \bibinfo {author}
  {\bibfnamefont {C.}~\bibnamefont {Monroe}},\ and\ \bibinfo {author}
  {\bibfnamefont {D.~J.}\ \bibnamefont {Wineland}},\ }\href
  {https://doi.org/10.1038/35057215} {\bibfield  {journal} {\bibinfo  {journal}
  {Nature}\ }\textbf {\bibinfo {volume} {409}},\ \bibinfo {pages} {791}
  (\bibinfo {year} {2001})}\BibitemShut {NoStop}%
\bibitem [{\citenamefont {Hensen}\ \emph {et~al.}(2015)\citenamefont {Hensen},
  \citenamefont {Bernien}, \citenamefont {Dr{\'e}au}, \citenamefont {Reiserer},
  \citenamefont {Kalb}, \citenamefont {Blok}, \citenamefont {Ruitenberg},
  \citenamefont {Vermeulen}, \citenamefont {Schouten}, \citenamefont
  {Abell{\'a}n}, \citenamefont {Amaya}, \citenamefont {Pruneri}, \citenamefont
  {Mitchell}, \citenamefont {Markham}, \citenamefont {Twitchen}, \citenamefont
  {Elkouss}, \citenamefont {Wehner}, \citenamefont {Taminiau},\ and\
  \citenamefont {Hanson}}]{hensen2015LoopholefreeBellInequality}%
  \BibitemOpen
  \bibfield  {author} {\bibinfo {author} {\bibfnamefont {B.}~\bibnamefont
  {Hensen}}, \bibinfo {author} {\bibfnamefont {H.}~\bibnamefont {Bernien}},
  \bibinfo {author} {\bibfnamefont {A.~E.}\ \bibnamefont {Dr{\'e}au}}, \bibinfo
  {author} {\bibfnamefont {A.}~\bibnamefont {Reiserer}}, \bibinfo {author}
  {\bibfnamefont {N.}~\bibnamefont {Kalb}}, \bibinfo {author} {\bibfnamefont
  {M.~S.}\ \bibnamefont {Blok}}, \bibinfo {author} {\bibfnamefont
  {J.}~\bibnamefont {Ruitenberg}}, \bibinfo {author} {\bibfnamefont {R.~F.~L.}\
  \bibnamefont {Vermeulen}}, \bibinfo {author} {\bibfnamefont {R.~N.}\
  \bibnamefont {Schouten}}, \bibinfo {author} {\bibfnamefont {C.}~\bibnamefont
  {Abell{\'a}n}}, \bibinfo {author} {\bibfnamefont {W.}~\bibnamefont {Amaya}},
  \bibinfo {author} {\bibfnamefont {V.}~\bibnamefont {Pruneri}}, \bibinfo
  {author} {\bibfnamefont {M.~W.}\ \bibnamefont {Mitchell}}, \bibinfo {author}
  {\bibfnamefont {M.}~\bibnamefont {Markham}}, \bibinfo {author} {\bibfnamefont
  {D.~J.}\ \bibnamefont {Twitchen}}, \bibinfo {author} {\bibfnamefont
  {D.}~\bibnamefont {Elkouss}}, \bibinfo {author} {\bibfnamefont
  {S.}~\bibnamefont {Wehner}}, \bibinfo {author} {\bibfnamefont {T.~H.}\
  \bibnamefont {Taminiau}},\ and\ \bibinfo {author} {\bibfnamefont
  {R.}~\bibnamefont {Hanson}},\ }\href {https://doi.org/10.1038/nature15759}
  {\bibfield  {journal} {\bibinfo  {journal} {Nature}\ }\textbf {\bibinfo
  {volume} {526}},\ \bibinfo {pages} {682} (\bibinfo {year}
  {2015})}\BibitemShut {NoStop}%
\bibitem [{\citenamefont {White}\ \emph {et~al.}(2016)\citenamefont {White},
  \citenamefont {Mutus}, \citenamefont {Dressel}, \citenamefont {Kelly},
  \citenamefont {Barends}, \citenamefont {Jeffrey}, \citenamefont {Sank},
  \citenamefont {Megrant}, \citenamefont {Campbell}, \citenamefont {Chen},
  \citenamefont {Chen}, \citenamefont {Chiaro}, \citenamefont {Dunsworth},
  \citenamefont {Hoi}, \citenamefont {Neill}, \citenamefont {O'Malley},
  \citenamefont {Roushan}, \citenamefont {Vainsencher}, \citenamefont {Wenner},
  \citenamefont {Korotkov},\ and\ \citenamefont
  {Martinis}}]{white2016PreservingEntanglementWeak}%
  \BibitemOpen
  \bibfield  {author} {\bibinfo {author} {\bibfnamefont {T.~C.}\ \bibnamefont
  {White}}, \bibinfo {author} {\bibfnamefont {J.~Y.}\ \bibnamefont {Mutus}},
  \bibinfo {author} {\bibfnamefont {J.}~\bibnamefont {Dressel}}, \bibinfo
  {author} {\bibfnamefont {J.}~\bibnamefont {Kelly}}, \bibinfo {author}
  {\bibfnamefont {R.}~\bibnamefont {Barends}}, \bibinfo {author} {\bibfnamefont
  {E.}~\bibnamefont {Jeffrey}}, \bibinfo {author} {\bibfnamefont
  {D.}~\bibnamefont {Sank}}, \bibinfo {author} {\bibfnamefont {A.}~\bibnamefont
  {Megrant}}, \bibinfo {author} {\bibfnamefont {B.}~\bibnamefont {Campbell}},
  \bibinfo {author} {\bibfnamefont {Y.}~\bibnamefont {Chen}}, \bibinfo {author}
  {\bibfnamefont {Z.}~\bibnamefont {Chen}}, \bibinfo {author} {\bibfnamefont
  {B.}~\bibnamefont {Chiaro}}, \bibinfo {author} {\bibfnamefont
  {A.}~\bibnamefont {Dunsworth}}, \bibinfo {author} {\bibfnamefont {I.-C.}\
  \bibnamefont {Hoi}}, \bibinfo {author} {\bibfnamefont {C.}~\bibnamefont
  {Neill}}, \bibinfo {author} {\bibfnamefont {P.~J.~J.}\ \bibnamefont
  {O'Malley}}, \bibinfo {author} {\bibfnamefont {P.}~\bibnamefont {Roushan}},
  \bibinfo {author} {\bibfnamefont {A.}~\bibnamefont {Vainsencher}}, \bibinfo
  {author} {\bibfnamefont {J.}~\bibnamefont {Wenner}}, \bibinfo {author}
  {\bibfnamefont {A.~N.}\ \bibnamefont {Korotkov}},\ and\ \bibinfo {author}
  {\bibfnamefont {J.~M.}\ \bibnamefont {Martinis}},\ }\href
  {https://doi.org/10.1038/npjqi.2015.22} {\bibfield  {journal} {\bibinfo
  {journal} {npj Quantum Inf}\ }\textbf {\bibinfo {volume} {2}},\ \bibinfo
  {pages} {15022} (\bibinfo {year} {2016})}\BibitemShut {NoStop}%
\bibitem [{\citenamefont {Shin}\ \emph {et~al.}(2019)\citenamefont {Shin},
  \citenamefont {Henson}, \citenamefont {Hodgman}, \citenamefont {Wasak},
  \citenamefont {Chwede{\'n}czuk},\ and\ \citenamefont
  {Truscott}}]{shin2019BellCorrelationsSpatially}%
  \BibitemOpen
  \bibfield  {author} {\bibinfo {author} {\bibfnamefont {D.~K.}\ \bibnamefont
  {Shin}}, \bibinfo {author} {\bibfnamefont {B.~M.}\ \bibnamefont {Henson}},
  \bibinfo {author} {\bibfnamefont {S.~S.}\ \bibnamefont {Hodgman}}, \bibinfo
  {author} {\bibfnamefont {T.}~\bibnamefont {Wasak}}, \bibinfo {author}
  {\bibfnamefont {J.}~\bibnamefont {Chwede{\'n}czuk}},\ and\ \bibinfo {author}
  {\bibfnamefont {A.~G.}\ \bibnamefont {Truscott}},\ }\href
  {https://doi.org/10.1038/s41467-019-12192-8} {\bibfield  {journal} {\bibinfo
  {journal} {Nat Commun}\ }\textbf {\bibinfo {volume} {10}},\ \bibinfo {pages}
  {4447} (\bibinfo {year} {2019})}\BibitemShut {NoStop}%
\bibitem [{\citenamefont {Shalm}\ \emph {et~al.}(2015)\citenamefont {Shalm},
  \citenamefont {{Meyer-Scott}}, \citenamefont {Christensen}, \citenamefont
  {Bierhorst}, \citenamefont {Wayne}, \citenamefont {Stevens}, \citenamefont
  {Gerrits}, \citenamefont {Glancy}, \citenamefont {Hamel}, \citenamefont
  {Allman}, \citenamefont {Coakley}, \citenamefont {Dyer}, \citenamefont
  {Hodge}, \citenamefont {Lita}, \citenamefont {Verma}, \citenamefont
  {Lambrocco}, \citenamefont {Tortorici}, \citenamefont {Migdall},
  \citenamefont {Zhang}, \citenamefont {Kumor}, \citenamefont {Farr},
  \citenamefont {Marsili}, \citenamefont {Shaw}, \citenamefont {Stern},
  \citenamefont {Abell{\'a}n}, \citenamefont {Amaya}, \citenamefont {Pruneri},
  \citenamefont {Jennewein}, \citenamefont {Mitchell}, \citenamefont {Kwiat},
  \citenamefont {Bienfang}, \citenamefont {Mirin}, \citenamefont {Knill},\ and\
  \citenamefont {Nam}}]{shalm2015StrongLoopholeFreeTest}%
  \BibitemOpen
  \bibfield  {author} {\bibinfo {author} {\bibfnamefont {L.~K.}\ \bibnamefont
  {Shalm}}, \bibinfo {author} {\bibfnamefont {E.}~\bibnamefont
  {{Meyer-Scott}}}, \bibinfo {author} {\bibfnamefont {B.~G.}\ \bibnamefont
  {Christensen}}, \bibinfo {author} {\bibfnamefont {P.}~\bibnamefont
  {Bierhorst}}, \bibinfo {author} {\bibfnamefont {M.~A.}\ \bibnamefont
  {Wayne}}, \bibinfo {author} {\bibfnamefont {M.~J.}\ \bibnamefont {Stevens}},
  \bibinfo {author} {\bibfnamefont {T.}~\bibnamefont {Gerrits}}, \bibinfo
  {author} {\bibfnamefont {S.}~\bibnamefont {Glancy}}, \bibinfo {author}
  {\bibfnamefont {D.~R.}\ \bibnamefont {Hamel}}, \bibinfo {author}
  {\bibfnamefont {M.~S.}\ \bibnamefont {Allman}}, \bibinfo {author}
  {\bibfnamefont {K.~J.}\ \bibnamefont {Coakley}}, \bibinfo {author}
  {\bibfnamefont {S.~D.}\ \bibnamefont {Dyer}}, \bibinfo {author}
  {\bibfnamefont {C.}~\bibnamefont {Hodge}}, \bibinfo {author} {\bibfnamefont
  {A.~E.}\ \bibnamefont {Lita}}, \bibinfo {author} {\bibfnamefont {V.~B.}\
  \bibnamefont {Verma}}, \bibinfo {author} {\bibfnamefont {C.}~\bibnamefont
  {Lambrocco}}, \bibinfo {author} {\bibfnamefont {E.}~\bibnamefont
  {Tortorici}}, \bibinfo {author} {\bibfnamefont {A.~L.}\ \bibnamefont
  {Migdall}}, \bibinfo {author} {\bibfnamefont {Y.}~\bibnamefont {Zhang}},
  \bibinfo {author} {\bibfnamefont {D.~R.}\ \bibnamefont {Kumor}}, \bibinfo
  {author} {\bibfnamefont {W.~H.}\ \bibnamefont {Farr}}, \bibinfo {author}
  {\bibfnamefont {F.}~\bibnamefont {Marsili}}, \bibinfo {author} {\bibfnamefont
  {M.~D.}\ \bibnamefont {Shaw}}, \bibinfo {author} {\bibfnamefont {J.~A.}\
  \bibnamefont {Stern}}, \bibinfo {author} {\bibfnamefont {C.}~\bibnamefont
  {Abell{\'a}n}}, \bibinfo {author} {\bibfnamefont {W.}~\bibnamefont {Amaya}},
  \bibinfo {author} {\bibfnamefont {V.}~\bibnamefont {Pruneri}}, \bibinfo
  {author} {\bibfnamefont {T.}~\bibnamefont {Jennewein}}, \bibinfo {author}
  {\bibfnamefont {M.~W.}\ \bibnamefont {Mitchell}}, \bibinfo {author}
  {\bibfnamefont {P.~G.}\ \bibnamefont {Kwiat}}, \bibinfo {author}
  {\bibfnamefont {J.~C.}\ \bibnamefont {Bienfang}}, \bibinfo {author}
  {\bibfnamefont {R.~P.}\ \bibnamefont {Mirin}}, \bibinfo {author}
  {\bibfnamefont {E.}~\bibnamefont {Knill}},\ and\ \bibinfo {author}
  {\bibfnamefont {S.~W.}\ \bibnamefont {Nam}},\ }\href
  {https://doi.org/10.1103/PhysRevLett.115.250402} {\bibfield  {journal}
  {\bibinfo  {journal} {Phys. Rev. Lett.}\ }\textbf {\bibinfo {volume} {115}},\
  \bibinfo {pages} {250402} (\bibinfo {year} {2015})}\BibitemShut {NoStop}%
\bibitem [{\citenamefont {Giustina}\ \emph {et~al.}(2015)\citenamefont
  {Giustina}, \citenamefont {Versteegh}, \citenamefont {Wengerowsky},
  \citenamefont {Handsteiner}, \citenamefont {Hochrainer}, \citenamefont
  {Phelan}, \citenamefont {Steinlechner}, \citenamefont {Kofler}, \citenamefont
  {Larsson}, \citenamefont {Abell{\'a}n}, \citenamefont {Amaya}, \citenamefont
  {Pruneri}, \citenamefont {Mitchell}, \citenamefont {Beyer}, \citenamefont
  {Gerrits}, \citenamefont {Lita}, \citenamefont {Shalm}, \citenamefont {Nam},
  \citenamefont {Scheidl}, \citenamefont {Ursin}, \citenamefont {Wittmann},\
  and\ \citenamefont
  {Zeilinger}}]{giustina2015SignificantLoopholeFreeTestBell}%
  \BibitemOpen
  \bibfield  {author} {\bibinfo {author} {\bibfnamefont {M.}~\bibnamefont
  {Giustina}}, \bibinfo {author} {\bibfnamefont {M.~A.~M.}\ \bibnamefont
  {Versteegh}}, \bibinfo {author} {\bibfnamefont {S.}~\bibnamefont
  {Wengerowsky}}, \bibinfo {author} {\bibfnamefont {J.}~\bibnamefont
  {Handsteiner}}, \bibinfo {author} {\bibfnamefont {A.}~\bibnamefont
  {Hochrainer}}, \bibinfo {author} {\bibfnamefont {K.}~\bibnamefont {Phelan}},
  \bibinfo {author} {\bibfnamefont {F.}~\bibnamefont {Steinlechner}}, \bibinfo
  {author} {\bibfnamefont {J.}~\bibnamefont {Kofler}}, \bibinfo {author}
  {\bibfnamefont {J.-{\AA}.}\ \bibnamefont {Larsson}}, \bibinfo {author}
  {\bibfnamefont {C.}~\bibnamefont {Abell{\'a}n}}, \bibinfo {author}
  {\bibfnamefont {W.}~\bibnamefont {Amaya}}, \bibinfo {author} {\bibfnamefont
  {V.}~\bibnamefont {Pruneri}}, \bibinfo {author} {\bibfnamefont {M.~W.}\
  \bibnamefont {Mitchell}}, \bibinfo {author} {\bibfnamefont {J.}~\bibnamefont
  {Beyer}}, \bibinfo {author} {\bibfnamefont {T.}~\bibnamefont {Gerrits}},
  \bibinfo {author} {\bibfnamefont {A.~E.}\ \bibnamefont {Lita}}, \bibinfo
  {author} {\bibfnamefont {L.~K.}\ \bibnamefont {Shalm}}, \bibinfo {author}
  {\bibfnamefont {S.~W.}\ \bibnamefont {Nam}}, \bibinfo {author} {\bibfnamefont
  {T.}~\bibnamefont {Scheidl}}, \bibinfo {author} {\bibfnamefont
  {R.}~\bibnamefont {Ursin}}, \bibinfo {author} {\bibfnamefont
  {B.}~\bibnamefont {Wittmann}},\ and\ \bibinfo {author} {\bibfnamefont
  {A.}~\bibnamefont {Zeilinger}},\ }\href
  {https://doi.org/10.1103/PhysRevLett.115.250401} {\bibfield  {journal}
  {\bibinfo  {journal} {Phys. Rev. Lett.}\ }\textbf {\bibinfo {volume} {115}},\
  \bibinfo {pages} {250401} (\bibinfo {year} {2015})}\BibitemShut {NoStop}%
\bibitem [{\citenamefont {Rosenfeld}\ \emph {et~al.}(2017)\citenamefont
  {Rosenfeld}, \citenamefont {Burchardt}, \citenamefont {Garthoff},
  \citenamefont {Redeker}, \citenamefont {Ortegel}, \citenamefont {Rau},\ and\
  \citenamefont {Weinfurter}}]{rosenfeld2017EventReadyBellTest}%
  \BibitemOpen
  \bibfield  {author} {\bibinfo {author} {\bibfnamefont {W.}~\bibnamefont
  {Rosenfeld}}, \bibinfo {author} {\bibfnamefont {D.}~\bibnamefont
  {Burchardt}}, \bibinfo {author} {\bibfnamefont {R.}~\bibnamefont {Garthoff}},
  \bibinfo {author} {\bibfnamefont {K.}~\bibnamefont {Redeker}}, \bibinfo
  {author} {\bibfnamefont {N.}~\bibnamefont {Ortegel}}, \bibinfo {author}
  {\bibfnamefont {M.}~\bibnamefont {Rau}},\ and\ \bibinfo {author}
  {\bibfnamefont {H.}~\bibnamefont {Weinfurter}},\ }\href
  {https://doi.org/10.1103/PhysRevLett.119.010402} {\bibfield  {journal}
  {\bibinfo  {journal} {Phys. Rev. Lett.}\ }\textbf {\bibinfo {volume} {119}},\
  \bibinfo {pages} {010402} (\bibinfo {year} {2017})}\BibitemShut {NoStop}%
\bibitem [{\citenamefont {Galland}\ \emph {et~al.}(2014)\citenamefont
  {Galland}, \citenamefont {Sangouard}, \citenamefont {Piro}, \citenamefont
  {Gisin},\ and\ \citenamefont
  {Kippenberg}}]{galland2014HeraldedSinglePhononPreparation}%
  \BibitemOpen
  \bibfield  {author} {\bibinfo {author} {\bibfnamefont {C.}~\bibnamefont
  {Galland}}, \bibinfo {author} {\bibfnamefont {N.}~\bibnamefont {Sangouard}},
  \bibinfo {author} {\bibfnamefont {N.}~\bibnamefont {Piro}}, \bibinfo {author}
  {\bibfnamefont {N.}~\bibnamefont {Gisin}},\ and\ \bibinfo {author}
  {\bibfnamefont {T.~J.}\ \bibnamefont {Kippenberg}},\ }\href
  {https://doi.org/10.1103/PhysRevLett.112.143602} {\bibfield  {journal}
  {\bibinfo  {journal} {Phys. Rev. Lett.}\ }\textbf {\bibinfo {volume} {112}},\
  \bibinfo {pages} {143602} (\bibinfo {year} {2014})}\BibitemShut {NoStop}%
\bibitem [{\citenamefont {Qian}\ \emph {et~al.}(2012)\citenamefont {Qian},
  \citenamefont {Clerk}, \citenamefont {Hammerer},\ and\ \citenamefont
  {Marquardt}}]{qian2012QuantumSignaturesOptomechanical}%
  \BibitemOpen
  \bibfield  {author} {\bibinfo {author} {\bibfnamefont {J.}~\bibnamefont
  {Qian}}, \bibinfo {author} {\bibfnamefont {A.~A.}\ \bibnamefont {Clerk}},
  \bibinfo {author} {\bibfnamefont {K.}~\bibnamefont {Hammerer}},\ and\
  \bibinfo {author} {\bibfnamefont {F.}~\bibnamefont {Marquardt}},\ }\href
  {https://doi.org/10.1103/PhysRevLett.109.253601} {\bibfield  {journal}
  {\bibinfo  {journal} {Phys. Rev. Lett.}\ }\textbf {\bibinfo {volume} {109}},\
  \bibinfo {pages} {253601} (\bibinfo {year} {2012})}\BibitemShut {NoStop}%
\bibitem [{\citenamefont {{Safavi-Naeini}}\ \emph {et~al.}(2014)\citenamefont
  {{Safavi-Naeini}}, \citenamefont {Hill}, \citenamefont {Meenehan},
  \citenamefont {Chan}, \citenamefont {Gr{\"o}blacher},\ and\ \citenamefont
  {Painter}}]{safavi-naeini2014TwoDimensionalPhononicPhotonicBand}%
  \BibitemOpen
  \bibfield  {author} {\bibinfo {author} {\bibfnamefont {A.~H.}\ \bibnamefont
  {{Safavi-Naeini}}}, \bibinfo {author} {\bibfnamefont {J.~T.}\ \bibnamefont
  {Hill}}, \bibinfo {author} {\bibfnamefont {S.}~\bibnamefont {Meenehan}},
  \bibinfo {author} {\bibfnamefont {J.}~\bibnamefont {Chan}}, \bibinfo {author}
  {\bibfnamefont {S.}~\bibnamefont {Gr{\"o}blacher}},\ and\ \bibinfo {author}
  {\bibfnamefont {O.}~\bibnamefont {Painter}},\ }\href
  {https://doi.org/10.1103/PhysRevLett.112.153603} {\bibfield  {journal}
  {\bibinfo  {journal} {Phys. Rev. Lett.}\ }\textbf {\bibinfo {volume} {112}},\
  \bibinfo {pages} {153603} (\bibinfo {year} {2014})}\BibitemShut {NoStop}%
\bibitem [{\citenamefont {Pu}\ and\ \citenamefont
  {Meystre}(2000)}]{pu2000CreatingMacroscopicAtomic}%
  \BibitemOpen
  \bibfield  {author} {\bibinfo {author} {\bibfnamefont {H.}~\bibnamefont
  {Pu}}\ and\ \bibinfo {author} {\bibfnamefont {P.}~\bibnamefont {Meystre}},\
  }\href {https://doi.org/10.1103/PhysRevLett.85.3987} {\bibfield  {journal}
  {\bibinfo  {journal} {Phys. Rev. Lett.}\ }\textbf {\bibinfo {volume} {85}},\
  \bibinfo {pages} {3987} (\bibinfo {year} {2000})}\BibitemShut {NoStop}%
\bibitem [{\citenamefont {Duan}\ \emph {et~al.}(2000)\citenamefont {Duan},
  \citenamefont {S{\o}rensen}, \citenamefont {Cirac},\ and\ \citenamefont
  {Zoller}}]{duan2000SqueezingEntanglementAtomic}%
  \BibitemOpen
  \bibfield  {author} {\bibinfo {author} {\bibfnamefont {L.-M.}\ \bibnamefont
  {Duan}}, \bibinfo {author} {\bibfnamefont {A.}~\bibnamefont {S{\o}rensen}},
  \bibinfo {author} {\bibfnamefont {J.~I.}\ \bibnamefont {Cirac}},\ and\
  \bibinfo {author} {\bibfnamefont {P.}~\bibnamefont {Zoller}},\ }\href
  {https://doi.org/10.1103/PhysRevLett.85.3991} {\bibfield  {journal} {\bibinfo
   {journal} {Phys. Rev. Lett.}\ }\textbf {\bibinfo {volume} {85}},\ \bibinfo
  {pages} {3991} (\bibinfo {year} {2000})}\BibitemShut {NoStop}%
\bibitem [{\citenamefont {Byrnes}\ \emph {et~al.}(2012)\citenamefont {Byrnes},
  \citenamefont {Wen},\ and\ \citenamefont
  {Yamamoto}}]{byrnes2012MacroscopicQuantumComputation}%
  \BibitemOpen
  \bibfield  {author} {\bibinfo {author} {\bibfnamefont {T.}~\bibnamefont
  {Byrnes}}, \bibinfo {author} {\bibfnamefont {K.}~\bibnamefont {Wen}},\ and\
  \bibinfo {author} {\bibfnamefont {Y.}~\bibnamefont {Yamamoto}},\ }\href
  {https://doi.org/10.1103/PhysRevA.85.040306} {\bibfield  {journal} {\bibinfo
  {journal} {Phys. Rev. A}\ }\textbf {\bibinfo {volume} {85}},\ \bibinfo
  {pages} {040306} (\bibinfo {year} {2012})}\BibitemShut {NoStop}%
\bibitem [{\citenamefont {Byrnes}\ \emph {et~al.}(2015)\citenamefont {Byrnes},
  \citenamefont {Rosseau}, \citenamefont {Khosla}, \citenamefont {Pyrkov},
  \citenamefont {Thomasen}, \citenamefont {Mukai}, \citenamefont {Koyama},
  \citenamefont {Abdelrahman},\ and\ \citenamefont
  {{Ilo-Okeke}}}]{byrnes2015MacroscopicQuantumInformation}%
  \BibitemOpen
  \bibfield  {author} {\bibinfo {author} {\bibfnamefont {T.}~\bibnamefont
  {Byrnes}}, \bibinfo {author} {\bibfnamefont {D.}~\bibnamefont {Rosseau}},
  \bibinfo {author} {\bibfnamefont {M.}~\bibnamefont {Khosla}}, \bibinfo
  {author} {\bibfnamefont {A.}~\bibnamefont {Pyrkov}}, \bibinfo {author}
  {\bibfnamefont {A.}~\bibnamefont {Thomasen}}, \bibinfo {author}
  {\bibfnamefont {T.}~\bibnamefont {Mukai}}, \bibinfo {author} {\bibfnamefont
  {S.}~\bibnamefont {Koyama}}, \bibinfo {author} {\bibfnamefont
  {A.}~\bibnamefont {Abdelrahman}},\ and\ \bibinfo {author} {\bibfnamefont
  {E.}~\bibnamefont {{Ilo-Okeke}}},\ }\href
  {https://doi.org/10.1016/j.optcom.2014.08.017} {\bibfield  {journal}
  {\bibinfo  {journal} {Optics Communications}\ }\textbf {\bibinfo {volume}
  {337}},\ \bibinfo {pages} {102} (\bibinfo {year} {2015})}\BibitemShut
  {NoStop}%
\bibitem [{\citenamefont {Pyrkov}\ and\ \citenamefont
  {Byrnes}(2014{\natexlab{a}})}]{pyrkov2014QuantumTeleportationSpin}%
  \BibitemOpen
  \bibfield  {author} {\bibinfo {author} {\bibfnamefont {A.~N.}\ \bibnamefont
  {Pyrkov}}\ and\ \bibinfo {author} {\bibfnamefont {T.}~\bibnamefont
  {Byrnes}},\ }\href {https://doi.org/10.1088/1367-2630/16/7/073038} {\bibfield
   {journal} {\bibinfo  {journal} {New J. Phys.}\ }\textbf {\bibinfo {volume}
  {16}},\ \bibinfo {pages} {073038} (\bibinfo {year}
  {2014}{\natexlab{a}})}\BibitemShut {NoStop}%
\bibitem [{\citenamefont {Pyrkov}\ and\ \citenamefont
  {Byrnes}(2014{\natexlab{b}})}]{pyrkov2014FullBlochsphereTeleportationSpinor}%
  \BibitemOpen
  \bibfield  {author} {\bibinfo {author} {\bibfnamefont {A.~N.}\ \bibnamefont
  {Pyrkov}}\ and\ \bibinfo {author} {\bibfnamefont {T.}~\bibnamefont
  {Byrnes}},\ }\href {https://doi.org/10.1103/PhysRevA.90.062336} {\bibfield
  {journal} {\bibinfo  {journal} {Phys. Rev. A}\ }\textbf {\bibinfo {volume}
  {90}},\ \bibinfo {pages} {062336} (\bibinfo {year}
  {2014}{\natexlab{b}})}\BibitemShut {NoStop}%
\bibitem [{\citenamefont {Appel}\ \emph {et~al.}(2009)\citenamefont {Appel},
  \citenamefont {Windpassinger}, \citenamefont {Oblak}, \citenamefont {Hoff},
  \citenamefont {Kjaergaard},\ and\ \citenamefont
  {Polzik}}]{appel2009MesoscopicAtomicEntanglement}%
  \BibitemOpen
  \bibfield  {author} {\bibinfo {author} {\bibfnamefont {J.}~\bibnamefont
  {Appel}}, \bibinfo {author} {\bibfnamefont {P.~J.}\ \bibnamefont
  {Windpassinger}}, \bibinfo {author} {\bibfnamefont {D.}~\bibnamefont
  {Oblak}}, \bibinfo {author} {\bibfnamefont {U.~B.}\ \bibnamefont {Hoff}},
  \bibinfo {author} {\bibfnamefont {N.}~\bibnamefont {Kjaergaard}},\ and\
  \bibinfo {author} {\bibfnamefont {E.~S.}\ \bibnamefont {Polzik}},\ }\href
  {https://doi.org/10.1073/pnas.0901550106} {\bibfield  {journal} {\bibinfo
  {journal} {Proceedings of the National Academy of Sciences}\ }\textbf
  {\bibinfo {volume} {106}},\ \bibinfo {pages} {10960} (\bibinfo {year}
  {2009})}\BibitemShut {NoStop}%
\bibitem [{\citenamefont {Krauter}\ \emph {et~al.}(2013)\citenamefont
  {Krauter}, \citenamefont {Salart}, \citenamefont {Muschik}, \citenamefont
  {Petersen}, \citenamefont {Shen}, \citenamefont {Fernholz},\ and\
  \citenamefont {Polzik}}]{krauter2013DeterministicQuantumTeleportation}%
  \BibitemOpen
  \bibfield  {author} {\bibinfo {author} {\bibfnamefont {H.}~\bibnamefont
  {Krauter}}, \bibinfo {author} {\bibfnamefont {D.}~\bibnamefont {Salart}},
  \bibinfo {author} {\bibfnamefont {C.~A.}\ \bibnamefont {Muschik}}, \bibinfo
  {author} {\bibfnamefont {J.~M.}\ \bibnamefont {Petersen}}, \bibinfo {author}
  {\bibfnamefont {H.}~\bibnamefont {Shen}}, \bibinfo {author} {\bibfnamefont
  {T.}~\bibnamefont {Fernholz}},\ and\ \bibinfo {author} {\bibfnamefont
  {E.~S.}\ \bibnamefont {Polzik}},\ }\href {https://doi.org/10.1038/nphys2631}
  {\bibfield  {journal} {\bibinfo  {journal} {Nature Phys}\ }\textbf {\bibinfo
  {volume} {9}},\ \bibinfo {pages} {400} (\bibinfo {year} {2013})}\BibitemShut
  {NoStop}%
\bibitem [{\citenamefont {Moxley}\ \emph {et~al.}(2016)\citenamefont {Moxley},
  \citenamefont {Dowling}, \citenamefont {Dai},\ and\ \citenamefont
  {Byrnes}}]{moxley2016SagnacInterferometryCoherent}%
  \BibitemOpen
  \bibfield  {author} {\bibinfo {author} {\bibfnamefont {F.~I.}\ \bibnamefont
  {Moxley}}, \bibinfo {author} {\bibfnamefont {J.~P.}\ \bibnamefont {Dowling}},
  \bibinfo {author} {\bibfnamefont {W.}~\bibnamefont {Dai}},\ and\ \bibinfo
  {author} {\bibfnamefont {T.}~\bibnamefont {Byrnes}},\ }\href
  {https://doi.org/10.1103/PhysRevA.93.053603} {\bibfield  {journal} {\bibinfo
  {journal} {Phys. Rev. A}\ }\textbf {\bibinfo {volume} {93}},\ \bibinfo
  {pages} {053603} (\bibinfo {year} {2016})}\BibitemShut {NoStop}%
\bibitem [{\citenamefont {Pezz{\`e}}\ \emph {et~al.}(2018)\citenamefont
  {Pezz{\`e}}, \citenamefont {Smerzi}, \citenamefont {Oberthaler},
  \citenamefont {Schmied},\ and\ \citenamefont
  {Treutlein}}]{pezze2018QuantumMetrologyNonclassical}%
  \BibitemOpen
  \bibfield  {author} {\bibinfo {author} {\bibfnamefont {L.}~\bibnamefont
  {Pezz{\`e}}}, \bibinfo {author} {\bibfnamefont {A.}~\bibnamefont {Smerzi}},
  \bibinfo {author} {\bibfnamefont {M.~K.}\ \bibnamefont {Oberthaler}},
  \bibinfo {author} {\bibfnamefont {R.}~\bibnamefont {Schmied}},\ and\ \bibinfo
  {author} {\bibfnamefont {P.}~\bibnamefont {Treutlein}},\ }\href
  {https://doi.org/10.1103/RevModPhys.90.035005} {\bibfield  {journal}
  {\bibinfo  {journal} {Rev. Mod. Phys.}\ }\textbf {\bibinfo {volume} {90}},\
  \bibinfo {pages} {035005} (\bibinfo {year} {2018})}\BibitemShut {NoStop}%
\bibitem [{\citenamefont {{Ilo-Okeke}}\ \emph {et~al.}(2018)\citenamefont
  {{Ilo-Okeke}}, \citenamefont {Tessler}, \citenamefont {Dowling},\ and\
  \citenamefont {Byrnes}}]{ilo-okeke2018RemoteQuantumClock}%
  \BibitemOpen
  \bibfield  {author} {\bibinfo {author} {\bibfnamefont {E.~O.}\ \bibnamefont
  {{Ilo-Okeke}}}, \bibinfo {author} {\bibfnamefont {L.}~\bibnamefont
  {Tessler}}, \bibinfo {author} {\bibfnamefont {J.~P.}\ \bibnamefont
  {Dowling}},\ and\ \bibinfo {author} {\bibfnamefont {T.}~\bibnamefont
  {Byrnes}},\ }\href {https://doi.org/10.1038/s41534-018-0090-2} {\bibfield
  {journal} {\bibinfo  {journal} {npj Quantum Inf}\ }\textbf {\bibinfo {volume}
  {4}},\ \bibinfo {pages} {40} (\bibinfo {year} {2018})}\BibitemShut {NoStop}%
\bibitem [{\citenamefont {Kitagawa}\ and\ \citenamefont
  {Ueda}(1993)}]{kitagawa1993SqueezedSpinStates}%
  \BibitemOpen
  \bibfield  {author} {\bibinfo {author} {\bibfnamefont {M.}~\bibnamefont
  {Kitagawa}}\ and\ \bibinfo {author} {\bibfnamefont {M.}~\bibnamefont
  {Ueda}},\ }\href {https://doi.org/10.1103/PhysRevA.47.5138} {\bibfield
  {journal} {\bibinfo  {journal} {Phys. Rev. A}\ }\textbf {\bibinfo {volume}
  {47}},\ \bibinfo {pages} {5138} (\bibinfo {year} {1993})}\BibitemShut
  {NoStop}%
\bibitem [{\citenamefont {L{\"u}cke}\ \emph {et~al.}(2011)\citenamefont
  {L{\"u}cke}, \citenamefont {Scherer}, \citenamefont {Kruse}, \citenamefont
  {Pezz{\'e}}, \citenamefont {Deuretzbacher}, \citenamefont {Hyllus},
  \citenamefont {Topic}, \citenamefont {Peise}, \citenamefont {Ertmer},
  \citenamefont {Arlt}, \citenamefont {Santos}, \citenamefont {Smerzi},\ and\
  \citenamefont {Klempt}}]{lucke2011TwinMatterWaves}%
  \BibitemOpen
  \bibfield  {author} {\bibinfo {author} {\bibfnamefont {B.}~\bibnamefont
  {L{\"u}cke}}, \bibinfo {author} {\bibfnamefont {M.}~\bibnamefont {Scherer}},
  \bibinfo {author} {\bibfnamefont {J.}~\bibnamefont {Kruse}}, \bibinfo
  {author} {\bibfnamefont {L.}~\bibnamefont {Pezz{\'e}}}, \bibinfo {author}
  {\bibfnamefont {F.}~\bibnamefont {Deuretzbacher}}, \bibinfo {author}
  {\bibfnamefont {P.}~\bibnamefont {Hyllus}}, \bibinfo {author} {\bibfnamefont
  {O.}~\bibnamefont {Topic}}, \bibinfo {author} {\bibfnamefont
  {J.}~\bibnamefont {Peise}}, \bibinfo {author} {\bibfnamefont
  {W.}~\bibnamefont {Ertmer}}, \bibinfo {author} {\bibfnamefont
  {J.}~\bibnamefont {Arlt}}, \bibinfo {author} {\bibfnamefont {L.}~\bibnamefont
  {Santos}}, \bibinfo {author} {\bibfnamefont {A.}~\bibnamefont {Smerzi}},\
  and\ \bibinfo {author} {\bibfnamefont {C.}~\bibnamefont {Klempt}},\ }\href
  {https://doi.org/10.1126/science.1208798} {\bibfield  {journal} {\bibinfo
  {journal} {Science}\ }\textbf {\bibinfo {volume} {334}},\ \bibinfo {pages}
  {773} (\bibinfo {year} {2011})}\BibitemShut {NoStop}%
\bibitem [{\citenamefont {Baumgarten}\ \emph {et~al.}(2008)\citenamefont
  {Baumgarten}, \citenamefont {Braun}, \citenamefont {Capiluppi}, \citenamefont
  {Ciullo}, \citenamefont {Dalpiaz}, \citenamefont {Kolster}, \citenamefont
  {Lenisa}, \citenamefont {Marukyan}, \citenamefont {Nass}, \citenamefont
  {Reggiani}, \citenamefont {Stancari},\ and\ \citenamefont
  {Steffens}}]{baumgarten2008FirstMeasurementHydrogen}%
  \BibitemOpen
  \bibfield  {author} {\bibinfo {author} {\bibfnamefont {C.}~\bibnamefont
  {Baumgarten}}, \bibinfo {author} {\bibfnamefont {B.}~\bibnamefont {Braun}},
  \bibinfo {author} {\bibfnamefont {M.}~\bibnamefont {Capiluppi}}, \bibinfo
  {author} {\bibfnamefont {G.}~\bibnamefont {Ciullo}}, \bibinfo {author}
  {\bibfnamefont {P.~F.}\ \bibnamefont {Dalpiaz}}, \bibinfo {author}
  {\bibfnamefont {H.}~\bibnamefont {Kolster}}, \bibinfo {author} {\bibfnamefont
  {P.}~\bibnamefont {Lenisa}}, \bibinfo {author} {\bibfnamefont
  {H.}~\bibnamefont {Marukyan}}, \bibinfo {author} {\bibfnamefont
  {A.}~\bibnamefont {Nass}}, \bibinfo {author} {\bibfnamefont {D.}~\bibnamefont
  {Reggiani}}, \bibinfo {author} {\bibfnamefont {M.}~\bibnamefont {Stancari}},\
  and\ \bibinfo {author} {\bibfnamefont {E.}~\bibnamefont {Steffens}},\ }\href
  {https://doi.org/10.1140/epjd/e2008-00124-1} {\bibfield  {journal} {\bibinfo
  {journal} {Eur. Phys. J. D}\ }\textbf {\bibinfo {volume} {48}},\ \bibinfo
  {pages} {343} (\bibinfo {year} {2008})}\BibitemShut {NoStop}%
\bibitem [{\citenamefont {Est{\`e}ve}\ \emph {et~al.}(2008)\citenamefont
  {Est{\`e}ve}, \citenamefont {Gross}, \citenamefont {Weller}, \citenamefont
  {Giovanazzi},\ and\ \citenamefont
  {Oberthaler}}]{esteve2008SqueezingEntanglementBose}%
  \BibitemOpen
  \bibfield  {author} {\bibinfo {author} {\bibfnamefont {J.}~\bibnamefont
  {Est{\`e}ve}}, \bibinfo {author} {\bibfnamefont {C.}~\bibnamefont {Gross}},
  \bibinfo {author} {\bibfnamefont {A.}~\bibnamefont {Weller}}, \bibinfo
  {author} {\bibfnamefont {S.}~\bibnamefont {Giovanazzi}},\ and\ \bibinfo
  {author} {\bibfnamefont {M.~K.}\ \bibnamefont {Oberthaler}},\ }\href
  {https://doi.org/10.1038/nature07332} {\bibfield  {journal} {\bibinfo
  {journal} {Nature}\ }\textbf {\bibinfo {volume} {455}},\ \bibinfo {pages}
  {1216} (\bibinfo {year} {2008})}\BibitemShut {NoStop}%
\bibitem [{\citenamefont {Riedel}\ \emph {et~al.}(2010)\citenamefont {Riedel},
  \citenamefont {B{\"o}hi}, \citenamefont {Li}, \citenamefont {H{\"a}nsch},
  \citenamefont {Sinatra},\ and\ \citenamefont
  {Treutlein}}]{riedel2010AtomchipbasedGenerationEntanglement}%
  \BibitemOpen
  \bibfield  {author} {\bibinfo {author} {\bibfnamefont {M.~F.}\ \bibnamefont
  {Riedel}}, \bibinfo {author} {\bibfnamefont {P.}~\bibnamefont {B{\"o}hi}},
  \bibinfo {author} {\bibfnamefont {Y.}~\bibnamefont {Li}}, \bibinfo {author}
  {\bibfnamefont {T.~W.}\ \bibnamefont {H{\"a}nsch}}, \bibinfo {author}
  {\bibfnamefont {A.}~\bibnamefont {Sinatra}},\ and\ \bibinfo {author}
  {\bibfnamefont {P.}~\bibnamefont {Treutlein}},\ }\href
  {https://doi.org/10.1038/nature08988} {\bibfield  {journal} {\bibinfo
  {journal} {Nature}\ }\textbf {\bibinfo {volume} {464}},\ \bibinfo {pages}
  {1170} (\bibinfo {year} {2010})}\BibitemShut {NoStop}%
\bibitem [{\citenamefont {Krauter}\ \emph {et~al.}(2011)\citenamefont
  {Krauter}, \citenamefont {Muschik}, \citenamefont {Jensen}, \citenamefont
  {Wasilewski}, \citenamefont {Petersen}, \citenamefont {Cirac},\ and\
  \citenamefont {Polzik}}]{krauter2011EntanglementGeneratedDissipation}%
  \BibitemOpen
  \bibfield  {author} {\bibinfo {author} {\bibfnamefont {H.}~\bibnamefont
  {Krauter}}, \bibinfo {author} {\bibfnamefont {C.~A.}\ \bibnamefont
  {Muschik}}, \bibinfo {author} {\bibfnamefont {K.}~\bibnamefont {Jensen}},
  \bibinfo {author} {\bibfnamefont {W.}~\bibnamefont {Wasilewski}}, \bibinfo
  {author} {\bibfnamefont {J.~M.}\ \bibnamefont {Petersen}}, \bibinfo {author}
  {\bibfnamefont {J.~I.}\ \bibnamefont {Cirac}},\ and\ \bibinfo {author}
  {\bibfnamefont {E.~S.}\ \bibnamefont {Polzik}},\ }\href
  {https://doi.org/10.1103/PhysRevLett.107.080503} {\bibfield  {journal}
  {\bibinfo  {journal} {Phys. Rev. Lett.}\ }\textbf {\bibinfo {volume} {107}},\
  \bibinfo {pages} {080503} (\bibinfo {year} {2011})}\BibitemShut {NoStop}%
\bibitem [{\citenamefont {Peise}\ \emph {et~al.}(2015)\citenamefont {Peise},
  \citenamefont {Kruse}, \citenamefont {Lange}, \citenamefont {L{\"u}cke},
  \citenamefont {Pezz{\`e}}, \citenamefont {Arlt}, \citenamefont {Ertmer},
  \citenamefont {Hammerer}, \citenamefont {Santos}, \citenamefont {Smerzi},\
  and\ \citenamefont {Klempt}}]{peise2015SatisfyingEinsteinPodolsky}%
  \BibitemOpen
  \bibfield  {author} {\bibinfo {author} {\bibfnamefont {J.}~\bibnamefont
  {Peise}}, \bibinfo {author} {\bibfnamefont {I.}~\bibnamefont {Kruse}},
  \bibinfo {author} {\bibfnamefont {K.}~\bibnamefont {Lange}}, \bibinfo
  {author} {\bibfnamefont {B.}~\bibnamefont {L{\"u}cke}}, \bibinfo {author}
  {\bibfnamefont {L.}~\bibnamefont {Pezz{\`e}}}, \bibinfo {author}
  {\bibfnamefont {J.}~\bibnamefont {Arlt}}, \bibinfo {author} {\bibfnamefont
  {W.}~\bibnamefont {Ertmer}}, \bibinfo {author} {\bibfnamefont
  {K.}~\bibnamefont {Hammerer}}, \bibinfo {author} {\bibfnamefont
  {L.}~\bibnamefont {Santos}}, \bibinfo {author} {\bibfnamefont
  {A.}~\bibnamefont {Smerzi}},\ and\ \bibinfo {author} {\bibfnamefont
  {C.}~\bibnamefont {Klempt}},\ }\href {https://doi.org/10.1038/ncomms9984}
  {\bibfield  {journal} {\bibinfo  {journal} {Nat Commun}\ }\textbf {\bibinfo
  {volume} {6}},\ \bibinfo {pages} {8984} (\bibinfo {year} {2015})}\BibitemShut
  {NoStop}%
\bibitem [{\citenamefont {L{\"u}cke}\ \emph {et~al.}(2014)\citenamefont
  {L{\"u}cke}, \citenamefont {Peise}, \citenamefont {Vitagliano}, \citenamefont
  {Arlt}, \citenamefont {Santos}, \citenamefont {T{\'o}th},\ and\ \citenamefont
  {Klempt}}]{lucke2014DetectingMultiparticleEntanglement}%
  \BibitemOpen
  \bibfield  {author} {\bibinfo {author} {\bibfnamefont {B.}~\bibnamefont
  {L{\"u}cke}}, \bibinfo {author} {\bibfnamefont {J.}~\bibnamefont {Peise}},
  \bibinfo {author} {\bibfnamefont {G.}~\bibnamefont {Vitagliano}}, \bibinfo
  {author} {\bibfnamefont {J.}~\bibnamefont {Arlt}}, \bibinfo {author}
  {\bibfnamefont {L.}~\bibnamefont {Santos}}, \bibinfo {author} {\bibfnamefont
  {G.}~\bibnamefont {T{\'o}th}},\ and\ \bibinfo {author} {\bibfnamefont
  {C.}~\bibnamefont {Klempt}},\ }\href
  {https://doi.org/10.1103/PhysRevLett.112.155304} {\bibfield  {journal}
  {\bibinfo  {journal} {Phys. Rev. Lett.}\ }\textbf {\bibinfo {volume} {112}},\
  \bibinfo {pages} {155304} (\bibinfo {year} {2014})}\BibitemShut {NoStop}%
\bibitem [{\citenamefont {Lange}\ \emph {et~al.}(2018)\citenamefont {Lange},
  \citenamefont {Peise}, \citenamefont {L{\"u}cke}, \citenamefont {Kruse},
  \citenamefont {Vitagliano}, \citenamefont {Apellaniz}, \citenamefont
  {Kleinmann}, \citenamefont {T{\'o}th},\ and\ \citenamefont
  {Klempt}}]{lange2018EntanglementTwoSpatially}%
  \BibitemOpen
  \bibfield  {author} {\bibinfo {author} {\bibfnamefont {K.}~\bibnamefont
  {Lange}}, \bibinfo {author} {\bibfnamefont {J.}~\bibnamefont {Peise}},
  \bibinfo {author} {\bibfnamefont {B.}~\bibnamefont {L{\"u}cke}}, \bibinfo
  {author} {\bibfnamefont {I.}~\bibnamefont {Kruse}}, \bibinfo {author}
  {\bibfnamefont {G.}~\bibnamefont {Vitagliano}}, \bibinfo {author}
  {\bibfnamefont {I.}~\bibnamefont {Apellaniz}}, \bibinfo {author}
  {\bibfnamefont {M.}~\bibnamefont {Kleinmann}}, \bibinfo {author}
  {\bibfnamefont {G.}~\bibnamefont {T{\'o}th}},\ and\ \bibinfo {author}
  {\bibfnamefont {C.}~\bibnamefont {Klempt}},\ }\href
  {https://doi.org/10.1126/science.aao2035} {\bibfield  {journal} {\bibinfo
  {journal} {Science}\ }\textbf {\bibinfo {volume} {360}},\ \bibinfo {pages}
  {416} (\bibinfo {year} {2018})}\BibitemShut {NoStop}%
\bibitem [{\citenamefont {Kunkel}\ \emph {et~al.}(2018)\citenamefont {Kunkel},
  \citenamefont {Pr{\"u}fer}, \citenamefont {Strobel}, \citenamefont
  {Linnemann}, \citenamefont {Fr{\"o}lian}, \citenamefont {Gasenzer},
  \citenamefont {G{\"a}rttner},\ and\ \citenamefont
  {Oberthaler}}]{kunkel2018SpatiallyDistributedMultipartite}%
  \BibitemOpen
  \bibfield  {author} {\bibinfo {author} {\bibfnamefont {P.}~\bibnamefont
  {Kunkel}}, \bibinfo {author} {\bibfnamefont {M.}~\bibnamefont {Pr{\"u}fer}},
  \bibinfo {author} {\bibfnamefont {H.}~\bibnamefont {Strobel}}, \bibinfo
  {author} {\bibfnamefont {D.}~\bibnamefont {Linnemann}}, \bibinfo {author}
  {\bibfnamefont {A.}~\bibnamefont {Fr{\"o}lian}}, \bibinfo {author}
  {\bibfnamefont {T.}~\bibnamefont {Gasenzer}}, \bibinfo {author}
  {\bibfnamefont {M.}~\bibnamefont {G{\"a}rttner}},\ and\ \bibinfo {author}
  {\bibfnamefont {M.~K.}\ \bibnamefont {Oberthaler}},\ }\href
  {https://doi.org/10.1126/science.aao2254} {\bibfield  {journal} {\bibinfo
  {journal} {Science}\ }\textbf {\bibinfo {volume} {360}},\ \bibinfo {pages}
  {413} (\bibinfo {year} {2018})}\BibitemShut {NoStop}%
\bibitem [{\citenamefont {Fadel}\ \emph {et~al.}(2018)\citenamefont {Fadel},
  \citenamefont {Zibold}, \citenamefont {D{\'e}camps},\ and\ \citenamefont
  {Treutlein}}]{fadel2018SpatialEntanglementPatterns}%
  \BibitemOpen
  \bibfield  {author} {\bibinfo {author} {\bibfnamefont {M.}~\bibnamefont
  {Fadel}}, \bibinfo {author} {\bibfnamefont {T.}~\bibnamefont {Zibold}},
  \bibinfo {author} {\bibfnamefont {B.}~\bibnamefont {D{\'e}camps}},\ and\
  \bibinfo {author} {\bibfnamefont {P.}~\bibnamefont {Treutlein}},\ }\href
  {https://doi.org/10.1126/science.aao1850} {\bibfield  {journal} {\bibinfo
  {journal} {Science}\ }\textbf {\bibinfo {volume} {360}},\ \bibinfo {pages}
  {409} (\bibinfo {year} {2018})}\BibitemShut {NoStop}%
\bibitem [{\citenamefont {Tura}\ \emph {et~al.}(2014)\citenamefont {Tura},
  \citenamefont {Augusiak}, \citenamefont {Sainz}, \citenamefont {V{\'e}rtesi},
  \citenamefont {Lewenstein},\ and\ \citenamefont
  {Ac{\'i}n}}]{tura2014DetectingNonlocalityManybody}%
  \BibitemOpen
  \bibfield  {author} {\bibinfo {author} {\bibfnamefont {J.}~\bibnamefont
  {Tura}}, \bibinfo {author} {\bibfnamefont {R.}~\bibnamefont {Augusiak}},
  \bibinfo {author} {\bibfnamefont {A.~B.}\ \bibnamefont {Sainz}}, \bibinfo
  {author} {\bibfnamefont {T.}~\bibnamefont {V{\'e}rtesi}}, \bibinfo {author}
  {\bibfnamefont {M.}~\bibnamefont {Lewenstein}},\ and\ \bibinfo {author}
  {\bibfnamefont {A.}~\bibnamefont {Ac{\'i}n}},\ }\href
  {https://doi.org/10.1126/science.1247715} {\bibfield  {journal} {\bibinfo
  {journal} {Science}\ }\textbf {\bibinfo {volume} {344}},\ \bibinfo {pages}
  {1256} (\bibinfo {year} {2014})}\BibitemShut {NoStop}%
\bibitem [{\citenamefont {Schmied}\ \emph {et~al.}(2016)\citenamefont
  {Schmied}, \citenamefont {Bancal}, \citenamefont {Allard}, \citenamefont
  {Fadel}, \citenamefont {Scarani}, \citenamefont {Treutlein},\ and\
  \citenamefont {Sangouard}}]{schmied2016BellCorrelationsBoseEinstein}%
  \BibitemOpen
  \bibfield  {author} {\bibinfo {author} {\bibfnamefont {R.}~\bibnamefont
  {Schmied}}, \bibinfo {author} {\bibfnamefont {J.-D.}\ \bibnamefont {Bancal}},
  \bibinfo {author} {\bibfnamefont {B.}~\bibnamefont {Allard}}, \bibinfo
  {author} {\bibfnamefont {M.}~\bibnamefont {Fadel}}, \bibinfo {author}
  {\bibfnamefont {V.}~\bibnamefont {Scarani}}, \bibinfo {author} {\bibfnamefont
  {P.}~\bibnamefont {Treutlein}},\ and\ \bibinfo {author} {\bibfnamefont
  {N.}~\bibnamefont {Sangouard}},\ }\href
  {https://doi.org/10.1126/science.aad8665} {\bibfield  {journal} {\bibinfo
  {journal} {Science}\ }\textbf {\bibinfo {volume} {352}},\ \bibinfo {pages}
  {441} (\bibinfo {year} {2016})}\BibitemShut {NoStop}%
\bibitem [{\citenamefont {Engelsen}\ \emph {et~al.}(2017)\citenamefont
  {Engelsen}, \citenamefont {Krishnakumar}, \citenamefont {Hosten},\ and\
  \citenamefont {Kasevich}}]{engelsen2017BellCorrelationsSpinSqueezed}%
  \BibitemOpen
  \bibfield  {author} {\bibinfo {author} {\bibfnamefont {N.~J.}\ \bibnamefont
  {Engelsen}}, \bibinfo {author} {\bibfnamefont {R.}~\bibnamefont
  {Krishnakumar}}, \bibinfo {author} {\bibfnamefont {O.}~\bibnamefont
  {Hosten}},\ and\ \bibinfo {author} {\bibfnamefont {M.~A.}\ \bibnamefont
  {Kasevich}},\ }\href {https://doi.org/10.1103/PhysRevLett.118.140401}
  {\bibfield  {journal} {\bibinfo  {journal} {Phys. Rev. Lett.}\ }\textbf
  {\bibinfo {volume} {118}},\ \bibinfo {pages} {140401} (\bibinfo {year}
  {2017})}\BibitemShut {NoStop}%
\bibitem [{\citenamefont {Aloy}\ \emph {et~al.}(2019)\citenamefont {Aloy},
  \citenamefont {Tura}, \citenamefont {Baccari}, \citenamefont {Ac{\'i}n},
  \citenamefont {Lewenstein},\ and\ \citenamefont
  {Augusiak}}]{aloy2019DeviceIndependentWitnessesEntanglement}%
  \BibitemOpen
  \bibfield  {author} {\bibinfo {author} {\bibfnamefont {A.}~\bibnamefont
  {Aloy}}, \bibinfo {author} {\bibfnamefont {J.}~\bibnamefont {Tura}}, \bibinfo
  {author} {\bibfnamefont {F.}~\bibnamefont {Baccari}}, \bibinfo {author}
  {\bibfnamefont {A.}~\bibnamefont {Ac{\'i}n}}, \bibinfo {author}
  {\bibfnamefont {M.}~\bibnamefont {Lewenstein}},\ and\ \bibinfo {author}
  {\bibfnamefont {R.}~\bibnamefont {Augusiak}},\ }\href
  {https://doi.org/10.1103/PhysRevLett.123.100507} {\bibfield  {journal}
  {\bibinfo  {journal} {Phys. Rev. Lett.}\ }\textbf {\bibinfo {volume} {123}},\
  \bibinfo {pages} {100507} (\bibinfo {year} {2019})}\BibitemShut {NoStop}%
\bibitem [{\citenamefont {Wagner}\ \emph {et~al.}(2017)\citenamefont {Wagner},
  \citenamefont {Schmied}, \citenamefont {Fadel}, \citenamefont {Treutlein},
  \citenamefont {Sangouard},\ and\ \citenamefont
  {Bancal}}]{wagner2017BellCorrelationsManyBody}%
  \BibitemOpen
  \bibfield  {author} {\bibinfo {author} {\bibfnamefont {S.}~\bibnamefont
  {Wagner}}, \bibinfo {author} {\bibfnamefont {R.}~\bibnamefont {Schmied}},
  \bibinfo {author} {\bibfnamefont {M.}~\bibnamefont {Fadel}}, \bibinfo
  {author} {\bibfnamefont {P.}~\bibnamefont {Treutlein}}, \bibinfo {author}
  {\bibfnamefont {N.}~\bibnamefont {Sangouard}},\ and\ \bibinfo {author}
  {\bibfnamefont {J.-D.}\ \bibnamefont {Bancal}},\ }\href
  {https://doi.org/10.1103/PhysRevLett.119.170403} {\bibfield  {journal}
  {\bibinfo  {journal} {Phys. Rev. Lett.}\ }\textbf {\bibinfo {volume} {119}},\
  \bibinfo {pages} {170403} (\bibinfo {year} {2017})}\BibitemShut {NoStop}%
\bibitem [{\citenamefont {Oudot}\ \emph {et~al.}(2019)\citenamefont {Oudot},
  \citenamefont {Bancal}, \citenamefont {Sekatski},\ and\ \citenamefont
  {Sangouard}}]{oudot2019BipartiteNonlocalityManybody}%
  \BibitemOpen
  \bibfield  {author} {\bibinfo {author} {\bibfnamefont {E.}~\bibnamefont
  {Oudot}}, \bibinfo {author} {\bibfnamefont {J.-D.}\ \bibnamefont {Bancal}},
  \bibinfo {author} {\bibfnamefont {P.}~\bibnamefont {Sekatski}},\ and\
  \bibinfo {author} {\bibfnamefont {N.}~\bibnamefont {Sangouard}},\ }\href
  {https://doi.org/10.1088/1367-2630/ab4c7c} {\bibfield  {journal} {\bibinfo
  {journal} {New J. Phys.}\ }\textbf {\bibinfo {volume} {21}},\ \bibinfo
  {pages} {103043} (\bibinfo {year} {2019})}\BibitemShut {NoStop}%
\bibitem [{\citenamefont {Kitzinger}\ \emph {et~al.}(2020)\citenamefont
  {Kitzinger}, \citenamefont {Chaudhary}, \citenamefont {Kondappan},
  \citenamefont {Ivannikov},\ and\ \citenamefont
  {Byrnes}}]{kitzinger2020TwoaxisTwospinSqueezed}%
  \BibitemOpen
  \bibfield  {author} {\bibinfo {author} {\bibfnamefont {J.}~\bibnamefont
  {Kitzinger}}, \bibinfo {author} {\bibfnamefont {M.}~\bibnamefont
  {Chaudhary}}, \bibinfo {author} {\bibfnamefont {M.}~\bibnamefont
  {Kondappan}}, \bibinfo {author} {\bibfnamefont {V.}~\bibnamefont
  {Ivannikov}},\ and\ \bibinfo {author} {\bibfnamefont {T.}~\bibnamefont
  {Byrnes}},\ }\href {https://doi.org/10.1103/PhysRevResearch.2.033504}
  {\bibfield  {journal} {\bibinfo  {journal} {Phys. Rev. Research}\ }\textbf
  {\bibinfo {volume} {2}},\ \bibinfo {pages} {033504} (\bibinfo {year}
  {2020})}\BibitemShut {NoStop}%
\bibitem [{\citenamefont {Cavalcanti}\ \emph {et~al.}(2007)\citenamefont
  {Cavalcanti}, \citenamefont {Foster}, \citenamefont {Reid},\ and\
  \citenamefont {Drummond}}]{cavalcanti2007BellInequalitiesContinuousVariable}%
  \BibitemOpen
  \bibfield  {author} {\bibinfo {author} {\bibfnamefont {E.~G.}\ \bibnamefont
  {Cavalcanti}}, \bibinfo {author} {\bibfnamefont {C.~J.}\ \bibnamefont
  {Foster}}, \bibinfo {author} {\bibfnamefont {M.~D.}\ \bibnamefont {Reid}},\
  and\ \bibinfo {author} {\bibfnamefont {P.~D.}\ \bibnamefont {Drummond}},\
  }\href {https://doi.org/10.1103/PhysRevLett.99.210405} {\bibfield  {journal}
  {\bibinfo  {journal} {Phys. Rev. Lett.}\ }\textbf {\bibinfo {volume} {99}},\
  \bibinfo {pages} {210405} (\bibinfo {year} {2007})}\BibitemShut {NoStop}%
\bibitem [{\citenamefont {Braunstein}\ and\ \citenamefont {{van
  Loock}}(2005)}]{braunstein2005QuantumInformationContinuous}%
  \BibitemOpen
  \bibfield  {author} {\bibinfo {author} {\bibfnamefont {S.~L.}\ \bibnamefont
  {Braunstein}}\ and\ \bibinfo {author} {\bibfnamefont {P.}~\bibnamefont {{van
  Loock}}},\ }\href {https://doi.org/10.1103/RevModPhys.77.513} {\bibfield
  {journal} {\bibinfo  {journal} {Rev. Mod. Phys.}\ }\textbf {\bibinfo {volume}
  {77}},\ \bibinfo {pages} {513} (\bibinfo {year} {2005})}\BibitemShut
  {NoStop}%
\bibitem [{\citenamefont {Ralph}\ \emph {et~al.}(2000)\citenamefont {Ralph},
  \citenamefont {Munro},\ and\ \citenamefont
  {Polkinghorne}}]{ralph2000ProposalMeasurementBellType}%
  \BibitemOpen
  \bibfield  {author} {\bibinfo {author} {\bibfnamefont {T.~C.}\ \bibnamefont
  {Ralph}}, \bibinfo {author} {\bibfnamefont {W.~J.}\ \bibnamefont {Munro}},\
  and\ \bibinfo {author} {\bibfnamefont {R.~E.~S.}\ \bibnamefont
  {Polkinghorne}},\ }\href {https://doi.org/10.1103/PhysRevLett.85.2035}
  {\bibfield  {journal} {\bibinfo  {journal} {Phys. Rev. Lett.}\ }\textbf
  {\bibinfo {volume} {85}},\ \bibinfo {pages} {2035} (\bibinfo {year}
  {2000})}\BibitemShut {NoStop}%
\bibitem [{\citenamefont {Thearle}\ \emph {et~al.}(2018)\citenamefont
  {Thearle}, \citenamefont {Janousek}, \citenamefont {Armstrong}, \citenamefont
  {Hosseini}, \citenamefont {Sch{\"u}nemann~(Mraz)}, \citenamefont {Assad},
  \citenamefont {Symul}, \citenamefont {James}, \citenamefont {Huntington},
  \citenamefont {Ralph},\ and\ \citenamefont
  {Lam}}]{thearle2018ViolationBellInequality}%
  \BibitemOpen
  \bibfield  {author} {\bibinfo {author} {\bibfnamefont {O.}~\bibnamefont
  {Thearle}}, \bibinfo {author} {\bibfnamefont {J.}~\bibnamefont {Janousek}},
  \bibinfo {author} {\bibfnamefont {S.}~\bibnamefont {Armstrong}}, \bibinfo
  {author} {\bibfnamefont {S.}~\bibnamefont {Hosseini}}, \bibinfo {author}
  {\bibfnamefont {M.}~\bibnamefont {Sch{\"u}nemann~(Mraz)}}, \bibinfo {author}
  {\bibfnamefont {S.}~\bibnamefont {Assad}}, \bibinfo {author} {\bibfnamefont
  {T.}~\bibnamefont {Symul}}, \bibinfo {author} {\bibfnamefont {M.~R.}\
  \bibnamefont {James}}, \bibinfo {author} {\bibfnamefont {E.}~\bibnamefont
  {Huntington}}, \bibinfo {author} {\bibfnamefont {T.~C.}\ \bibnamefont
  {Ralph}},\ and\ \bibinfo {author} {\bibfnamefont {P.~K.}\ \bibnamefont
  {Lam}},\ }\href {https://doi.org/10.1103/PhysRevLett.120.040406} {\bibfield
  {journal} {\bibinfo  {journal} {Phys. Rev. Lett.}\ }\textbf {\bibinfo
  {volume} {120}},\ \bibinfo {pages} {040406} (\bibinfo {year}
  {2018})}\BibitemShut {NoStop}%
\bibitem [{\citenamefont {Jing}\ \emph {et~al.}(2019)\citenamefont {Jing},
  \citenamefont {Fadel}, \citenamefont {Ivannikov},\ and\ \citenamefont
  {Byrnes}}]{jing2019SplitSpinsqueezedBose}%
  \BibitemOpen
  \bibfield  {author} {\bibinfo {author} {\bibfnamefont {Y.}~\bibnamefont
  {Jing}}, \bibinfo {author} {\bibfnamefont {M.}~\bibnamefont {Fadel}},
  \bibinfo {author} {\bibfnamefont {V.}~\bibnamefont {Ivannikov}},\ and\
  \bibinfo {author} {\bibfnamefont {T.}~\bibnamefont {Byrnes}},\ }\href
  {https://doi.org/10.1088/1367-2630/ab3fcf} {\bibfield  {journal} {\bibinfo
  {journal} {New J. Phys.}\ }\textbf {\bibinfo {volume} {21}},\ \bibinfo
  {pages} {093038} (\bibinfo {year} {2019})}\BibitemShut {NoStop}%
\bibitem [{\citenamefont {Fadel}\ and\ \citenamefont
  {Gessner}(2020)}]{fadel2020RelatingSpinSqueezing}%
  \BibitemOpen
  \bibfield  {author} {\bibinfo {author} {\bibfnamefont {M.}~\bibnamefont
  {Fadel}}\ and\ \bibinfo {author} {\bibfnamefont {M.}~\bibnamefont
  {Gessner}},\ }\href {https://doi.org/10.1103/PhysRevA.102.012412} {\bibfield
  {journal} {\bibinfo  {journal} {Phys. Rev. A}\ }\textbf {\bibinfo {volume}
  {102}},\ \bibinfo {pages} {012412} (\bibinfo {year} {2020})}\BibitemShut
  {NoStop}%
\bibitem [{\citenamefont {Chang}\ \emph {et~al.}(2004)\citenamefont {Chang},
  \citenamefont {Hamley}, \citenamefont {Barrett}, \citenamefont {Sauer},
  \citenamefont {Fortier}, \citenamefont {Zhang}, \citenamefont {You},\ and\
  \citenamefont {Chapman}}]{chang2004ObservationSpinorDynamics}%
  \BibitemOpen
  \bibfield  {author} {\bibinfo {author} {\bibfnamefont {M.-S.}\ \bibnamefont
  {Chang}}, \bibinfo {author} {\bibfnamefont {C.~D.}\ \bibnamefont {Hamley}},
  \bibinfo {author} {\bibfnamefont {M.~D.}\ \bibnamefont {Barrett}}, \bibinfo
  {author} {\bibfnamefont {J.~A.}\ \bibnamefont {Sauer}}, \bibinfo {author}
  {\bibfnamefont {K.~M.}\ \bibnamefont {Fortier}}, \bibinfo {author}
  {\bibfnamefont {W.}~\bibnamefont {Zhang}}, \bibinfo {author} {\bibfnamefont
  {L.}~\bibnamefont {You}},\ and\ \bibinfo {author} {\bibfnamefont {M.~S.}\
  \bibnamefont {Chapman}},\ }\href
  {https://doi.org/10.1103/PhysRevLett.92.140403} {\bibfield  {journal}
  {\bibinfo  {journal} {Phys. Rev. Lett.}\ }\textbf {\bibinfo {volume} {92}},\
  \bibinfo {pages} {140403} (\bibinfo {year} {2004})}\BibitemShut {NoStop}%
\bibitem [{\citenamefont {Klempt}\ \emph {et~al.}(2010)\citenamefont {Klempt},
  \citenamefont {Topic}, \citenamefont {Gebreyesus}, \citenamefont {Scherer},
  \citenamefont {Henninger}, \citenamefont {Hyllus}, \citenamefont {Ertmer},
  \citenamefont {Santos},\ and\ \citenamefont
  {Arlt}}]{klempt2010ParametricAmplificationVacuum}%
  \BibitemOpen
  \bibfield  {author} {\bibinfo {author} {\bibfnamefont {C.}~\bibnamefont
  {Klempt}}, \bibinfo {author} {\bibfnamefont {O.}~\bibnamefont {Topic}},
  \bibinfo {author} {\bibfnamefont {G.}~\bibnamefont {Gebreyesus}}, \bibinfo
  {author} {\bibfnamefont {M.}~\bibnamefont {Scherer}}, \bibinfo {author}
  {\bibfnamefont {T.}~\bibnamefont {Henninger}}, \bibinfo {author}
  {\bibfnamefont {P.}~\bibnamefont {Hyllus}}, \bibinfo {author} {\bibfnamefont
  {W.}~\bibnamefont {Ertmer}}, \bibinfo {author} {\bibfnamefont
  {L.}~\bibnamefont {Santos}},\ and\ \bibinfo {author} {\bibfnamefont {J.~J.}\
  \bibnamefont {Arlt}},\ }\href
  {https://doi.org/10.1103/PhysRevLett.104.195303} {\bibfield  {journal}
  {\bibinfo  {journal} {Phys. Rev. Lett.}\ }\textbf {\bibinfo {volume} {104}},\
  \bibinfo {pages} {195303} (\bibinfo {year} {2010})}\BibitemShut {NoStop}%
\bibitem [{\citenamefont {Law}\ \emph {et~al.}(1998)\citenamefont {Law},
  \citenamefont {Pu},\ and\ \citenamefont
  {Bigelow}}]{law1998QuantumSpinsMixing}%
  \BibitemOpen
  \bibfield  {author} {\bibinfo {author} {\bibfnamefont {C.~K.}\ \bibnamefont
  {Law}}, \bibinfo {author} {\bibfnamefont {H.}~\bibnamefont {Pu}},\ and\
  \bibinfo {author} {\bibfnamefont {N.~P.}\ \bibnamefont {Bigelow}},\ }\href
  {https://doi.org/10.1103/PhysRevLett.81.5257} {\bibfield  {journal} {\bibinfo
   {journal} {Phys. Rev. Lett.}\ }\textbf {\bibinfo {volume} {81}},\ \bibinfo
  {pages} {5257} (\bibinfo {year} {1998})}\BibitemShut {NoStop}%
\bibitem [{\citenamefont {Gabbrielli}\ \emph {et~al.}(2015)\citenamefont
  {Gabbrielli}, \citenamefont {Pezz{\`e}},\ and\ \citenamefont
  {Smerzi}}]{gabbrielli2015SpinMixingInterferometryBoseEinstein}%
  \BibitemOpen
  \bibfield  {author} {\bibinfo {author} {\bibfnamefont {M.}~\bibnamefont
  {Gabbrielli}}, \bibinfo {author} {\bibfnamefont {L.}~\bibnamefont
  {Pezz{\`e}}},\ and\ \bibinfo {author} {\bibfnamefont {A.}~\bibnamefont
  {Smerzi}},\ }\href {https://doi.org/10.1103/PhysRevLett.115.163002}
  {\bibfield  {journal} {\bibinfo  {journal} {Phys. Rev. Lett.}\ }\textbf
  {\bibinfo {volume} {115}},\ \bibinfo {pages} {163002} (\bibinfo {year}
  {2015})}\BibitemShut {NoStop}%
\bibitem [{\citenamefont
  {Carmichael}(1999)}]{carmichael1999StatisticalMethodsQuantum}%
  \BibitemOpen
  \bibfield  {author} {\bibinfo {author} {\bibfnamefont {H.~J.}\ \bibnamefont
  {Carmichael}},\ }\href {https://doi.org/10.1007/978-3-662-03875-8} {\emph
  {\bibinfo {title} {Statistical {{Methods}} in {{Quantum Optics}} 1}}}\
  (\bibinfo  {publisher} {{Springer Berlin Heidelberg}},\ \bibinfo {address}
  {{Berlin, Heidelberg}},\ \bibinfo {year} {1999})\BibitemShut {NoStop}%
\bibitem [{\citenamefont {Walls}\ and\ \citenamefont
  {Milburn}(1994)}]{walls1994QuantumOptics}%
  \BibitemOpen
  \bibfield  {author} {\bibinfo {author} {\bibfnamefont {D.~F.}\ \bibnamefont
  {Walls}}\ and\ \bibinfo {author} {\bibfnamefont {G.~J.}\ \bibnamefont
  {Milburn}},\ }\href {https://doi.org/10.1007/978-3-642-79504-6} {\emph
  {\bibinfo {title} {Quantum {{Optics}}}}}\ (\bibinfo  {publisher} {{Springer
  Berlin Heidelberg}},\ \bibinfo {address} {{Berlin, Heidelberg}},\ \bibinfo
  {year} {1994})\BibitemShut {NoStop}%
\bibitem [{\citenamefont {Gerry}\ and\ \citenamefont
  {Knight}(2004)}]{gerry2004IntroductoryQuantumOptics}%
  \BibitemOpen
  \bibfield  {author} {\bibinfo {author} {\bibfnamefont {C.}~\bibnamefont
  {Gerry}}\ and\ \bibinfo {author} {\bibfnamefont {P.}~\bibnamefont {Knight}},\
  }\href {https://doi.org/10.1017/CBO9780511791239} {\emph {\bibinfo {title}
  {Introductory {{Quantum Optics}}}}},\ \bibinfo {edition} {1st}\ ed.\
  (\bibinfo  {publisher} {{Cambridge University Press}},\ \bibinfo {year}
  {2004})\BibitemShut {NoStop}%
\bibitem [{\citenamefont {Scully}\ and\ \citenamefont
  {Zubairy}(1997)}]{scully1997QuantumOpticsa}%
  \BibitemOpen
  \bibfield  {author} {\bibinfo {author} {\bibfnamefont {M.~O.}\ \bibnamefont
  {Scully}}\ and\ \bibinfo {author} {\bibfnamefont {M.~S.}\ \bibnamefont
  {Zubairy}},\ }\href {https://doi.org/10.1017/CBO9780511813993} {\emph
  {\bibinfo {title} {Quantum {{Optics}}}}},\ \bibinfo {edition} {1st}\ ed.\
  (\bibinfo  {publisher} {{Cambridge University Press}},\ \bibinfo {year}
  {1997})\BibitemShut {NoStop}%
\bibitem [{\citenamefont {Byrnes}\ and\ \citenamefont
  {{Ilo-Okeke}}(2021)}]{byrnes2021QuantumAtomOptics}%
  \BibitemOpen
  \bibfield  {author} {\bibinfo {author} {\bibfnamefont {T.}~\bibnamefont
  {Byrnes}}\ and\ \bibinfo {author} {\bibfnamefont {E.~O.}\ \bibnamefont
  {{Ilo-Okeke}}},\ }\href {https://doi.org/10.1017/9781108975353} {\emph
  {\bibinfo {title} {Quantum {{Atom Optics}}: Theory and {{Applications}} to
  {{Quantum Technology}}}}},\ \bibinfo {edition} {1st}\ ed.\ (\bibinfo
  {publisher} {{Cambridge University Press}},\ \bibinfo {year}
  {2021})\BibitemShut {NoStop}%
\bibitem [{\citenamefont
  {Pearle}(1970)}]{pearle1970HiddenVariableExampleBased}%
  \BibitemOpen
  \bibfield  {author} {\bibinfo {author} {\bibfnamefont {P.~M.}\ \bibnamefont
  {Pearle}},\ }\href {https://doi.org/10.1103/PhysRevD.2.1418} {\bibfield
  {journal} {\bibinfo  {journal} {Phys. Rev. D}\ }\textbf {\bibinfo {volume}
  {2}},\ \bibinfo {pages} {1418} (\bibinfo {year} {1970})}\BibitemShut
  {NoStop}%
\bibitem [{\citenamefont
  {Peres}(1997)}]{peres1997BellInequalitiesPostselection}%
  \BibitemOpen
  \bibfield  {author} {\bibinfo {author} {\bibfnamefont {A.}~\bibnamefont
  {Peres}},\ }in\ \href {https://doi.org/10.1007/978-94-017-2732-7_14} {\emph
  {\bibinfo {booktitle} {Potentiality, {{Entanglement}} and
  {{Passion}}-at-a-{{Distance}}}}},\ Vol.\ \bibinfo {volume} {194},\ \bibinfo
  {editor} {edited by\ \bibinfo {editor} {\bibfnamefont {R.~S.}\ \bibnamefont
  {Cohen}}, \bibinfo {editor} {\bibfnamefont {M.~W.}\ \bibnamefont
  {Wartofsky}}, \bibinfo {editor} {\bibfnamefont {R.~S.}\ \bibnamefont
  {Cohen}}, \bibinfo {editor} {\bibfnamefont {M.}~\bibnamefont {Horne}},\ and\
  \bibinfo {editor} {\bibfnamefont {J.}~\bibnamefont {Stachel}}}\ (\bibinfo
  {publisher} {{Springer Netherlands}},\ \bibinfo {address} {{Dordrecht}},\
  \bibinfo {year} {1997})\ pp.\ \bibinfo {pages} {191--196}\BibitemShut
  {NoStop}%
\bibitem [{\citenamefont {De~Caro}\ and\ \citenamefont
  {Garuccio}(1994)}]{decaro1994ReliabilityBellinequalityMeasurements}%
  \BibitemOpen
  \bibfield  {author} {\bibinfo {author} {\bibfnamefont {L.}~\bibnamefont
  {De~Caro}}\ and\ \bibinfo {author} {\bibfnamefont {A.}~\bibnamefont
  {Garuccio}},\ }\href {https://doi.org/10.1103/PhysRevA.50.R2803} {\bibfield
  {journal} {\bibinfo  {journal} {Phys. Rev. A}\ }\textbf {\bibinfo {volume}
  {50}},\ \bibinfo {pages} {R2803} (\bibinfo {year} {1994})}\BibitemShut
  {NoStop}%
\bibitem [{\citenamefont {Bonneau}\ \emph {et~al.}(2018)\citenamefont
  {Bonneau}, \citenamefont {Munro}, \citenamefont {Nemoto},\ and\ \citenamefont
  {Schmiedmayer}}]{bonneau2018CharacterizingTwinparticleEntanglement}%
  \BibitemOpen
  \bibfield  {author} {\bibinfo {author} {\bibfnamefont {M.}~\bibnamefont
  {Bonneau}}, \bibinfo {author} {\bibfnamefont {W.~J.}\ \bibnamefont {Munro}},
  \bibinfo {author} {\bibfnamefont {K.}~\bibnamefont {Nemoto}},\ and\ \bibinfo
  {author} {\bibfnamefont {J.}~\bibnamefont {Schmiedmayer}},\ }\href
  {https://doi.org/10.1103/PhysRevA.98.033608} {\bibfield  {journal} {\bibinfo
  {journal} {Phys. Rev. A}\ }\textbf {\bibinfo {volume} {98}},\ \bibinfo
  {pages} {033608} (\bibinfo {year} {2018})}\BibitemShut {NoStop}%
\bibitem [{\citenamefont {Gilchrist}\ \emph {et~al.}(1998)\citenamefont
  {Gilchrist}, \citenamefont {Deuar},\ and\ \citenamefont
  {Reid}}]{gilchrist1998ContradictionQuantumMechanics}%
  \BibitemOpen
  \bibfield  {author} {\bibinfo {author} {\bibfnamefont {A.}~\bibnamefont
  {Gilchrist}}, \bibinfo {author} {\bibfnamefont {P.}~\bibnamefont {Deuar}},\
  and\ \bibinfo {author} {\bibfnamefont {M.~D.}\ \bibnamefont {Reid}},\ }\href
  {https://doi.org/10.1103/PhysRevLett.80.3169} {\bibfield  {journal} {\bibinfo
   {journal} {Phys. Rev. Lett.}\ }\textbf {\bibinfo {volume} {80}},\ \bibinfo
  {pages} {3169} (\bibinfo {year} {1998})}\BibitemShut {NoStop}%
\bibitem [{\citenamefont {Munro}(1999)}]{munro1999OptimalStatesBellinequality}%
  \BibitemOpen
  \bibfield  {author} {\bibinfo {author} {\bibfnamefont {W.~J.}\ \bibnamefont
  {Munro}},\ }\href {https://doi.org/10.1103/PhysRevA.59.4197} {\bibfield
  {journal} {\bibinfo  {journal} {Phys. Rev. A}\ }\textbf {\bibinfo {volume}
  {59}},\ \bibinfo {pages} {4197} (\bibinfo {year} {1999})}\BibitemShut
  {NoStop}%
\bibitem [{\citenamefont {Wenger}\ \emph {et~al.}(2003)\citenamefont {Wenger},
  \citenamefont {Hafezi}, \citenamefont {Grosshans}, \citenamefont
  {{Tualle-Brouri}},\ and\ \citenamefont
  {Grangier}}]{wenger2003MaximalViolationBell}%
  \BibitemOpen
  \bibfield  {author} {\bibinfo {author} {\bibfnamefont {J.}~\bibnamefont
  {Wenger}}, \bibinfo {author} {\bibfnamefont {M.}~\bibnamefont {Hafezi}},
  \bibinfo {author} {\bibfnamefont {F.}~\bibnamefont {Grosshans}}, \bibinfo
  {author} {\bibfnamefont {R.}~\bibnamefont {{Tualle-Brouri}}},\ and\ \bibinfo
  {author} {\bibfnamefont {P.}~\bibnamefont {Grangier}},\ }\href
  {https://doi.org/10.1103/PhysRevA.67.012105} {\bibfield  {journal} {\bibinfo
  {journal} {Phys. Rev. A}\ }\textbf {\bibinfo {volume} {67}},\ \bibinfo
  {pages} {012105} (\bibinfo {year} {2003})}\BibitemShut {NoStop}%
\bibitem [{\citenamefont {Nha}\ and\ \citenamefont
  {Carmichael}(2004)}]{nha2004ProposedTestQuantum}%
  \BibitemOpen
  \bibfield  {author} {\bibinfo {author} {\bibfnamefont {H.}~\bibnamefont
  {Nha}}\ and\ \bibinfo {author} {\bibfnamefont {H.~J.}\ \bibnamefont
  {Carmichael}},\ }\href {https://doi.org/10.1103/PhysRevLett.93.020401}
  {\bibfield  {journal} {\bibinfo  {journal} {Phys. Rev. Lett.}\ }\textbf
  {\bibinfo {volume} {93}},\ \bibinfo {pages} {020401} (\bibinfo {year}
  {2004})}\BibitemShut {NoStop}%
\bibitem [{\citenamefont {Caprara~Vivoli}\ \emph {et~al.}(2015)\citenamefont
  {Caprara~Vivoli}, \citenamefont {Sekatski}, \citenamefont {Bancal},
  \citenamefont {Lim}, \citenamefont {Christensen}, \citenamefont {Martin},
  \citenamefont {Thew}, \citenamefont {Zbinden}, \citenamefont {Gisin},\ and\
  \citenamefont {Sangouard}}]{capraravivoli2015ChallengingPreconceptionsBella}%
  \BibitemOpen
  \bibfield  {author} {\bibinfo {author} {\bibfnamefont {V.}~\bibnamefont
  {Caprara~Vivoli}}, \bibinfo {author} {\bibfnamefont {P.}~\bibnamefont
  {Sekatski}}, \bibinfo {author} {\bibfnamefont {J.-D.}\ \bibnamefont
  {Bancal}}, \bibinfo {author} {\bibfnamefont {C.~C.~W.}\ \bibnamefont {Lim}},
  \bibinfo {author} {\bibfnamefont {B.~G.}\ \bibnamefont {Christensen}},
  \bibinfo {author} {\bibfnamefont {A.}~\bibnamefont {Martin}}, \bibinfo
  {author} {\bibfnamefont {R.~T.}\ \bibnamefont {Thew}}, \bibinfo {author}
  {\bibfnamefont {H.}~\bibnamefont {Zbinden}}, \bibinfo {author} {\bibfnamefont
  {N.}~\bibnamefont {Gisin}},\ and\ \bibinfo {author} {\bibfnamefont
  {N.}~\bibnamefont {Sangouard}},\ }\href
  {https://doi.org/10.1103/PhysRevA.91.012107} {\bibfield  {journal} {\bibinfo
  {journal} {Phys. Rev. A}\ }\textbf {\bibinfo {volume} {91}},\ \bibinfo
  {pages} {012107} (\bibinfo {year} {2015})}\BibitemShut {NoStop}%
\bibitem [{\citenamefont {Zhao}\ \emph {et~al.}(2019)\citenamefont {Zhao},
  \citenamefont {Cao}, \citenamefont {Zhen}, \citenamefont {Chen},
  \citenamefont {Li}, \citenamefont {Liu}, \citenamefont {Xu},\ and\
  \citenamefont {Chen}}]{zhao2019HigherAmountsLoopholefree}%
  \BibitemOpen
  \bibfield  {author} {\bibinfo {author} {\bibfnamefont {S.}~\bibnamefont
  {Zhao}}, \bibinfo {author} {\bibfnamefont {W.-F.}\ \bibnamefont {Cao}},
  \bibinfo {author} {\bibfnamefont {Y.-Z.}\ \bibnamefont {Zhen}}, \bibinfo
  {author} {\bibfnamefont {C.}~\bibnamefont {Chen}}, \bibinfo {author}
  {\bibfnamefont {L.}~\bibnamefont {Li}}, \bibinfo {author} {\bibfnamefont
  {N.-L.}\ \bibnamefont {Liu}}, \bibinfo {author} {\bibfnamefont
  {F.}~\bibnamefont {Xu}},\ and\ \bibinfo {author} {\bibfnamefont
  {K.}~\bibnamefont {Chen}},\ }\href {https://doi.org/10.1088/1367-2630/ab4538}
  {\bibfield  {journal} {\bibinfo  {journal} {New J. Phys.}\ }\textbf {\bibinfo
  {volume} {21}},\ \bibinfo {pages} {103008} (\bibinfo {year}
  {2019})}\BibitemShut {NoStop}%
\bibitem [{\citenamefont {Clauser}\ and\ \citenamefont
  {Shimony}(1978)}]{clauser1978BellTheoremExperimentala}%
  \BibitemOpen
  \bibfield  {author} {\bibinfo {author} {\bibfnamefont {J.~F.}\ \bibnamefont
  {Clauser}}\ and\ \bibinfo {author} {\bibfnamefont {A.}~\bibnamefont
  {Shimony}},\ }\href {https://doi.org/10.1088/0034-4885/41/12/002} {\bibfield
  {journal} {\bibinfo  {journal} {Rep. Prog. Phys.}\ }\textbf {\bibinfo
  {volume} {41}},\ \bibinfo {pages} {1881} (\bibinfo {year}
  {1978})}\BibitemShut {NoStop}%
\bibitem [{\citenamefont {Clauser}\ and\ \citenamefont
  {Horne}(1974)}]{clauser1974ExperimentalConsequencesObjective}%
  \BibitemOpen
  \bibfield  {author} {\bibinfo {author} {\bibfnamefont {J.~F.}\ \bibnamefont
  {Clauser}}\ and\ \bibinfo {author} {\bibfnamefont {M.~A.}\ \bibnamefont
  {Horne}},\ }\href {https://doi.org/10.1103/PhysRevD.10.526} {\bibfield
  {journal} {\bibinfo  {journal} {Phys. Rev. D}\ }\textbf {\bibinfo {volume}
  {10}},\ \bibinfo {pages} {526} (\bibinfo {year} {1974})}\BibitemShut
  {NoStop}%
\bibitem [{\citenamefont {Nielsen}\ and\ \citenamefont
  {Chuang}(2010)}]{nielsen2010QuantumComputationQuantuma}%
  \BibitemOpen
  \bibfield  {author} {\bibinfo {author} {\bibfnamefont {M.~A.}\ \bibnamefont
  {Nielsen}}\ and\ \bibinfo {author} {\bibfnamefont {I.~L.}\ \bibnamefont
  {Chuang}},\ }\href {https://doi.org/10.1017/CBO9780511976667} {\emph
  {\bibinfo {title} {Quantum {{Computation}} and {{Quantum Information}}: 10th
  {{Anniversary Edition}}}}}\ (\bibinfo  {publisher} {{Cambridge University
  Press}},\ \bibinfo {address} {{Cambridge}},\ \bibinfo {year}
  {2010})\BibitemShut {NoStop}%
\bibitem [{\citenamefont {Barnett}\ \emph {et~al.}(1998)\citenamefont
  {Barnett}, \citenamefont {Phillips},\ and\ \citenamefont
  {Pegg}}]{barnett1998ImperfectPhotodetectionProjection}%
  \BibitemOpen
  \bibfield  {author} {\bibinfo {author} {\bibfnamefont {S.~M.}\ \bibnamefont
  {Barnett}}, \bibinfo {author} {\bibfnamefont {L.~S.}\ \bibnamefont
  {Phillips}},\ and\ \bibinfo {author} {\bibfnamefont {D.~T.}\ \bibnamefont
  {Pegg}},\ }\href {https://doi.org/10.1016/S0030-4018(98)00511-2} {\bibfield
  {journal} {\bibinfo  {journal} {Optics Communications}\ }\textbf {\bibinfo
  {volume} {158}},\ \bibinfo {pages} {45} (\bibinfo {year} {1998})}\BibitemShut
  {NoStop}%
\bibitem [{\citenamefont {Hume}\ \emph {et~al.}(2013)\citenamefont {Hume},
  \citenamefont {Stroescu}, \citenamefont {Joos}, \citenamefont {Muessel},
  \citenamefont {Strobel},\ and\ \citenamefont
  {Oberthaler}}]{hume2013AccurateAtomCounting}%
  \BibitemOpen
  \bibfield  {author} {\bibinfo {author} {\bibfnamefont {D.~B.}\ \bibnamefont
  {Hume}}, \bibinfo {author} {\bibfnamefont {I.}~\bibnamefont {Stroescu}},
  \bibinfo {author} {\bibfnamefont {M.}~\bibnamefont {Joos}}, \bibinfo {author}
  {\bibfnamefont {W.}~\bibnamefont {Muessel}}, \bibinfo {author} {\bibfnamefont
  {H.}~\bibnamefont {Strobel}},\ and\ \bibinfo {author} {\bibfnamefont {M.~K.}\
  \bibnamefont {Oberthaler}},\ }\href
  {https://doi.org/10.1103/PhysRevLett.111.253001} {\bibfield  {journal}
  {\bibinfo  {journal} {Phys. Rev. Lett.}\ }\textbf {\bibinfo {volume} {111}},\
  \bibinfo {pages} {253001} (\bibinfo {year} {2013})}\BibitemShut {NoStop}%
\bibitem [{\citenamefont {H{\"u}per}\ \emph {et~al.}(2020)\citenamefont
  {H{\"u}per}, \citenamefont {P{\"u}r}, \citenamefont {Hetzel}, \citenamefont
  {Geng}, \citenamefont {Peise}, \citenamefont {Kruse}, \citenamefont
  {Kristensen}, \citenamefont {Ertmer}, \citenamefont {Arlt},\ and\
  \citenamefont {Klempt}}]{huper2020NumberresolvedPreparationMesoscopic}%
  \BibitemOpen
  \bibfield  {author} {\bibinfo {author} {\bibfnamefont {A.}~\bibnamefont
  {H{\"u}per}}, \bibinfo {author} {\bibfnamefont {C.}~\bibnamefont {P{\"u}r}},
  \bibinfo {author} {\bibfnamefont {M.}~\bibnamefont {Hetzel}}, \bibinfo
  {author} {\bibfnamefont {J.}~\bibnamefont {Geng}}, \bibinfo {author}
  {\bibfnamefont {J.}~\bibnamefont {Peise}}, \bibinfo {author} {\bibfnamefont
  {I.}~\bibnamefont {Kruse}}, \bibinfo {author} {\bibfnamefont {M.~A.}\
  \bibnamefont {Kristensen}}, \bibinfo {author} {\bibfnamefont
  {W.}~\bibnamefont {Ertmer}}, \bibinfo {author} {\bibfnamefont
  {J.}~\bibnamefont {Arlt}},\ and\ \bibinfo {author} {\bibfnamefont
  {C.}~\bibnamefont {Klempt}},\ }\bibfield  {journal} {\bibinfo  {journal} {New
  J. Phys.}\ }\href {https://doi.org/10.1088/1367-2630/abd058}
  {10.1088/1367-2630/abd058} (\bibinfo {year} {2020})\BibitemShut {NoStop}%
\bibitem [{\citenamefont {Qu}\ \emph {et~al.}(2020)\citenamefont {Qu},
  \citenamefont {Evrard}, \citenamefont {Dalibard},\ and\ \citenamefont
  {Gerbier}}]{qu2020ProbingSpinCorrelations}%
  \BibitemOpen
  \bibfield  {author} {\bibinfo {author} {\bibfnamefont {A.}~\bibnamefont
  {Qu}}, \bibinfo {author} {\bibfnamefont {B.}~\bibnamefont {Evrard}}, \bibinfo
  {author} {\bibfnamefont {J.}~\bibnamefont {Dalibard}},\ and\ \bibinfo
  {author} {\bibfnamefont {F.}~\bibnamefont {Gerbier}},\ }\href
  {https://doi.org/10.1103/PhysRevLett.125.033401} {\bibfield  {journal}
  {\bibinfo  {journal} {Phys. Rev. Lett.}\ }\textbf {\bibinfo {volume} {125}},\
  \bibinfo {pages} {033401} (\bibinfo {year} {2020})}\BibitemShut {NoStop}%
\bibitem [{Note1()}]{Note1}%
  \BibitemOpen
  \bibinfo {note} {\protect \url
  {https://github.com/kitzingj/BEC-Bell-correlations}}\BibitemShut {NoStop}%
\bibitem [{\citenamefont {Rackauckas}\ and\ \citenamefont
  {Nie}(2017)}]{rackauckas2017DifferentialEquationsJlPerformant}%
  \BibitemOpen
  \bibfield  {author} {\bibinfo {author} {\bibfnamefont {C.}~\bibnamefont
  {Rackauckas}}\ and\ \bibinfo {author} {\bibfnamefont {Q.}~\bibnamefont
  {Nie}},\ }\href {https://doi.org/10.5334/jors.151} {\bibfield  {journal}
  {\bibinfo  {journal} {Journal of Open Research Software}\ }\textbf {\bibinfo
  {volume} {5}},\ \bibinfo {pages} {15} (\bibinfo {year} {2017})}\BibitemShut
  {NoStop}%
\bibitem [{\citenamefont {Tsitouras}(2011)}]{tsitouras2011RungeKuttaPairs}%
  \BibitemOpen
  \bibfield  {author} {\bibinfo {author} {\bibfnamefont {C.}~\bibnamefont
  {Tsitouras}},\ }\href {https://doi.org/10.1016/j.camwa.2011.06.002}
  {\bibfield  {journal} {\bibinfo  {journal} {Computers \& Mathematics with
  Applications}\ }\textbf {\bibinfo {volume} {62}},\ \bibinfo {pages} {770}
  (\bibinfo {year} {2011})}\BibitemShut {NoStop}%
\end{thebibliography}
%

\end{document}